\documentclass[fleqn,usenatbib]{mnras}

\usepackage{newtxtext,newtxmath}
\newcommand{\colibre}{\textsc{colibre}}
\newcommand{\gsmf}{GSMF}
\newcommand{\gsmfs}{GSMFs}
\newcommand{\sfms}{SFMS}
\newcommand{\sfmss}{SFMSs}
\newcommand{\shmr}{SHMR}
\newcommand{\shmrs}{SHMRs}
\newcommand{\msevencolor}{dark-red}
\newcommand{\msevenhybridcolor}{yellow}
\newcommand{\msixcolor}{orange}
\newcommand{\mfivecolor}{light-blue}
\usepackage[T1]{fontenc}

\DeclareRobustCommand{\VAN}[3]{#2}
\let\VANthebibliography\thebibliography
\def\thebibliography{\DeclareRobustCommand{\VAN}[3]{##3}\VANthebibliography}

\usepackage{booktabs}
\usepackage{graphicx}	% Including figure files
\usepackage{amsmath}	% Advanced maths commands
\title[Evolution of the GSMF and SFRs in COLIBRE]{The evolution of the galaxy stellar mass function and star formation rates in the COLIBRE simulations from redshift 17 to 0}

\author[E. Chaikin et al.]{Evgenii Chaikin,$^{1,2}$\thanks{E-mail: chaikin@strw.leidenuniv.nl} 
Joop Schaye,$^{1}$ 
Matthieu Schaller,$^{3,1}$ 
Sylvia Ploeckinger,$^{4}$ 
Alejandro Ben\'{i}tez-Llambay,$^{5}$ \newauthor 
Carlos S. Frenk,$^{2}$ 
Filip Hu\v{s}ko,$^{1}$ 
Robert J. McGibbon,$^{1}$ 
Alexander J. Richings,$^{6,7}$ and
James W. Trayford$^{8}$ \\
$^{1}$Leiden Observatory, Leiden University, PO Box 9513, 2300 RA Leiden, the Netherlands \\
$^{2}$Institute for Computational Cosmology, Department of Physics, University of Durham, South Road, Durham, DH1 3LE, UK \\ 
$^{3}$Lorentz Institute for Theoretical Physics, Leiden University, PO Box 9506, 2300 RA Leiden, the Netherlands \\
$^{4}$Department of Astrophysics, University of Vienna, T\"{u}rkenschanzstrasse 17, 1180 Vienna, Austria \\
$^{5}$Dipartimento di Fisica G. Occhialini, Universit\`{a} degli Studi di Milano Bicocca, Piazza della Scienza, 3 I-20126 Milano MI, Italy \\
$^{6}$Centre for Data Science, Artificial Intelligence and Modelling, University of Hull, Cottingham Road, Hull, HU6 7RX, UK \\
$^{7}$E. A. Milne Centre for Astrophysics, University of Hull, Cottingham Road, Hull, HU6 7RX, UK \\
$^{8}$Institute of Cosmology and Gravitation, University of Portsmouth, Dennis Sciama Building, Burnaby Road, Portsmouth PO1 3FX, UK \\
}

\date{Accepted XXX. Received YYY; in original form ZZZ}

\pubyear{2015}

\begin{document}
\label{firstpage}
\pagerange{\pageref{firstpage}--\pageref{lastpage}}
\maketitle

\begin{abstract}
We investigate the evolution of the galaxy stellar mass function (\gsmf) and star formation rates (SFRs) across cosmic time in the \colibre{} simulations of galaxy formation. \colibre{} includes a multiphase interstellar medium, radiative cooling rates coupled to a model for the evolution of dust grains, and employs prescriptions for stellar and AGN feedback calibrated to reproduce the $z=0$ observed \gsmf{} and stellar mass -- size relation. We present the evolution of the \gsmf{} from simulations at three resolutions: $m_{\rm gas}\approx m_{\rm dm}\sim 10^7$, $10^6$, and $10^5~\mathrm{M_\odot}$, in cosmological volumes of up to $400^3$, $200^3$, and $100^3$~cMpc$^3$, respectively. We demonstrate that \colibre{} is consistent with the observed \gsmf{} over the full redshift range for which there are observations to compare with ($0<z<12$), with maximum systematic deviations of $\approx 0.3$~dex reached at $2<z<4$. We also examine the evolution of the star-forming main sequence, cosmic SFR density, stellar mass density, and galaxy quenched fraction, making predictions for both the fiducial \colibre{} model with thermally-driven AGN feedback and its variant with hybrid (thermal + kinetic jet) AGN feedback, and finding good agreement with observations. Notably, we show that \colibre{} matches the number density of massive quiescent galaxies at high redshifts reported by \textit{JWST}, while predicting a stellar-to-halo mass relation that evolves little with redshift. We conclude that neither a redshift-dependent star formation efficiency, nor a variable stellar initial mass function, nor a deviation from $\Lambda$CDM is necessary to reproduce the high-redshift \textit{JWST} stellar masses and SFRs. 
\end{abstract}

\begin{keywords}
methods: numerical -- galaxies: general -- galaxies: formation -- galaxies: evolution
\end{keywords}

%%%%%%%%%%%%%%%%%%%%%%%%%%%%%%%%%%%%%%%%%%%%%%%%%%

%%%%%%%%%%%%%%%%% BODY OF PAPER %%%%%%%%%%%%%%%%%%

\section{Introduction}
\label{section: introduction}

The galaxy stellar mass function (\gsmf) is a fundamental statistic for studying the evolution of galaxies over cosmic time \citep[e.g.][]{2001MNRAS.326..255C,2009ApJ...696..620C,2010A&A...523A..13P}. Defined as the comoving number density of galaxies per logarithmic stellar mass bin, $\mathrm{d}n(M_*)/\mathrm{d \, log} M_*$, the observationally inferred \gsmf{} not only provides one of the most precise probes of the galaxy population in the observable Universe, but also serves as a critical constraint for state-of-the-art numerical simulations of galaxy formation \citep[e.g.][]{2014MNRAS.438.1985T,2015MNRAS.450.1937C}. 

The shape and normalization of the \gsmf{} are intrinsically linked to those of the halo mass function (HMF), which can be predicted with remarkable accuracy over a wide range of masses, redshifts, and cosmologies \citep[e.g.][]{2001MNRAS.321..372J,2008ApJ...688..709T,2016MNRAS.456.2361B,2020Natur.585...39W}. However, the slopes at the low- and high-mass ends of the \gsmf{} deviate significantly from that of the HMF due to the halo mass-dependent efficiency with which haloes convert their share of inflowing baryonic matter into stars. At the high-mass end, the conversion efficiency is controlled by the long cooling times of hot gas and energy feedback from active galactic nuclei (AGN), while in low-mass haloes the dominant controlling mechanism is energy feedback from stellar winds, radiation, and supernovae (SNe) \citep[e.g.][]{2003ApJ...599...38B,2006MNRAS.370..645B, 2008MNRAS.391..481S,10.1093/mnras/stt1789,2015MNRAS.450.1937C,2018MNRAS.481.3573L,2019MNRAS.488.3143B}. Direct measurements, empirical modelling, and halo abundance matching and clustering  \citep[e.g.][]{2004MNRAS.355..769E,2010ApJ...710..903M,2010ApJ...717..379B} indicate that the ratio of stellar to halo mass peaks at around 3 per cent at the halo mass of $M_{\rm 200c}\sim 10^{12}~\mathrm{M_\odot}$ and drops steeply towards both higher and lower values of $M_{\rm 200c}$.

At low redshifts ($z \lesssim 1$), the \gsmf{} has been tightly constrained by large spectroscopic and photometric surveys, such as the Galaxy And Mass Assembly Survey \citep[GAMA,][]{Driver2011,2012MNRAS.421..621B,2015MNRAS.452.2087L,2022MNRAS.513..439D}, the Cosmic Evolution Survey  \citep[COSMOS,][]{2007ApJS..172....1S,2017A&A...605A..70D,2022ApJS..258...11W}, the Deep Extragalactic VIsible Legacy Survey  \citep[DEVILS,][]{2018MNRAS.480..768D,2021MNRAS.505..540T}, and, very recently, the Cosmic Dawn Survey \citep{2025arXiv250417867E,2025A&A...695A.229E}. These surveys, some of which also probe higher redshifts, have constructed galaxy samples at $z\lesssim 1$ that are complete down to stellar masses of $M_* \sim 10^7-10^9~\mathrm{M_\odot}$ (depending on redshift), encompassing up to $\sim 10^6$ objects over areas of sky of $\sim 1$ to $10^2$ deg$^2$. Multi-wavelength photometry -- spanning (far-)UV to (far-)IR wavelengths -- is typically available in $\approx 10-30$ bands per galaxy. These extensive data enable robust stellar mass determination via spectral energy distribution (SED) fitting, with systematic uncertainties generally within $\approx 0.15$~dex at $z\approx 0$ \citep[e.g.][]{2020MNRAS.495..905R}. Furthermore, very recently, \citet{2025MNRAS.540.1635X} probed the faint end of the $z\approx 0$ \gsmf{} down to $\sim 10^6~\mathrm{M_\odot}$ by applying the statistical Photometric Objects Around Cosmic-webs (PAC) method to the Dark Energy Spectroscopic Instrument (DESI; \citealt{2016arXiv161100036D}) survey data. PAC combines a large-scale (but relatively shallow) spectroscopic sample with deep photometric data to statistically associate faint galaxies (lacking spectroscopy) with nearby spectroscopic tracers, enabling redshift determination and stellar mass estimates for galaxies too faint to be targeted spectroscopically.

At higher redshifts, inferring the \gsmf{} becomes increasingly challenging due to the declining surface brightness of galaxies, which requires longer integration times to probe the low-mass end ($M_* \lesssim 10^9~\mathrm{M_\odot}$). Additionally, reliable stellar mass estimates from SED fitting require coverage of the rest-frame optical and near-IR regions of the spectrum \citep[e.g.][]{2011Ap&SS.331....1W}. At $z \gtrsim 3$, these wavelengths are redshifted into spectral bands that are difficult to observe with ground-based telescopes due to atmospheric absorption, a limitation that can only be circumvented with space-based observatories.

Prior to \textit{JWST}, estimates of the high-$z$ \gsmf{} were primarily based on data collected by space-based (far-)IR instruments such as \textit{Spitzer}/IRAC and \textit{HST}/WFC3, often supported by additional observations from ground-based facilities. For instance, \citet{2009ApJ...702.1393K} and \citet{2012A&A...538A..33S} utilized deep \textit{Spitzer}/IRAC and \textit{HST}/WFC3 IR imaging to estimate the \gsmf{} up to $z \approx 3$ and $z \approx 4.5$, respectively. Similarly, \citet{2014MNRAS.444.2960D}, \citet{2016ApJ...825....5S}, and \citet{2021ApJ...922...29S} combined data from \textit{HST}/ACS, \textit{HST}/WFC3, and \textit{Spitzer}/IRAC to derive the \gsmf{} for redshift ranges of $z \approx 4-7$, $z \approx 4-8$, and $z \approx 6-10$, respectively. These studies relied on galaxy samples of $\sim 10^3-10^4$ objects, covering areas of less than 1 deg$^2$, with most galaxies lacking spectroscopically confirmed redshifts.

\begin{table*}
\caption{\colibre{} simulations used in this work. Column (1): simulation name; column (2): size of the cosmological volume per dimension; column (3): the mean initial gas particle mass; column (4): the mean DM particle mass; column (5): number of gas particles in the initial conditions; column (6): number of DM particles in the initial conditions; column (7): Plummer-equivalent gravitational softening length used for baryons and DM; column (8): AGN feedback model used in the simulation; column (9): the redshift down to which the simulation has been run at the time of writing.}
	\centering
	\begin{tabular}{lrrrrrrrlllr} % four columns, alignment for each
   \hline
   Name & $L_{\rm box}$ [cMpc] & $m_{\rm gas}$ [$\rm M_\odot$]  & $m_{\rm dm}$ [$\rm M_\odot$] & $N_{\rm gas}$ & $N_{\rm dm}$  & $\varepsilon_{\rm soft}$ & AGN feedback model & $z$\\
	    \hline
    \\
    \multicolumn{2}{l}{\textbf{m7 resolution}}  \\
     \hline
    L400m7 & $400$ & $1.47 \times 10^7$  & $1.94 \times 10^7$ & $3008^3$ & $4\times 3008^3$ & $\mathrm{min(1.4 \, pkpc, 3.6   \,ckpc)}$ & thermal & $0$ \\
    L200m7 & $200$ & $1.47 \times 10^7$  & $1.94 \times 10^7$ & $1504^3$ & $4\times 1504^3$ & $\mathrm{min(1.4 \, pkpc, 3.6   \,ckpc)}$ & thermal & $0$  \\
    L100m7 & $100$ & $1.47 \times 10^7$  & $1.94 \times 10^7$ & $752^3$ & $4\times 752^3$ & $\mathrm{min(1.4 \, pkpc, 3.6   \,ckpc)}$ & thermal & $0$  \\
    L050m7 & $50$ & $1.47 \times 10^7$  & $1.94 \times 10^7$ & $376^3$ & $4\times 376^3$ & $\mathrm{min(1.4 \, pkpc, 3.6   \,ckpc)}$ & thermal & $0$  \\
    L025m7 & $25$ & $1.47 \times 10^7$  & $1.94 \times 10^7$ & $188^3$ & $4\times 188^3$ & $\mathrm{min(1.4 \, pkpc, 3.6   \,ckpc)}$ & thermal & $0$  \\ \\
    L200m7h & $200$ & $1.47 \times 10^7$  & $1.94 \times 10^7$ & $1504^3$ & $4\times 1504^3$ & $\mathrm{min(1.4 \, pkpc, 3.6 \, ckpc)}$  & hybrid & $0$  \\
    \\
    \multicolumn{2}{l}{\textbf{m6 resolution}}  \\
    \hline
    L200m6 & $200$ & $1.8 \times 10^6$  & $2.4 \times 10^6$ & $3008^3$ & $4\times 3008^3$  & $\mathrm{min(0.7 \, pkpc, 1.8  \,ckpc)}$  & thermal & $0$ \\
    L100m6 & $100$ & $1.8 \times 10^6$  & $2.4 \times 10^6$ & $1504^3$ & $4\times 1504^3$  & $\mathrm{min(0.7 \, pkpc, 1.8  \,ckpc)}$  & thermal & $0$  \\
    L050m6 & $50$ & $1.8 \times 10^6$  & $2.4 \times 10^6$ & $752^3$ & $4\times 752^3$  & $\mathrm{min(0.7 \, pkpc, 1.8  \,ckpc)}$  & thermal & $0$  \\
    L025m6 & $25$ & $1.8 \times 10^6$  & $2.4 \times 10^6$ & $376^3$ & $4\times 376^3$  & $\mathrm{min(0.7 \, pkpc, 1.8  \,ckpc)}$  & thermal & $0$  \\
    \\
    \multicolumn{2}{l}{\textbf{m5 resolution}}  \\
    \hline
    L100m5 & $100$ & $2.3 \times 10^5$  & $3.0 \times 10^5$ & $3008^3$ & $4\times 3008^3$ & $\mathrm{min(0.35 \, pkpc, 0.9 \, ckpc)}$  & thermal & $4.6$ \\
    L050m5 & $50$ & $2.3 \times 10^5$  & $3.0 \times 10^5$ & $1504^3$ & $4\times 1504^3$ & $\mathrm{min(0.35 \, pkpc, 0.9 \, ckpc)}$  & thermal & $0.9$ \\
    L025m5 & $25$ & $2.3 \times 10^5$  & $3.0 \times 10^5$ & $752^3$ & $4\times 752^3$ & $\mathrm{min(0.35 \, pkpc, 0.9 \, ckpc)}$  & thermal & $0$ \\
    \hline
\end{tabular}
\label{table: simulations}
\end{table*}

The unprecedented IR sensitivity and resolution of \textit{JWST} have prompted a re-evaluation of how stellar mass accumulated in the high-redshift Universe. In particular, \textit{JWST} observations at $z \gtrsim 4$ reveal a more rapid growth of massive galaxies than previously anticipated \citep[e.g.][]{2023A&A...677A..88B,2024MNRAS.533.1808W,2024MNRAS.534..325C,2025ApJ...988L..35W}, suggesting that galaxy formation efficiencies $\gtrsim 30$ per cent may be necessary to reach the observed stellar masses ($M_* \gtrsim 10^{10}~\mathrm{M_\odot}$) within the age of the Universe at the observed redshift \citep[e.g.][]{2023ApJS..265....5H,2024ApJ...965...98C}. The SEDs of these massive galaxies often indicate significant dust content and suggest that some of these galaxies are already quenched (i.e. no longer actively forming stars) at those high redshifts -- a challenge that most theoretical models struggle to reproduce \citep[e.g.][]{2023ApJ...947...20V,2025MNRAS.539..557B,2025ApJ...983...11W,2025NatAs...9..280D,2025MNRAS.536.2324L}. Some of these conclusions, however, may be affected by significant uncertainties in stellar mass estimates (within $\approx 0.5$~dex, e.g. \citealt{2025ApJ...978L..42C}, \citealt{2025ApJ...988L..35W}), cosmic variance due to the limited surface areas probed by high-$z$ surveys, and the reliance on photometric redshifts for most galaxies.

The pivotal role of the \gsmf{} in tracing the buildup of stellar mass and the hierarchical assembly of galaxies in the Universe, combined with the wealth of observational data available at both low and high redshifts, has established the \gsmf{} as one of the most widely used observables for calibrating and validating galaxy formation models. Numerous galaxy formation simulations, including \textsc{Illustris} \citep{2014MNRAS.444.1518V}, \textsc{eagle} \citep{2015MNRAS.446..521S}, \textsc{bahamas} \citep{2017MNRAS.465.2936M}, \textsc{IllustrisTNG} \citep{2018MNRAS.473.4077P}, \textsc{Simba} \citep{2019MNRAS.486.2827D}, and \textsc{flamingo} \citep{2023MNRAS.526.4978S}, as well as some semi-analytic models \citep[e.g.][]{2008MNRAS.391..481S,2011MNRAS.413..101G,2012MNRAS.423.1992S,2015MNRAS.451.2663H,2018MNRAS.481.3573L,2024MNRAS.531.3551L}, have been calibrated (via parameter tuning) to reproduce the $z\approx 0$ \gsmf{} or used it as a loose constraint on the model parameters. In contrast, high-redshift ($z\gtrsim 3$) \gsmfs{} are typically predictions of these simulations and semi-analytic models, as they are rarely adopted as observational constraints during calibration. This makes high-$z$ \gsmfs{} a powerful test of the underlying physics in galaxy formation models, particularly when compared to recent observations from \textit{JWST}.

We emphasize, however, that while galaxy stellar masses -- and thus the \gsmf{} -- can be directly predicted in simulations, they are not directly measurable in observations. Instead, they must be inferred from photometric and/or spectroscopic data using stellar population synthesis (SPS) models, which depend on several assumptions, including the stellar initial mass function (IMF), dust attenuation, metallicity, and star formation history (SFH), each of which can introduce systematic uncertainties \citep[e.g.][]{2012MNRAS.422.3285P,2015ApJ...808..101M,lee2025resolvedstellarmassestimation,2025MNRAS.540.2703B}.

In this work, we use the new \colibre{} cosmological hydrodynamical simulations of galaxy formation \citep{2026COLIBREproject,2026COLIBREcalibration,2026MNRAS.547ag324H} to investigate the buildup of stellar mass in the Universe across cosmic time. We compare simulation predictions with observational data from $z = 0$ up to the highest redshift with available observations ($z \approx 12$ for the \gsmf) and make predictions out to $z = 17$. Unlike previous galaxy formation models applied to cosmological volumes of comparable size, \colibre{} includes a multiphase interstellar medium (ISM) with non-equilibrium chemistry for hydrogen and helium \citep[][]{2025arXiv250615773P}, employs a detailed model for the formation and evolution of dust grains coupled to the chemistry \citep{2026MNRAS.545f2040T}, and suppresses spurious transfer of energy from dynamically-hot dark matter (DM) to dynamically-cooler stars by using four times as many DM particles as baryonic particles. In addition to the \gsmf, we examine several other statistics closely tied to stellar mass evolution: the cosmic star formation rate and stellar mass densities, the stellar-to-halo mass relation (\shmr), the star-forming main sequence (\sfms), the galaxy quenched fraction -- stellar mass relation, and the number density of massive quiescent galaxies in the high-redshift Universe. The paper is structured as follows: Section \ref{section: methods} details the simulation methodology, Section \ref{section: results} presents our findings, and Section \ref{section: conclusions} summarizes our key conclusions.

\section{Methods}
\label{section: methods}

\subsection{The COLIBRE simulations}

We use the \colibre{} simulations of galaxy formation, detailed in \citet{2026COLIBREproject}. All simulations were run using the astrophysical code \textsc{Swift} \citep{2024MNRAS.530.2378S}, which utilizes task-based parallelism. Gas hydrodynamics is modelled with the density-energy smoothed particle hydrodynamics (SPH) scheme \textsc{Sphenix} \citep{2022MNRAS.511.2367B}. Gravity is solved using a Fourier-space particle-mesh method for long-range forces and the Fast Multipole Method for short-range forces, employing fixed softening lengths for baryons and DM.

The initial conditions for the \colibre{} simulations were generated at $z=63$ by the \textsc{monofonIC} code \citep{2020ascl.soft08024H,2021MNRAS.500..663M} using second-order Lagrangian perturbation theory. The assumed cosmology for all simulations is the $\Lambda$CDM `3x2pt + all external constraints' cosmology from \citet{2022PhRvD.105b3520A}: $\Omega_{\rm m,0} = 0.306$, $\Omega_{\rm b, 0} = 0.0486$, $\sigma_8 = 0.807$, $h = 0.681$, $n_{s} = 0.967$, alongside a single massive neutrino species with a mass of $0.06$ eV.

The \colibre{} simulations are available at three resolutions: m7 (gas and DM particle masses of $m_{\rm gas} = 1.47 \times 10^7~\mathrm{M_\odot}$ and $m_{\rm dm} = 1.94 \times 10^7~\mathrm{M_\odot}$, respectively), m6 ($m_{\rm gas} = 1.8 \times 10^6~\mathrm{M_\odot}$, $m_{\rm dm} = 2.4 \times 10^6~\mathrm{M_\odot}$), and m5 ($m_{\rm gas} = 2.3 \times 10^5~\mathrm{M_\odot}$, $m_{\rm dm} = 3.0 \times 10^5~\mathrm{M_\odot}$). In contrast to previous cosmological simulations, which generally used similar numbers of DM particles and gas resolution elements in the initial conditions (with the DM particle mass of \mbox{$m_{\rm dm} = [(\Omega_{\rm m,0} - \Omega_{\rm b,0})/\Omega_{\rm b,0}] \, m_{\rm gas} \approx 5 \, m_{\rm gas}$}), \colibre{} uses four times as many DM particles as baryonic particles to reduce spurious energy transfer from DM to stars, which can have adverse effects on galaxy stellar components \citep{2019MNRAS.488.3663L, 2021MNRAS.508.5114L,2023MNRAS.525.5614L,2023MNRAS.519.5942W}. At m7 resolution, the Plummer-equivalent gravitational softening length of baryons and DM is set to the minimum of $1.4$~proper kpc (pkpc) and $3.6$~comoving kpc (ckpc), while at m6 and m5 resolutions it is equal to $\rm min(0.7 \, pkpc, 1.8 \, ckpc)$ and $\rm min(0.35 \, pkpc, 0.9 \, ckpc)$, respectively. 

In this work, we use the \colibre{} simulations at m7 resolution with comoving volumes of $400^3$ and $200^3$~cMpc$^3$, at m6 resolution with a volume of $200^3$~cMpc$^3$, and at m5 resolution with $100^3$, $50^3$, and $25^3$~cMpc$^3$. Additionally, to assess the separate convergence with resolution and with cosmological volume (see Appendix \ref{appendix:convergence}), we use smaller volumes at m7 and m6 resolutions: $100^3~\mathrm{cMpc}^3$, $50^3~\mathrm{cMpc}^3$, and $25^3~\mathrm{cMpc}^3$. Table \ref{table: simulations} provides a summary of all simulations used in this study. Unless specified otherwise, we use the largest available simulation at each resolution\footnote{At the time of writing, the m5 simulation in the $100^3$ ($50^3$) cMpc$^3$ volume is at $z = 4.6$ ($z = 0.9$). Therefore, unless stated otherwise, results at m5 resolution are shown from the largest available volume at a given redshift, while time-evolving quantities are interpolated around $z = 4.6$ and $z = 0.9$, switching from L100m5 to L050m5 and from L050m5 to L025m5, respectively.}.

\subsubsection{The subgrid model}

The radiative cooling and heating rates for primordial elements and their free electrons are computed using the non-equilibrium thermochemistry solver \textsc{chimes} \citep{2014MNRAS.440.3349R,2014MNRAS.442.2780R}. The rates for nine metals -- C, N, O, Ne, Mg, Si, S, Ca, and Fe -- are provided by \textsc{hybrid-chimes} \citep{2025arXiv250615773P}, which are based on species fractions calculated by \textsc{chimes} under the assumption of ionization equilibrium and steady state chemistry, but rescaled to account for the difference between the non-equilibrium and equilibrium free electron number densities. The radiative cooling and heating rates account for the presence of a redshift-dependent homogeneous radiation background from distant galaxies and quasars and -- in the interstellar medium -- an interstellar radiation field (ISRF), cosmic rays, dust, and shielding. Because \colibre{} does not include radiative transfer calculations, the strength of the ISRF, the cosmic ray rate, and the shielding column density are determined by a density- and temperature-dependent characteristic length scale, based on the local Jeans length. The details of the combined non-equilibrium and quasi-equilibrium radiative cooling model used in \colibre{} can be found in \citet{2025arXiv250615773P}.

The gas chemical composition is tracked by modelling the abundances of 12 individual elements\footnote{This set of elements differs from those used in the prescription for gas radiative cooling. In particular, Sr, Ba, and Eu are not included in the cooling calculations, as their contributions to the total cooling rate are negligible. Conversely, Ca and S -- elements not tracked in the \colibre{} chemistry but required for the radiative cooling -- are assumed to have solar mass ratios relative to Si \citep{doi:10.1146/annurev.astro.46.060407.145222}.}: H, He, C, N, O, Ne, Mg, Si, Fe, Sr, Ba, and Eu. These abundances are diffused among SPH neighbours using a velocity shear-based subgrid model for turbulent mixing \citep{2026MNRAS.tmp..607C}. Additionally, the \colibre{} simulations include a subgrid model for the evolution of dust grains, which is coupled to the chemistry and non-equilibrium cooling rates \citep{2026MNRAS.545f2040T}. Graphite and silicate grains are produced by asymptotic giant branch (AGB) stars and SNe. Dust grains grow by accreting mass from the gas phase, are destroyed via thermal sputtering and shocks, and undergo size modifications through shattering and coagulation. Two grain sizes are explicitly modelled: $0.01$ and $0.1$ $\mu$m. The dust model is coupled to the \textsc{chimes} solver, where evolving dust abundances influence gas heating and cooling rates and facilitate the formation of molecular hydrogen.

Following \citet{2024MNRAS.532.3299N}, star-forming gas particles are identified using the gravitational instability criterion. To assign an SFR to these particles, we apply the \citet{1959ApJ...129..243S} law, assuming a fixed star formation efficiency per free-fall time of $\varepsilon=0.01$. Newly formed stellar particles represent simple stellar populations characterized by a \citet{2003PASP..115..763C} IMF. Stellar particles enrich their surrounding gas with metals through six chemical enrichment channels: AGB stars, type-Ia SNe, type-II SNe, neutron star mergers, common envelope jet SNe, and collapsars \citep[see][for further details]{2026MNRAS.tmp..607C}.

Energy feedback from core-collapse supernovae (CC SNe) is implemented using a stochastic thermal-kinetic feedback prescription following \citet{2023MNRAS.523.3709C} with two modifications. First, the constant heating temperature of $\Delta T_{\rm SN} = 10^{7.5}$ K is replaced by a density-dependent function, $\Delta T_{\rm SN}(\rho_{\rm gas})\propto \rho_{\rm gas}^{2/3}$, where $\rho_{\rm gas}$ is the density of the gas around the SN estimated at the time of feedback. Second, the constant energy per single CC SN (in units of $10^{51}$ erg), $f_{\rm E}$, is replaced by a pressure-dependent function, $f_{\rm E}(P_{\rm birth})$, where $P_{\rm birth}$ represents the thermal pressure of the parent gas particles measured at the time when they were converted into stellar particles. For details, see \citet{2026COLIBREproject}.

In addition to CC SN feedback, \colibre{} incorporates stochastic thermal energy feedback from type-Ia SNe, applying the same implementation used for CC SN thermal energy injections, but with $f_{\rm E} = 1$. The number of individual type-Ia SNe per stellar particle is determined by an exponential delay time distribution function with a normalization of $1.54 \times 10^{-3}$ individual type-Ia SNe per solar mass and an exponential time-scale of $2$ Gyr (see Nobels et al., in preparation, for details).

\colibre{} also includes three early stellar feedback processes: H \textsc{ii} regions, stellar winds, and radiation pressure \citep{2026MNRAS.546ag268B}. These processes are implemented stochastically, similarly to the energy feedback from SNe. 

Supermassive black holes (SMBHs) are represented by black hole (BH) particles, seeded at high redshift by converting the densest gas particle in a halo into a BH particle. Seeding occurs when the halo mass, calculated on-the-fly using the Friends-of-Friends (FoF) algorithm, exceeds a threshold of $5 \times 10^{10}~\mathrm{M_\odot}$ at m7 resolution or $10^{10}~\mathrm{M_\odot}$ at m6 and m5 resolutions, provided the FoF halo does not already contain a BH particle. The mass accretion rate onto BHs is computed using the modified Bondi-Hoyle-Lyttleton formula, with turbulence and vorticity corrections from \citet{Krumholz_et_al_2006}. Dynamical friction and BH-BH mergers are modelled following the prescriptions from \citet{2022MNRAS.516..167B} and \citet{2026COLIBREproject}.

At each resolution, the \colibre{} simulations include models with purely thermal AGN feedback and hybrid AGN feedback (combining kinetic jets and thermal energy injections), with the largest \colibre{} volumes available only for the thermal models \citep[see table 2 in][]{2026COLIBREproject}. The primary focus of this work is on the \colibre{} simulations with purely thermal AGN feedback, though we provide a comparison between the hybrid and thermal models in $\S$\ref{subsection:thermal_vs_hybrid}.

Thermal AGN feedback is implemented deterministically, following \citet{2009MNRAS.398...53B}, but with BH particles heating their neighbouring gas by a temperature increment proportional to the BH (subgrid) mass, $\Delta T_{\rm AGN} \propto m_{\rm BH}$, rather than by a fixed temperature increment. The BH mass-dependent heating ensures that the frequency of AGN feedback events remains independent of BH mass at a fixed Eddington fraction, and that at the highest \colibre{} resolution, m5, the BH mass does not increase too steeply with galaxy stellar mass (see \citealt{2026COLIBREcalibration} for details). 

The hybrid AGN model detailed in \citet{2026MNRAS.547ag324H}, additionally tracks the evolution of BH spin and distinguishes between three accretion states: the thick, thin and slim discs. The spin of each BH particle evolves over time due to gas accretion, BH-BH mergers, and jet-induced spindown. The kinetic jet feedback mode is implemented deterministically: once a BH particle has accreted a mass whose energy equivalent exceeds that of a single jet event, two neighbouring gas particles are kicked in opposite directions, nearly aligned with the BH spin (within $7.5$ degrees of the spin axis), with the jet kick velocity proportional to the square root of the BH (subgrid) mass, such that the injected energy scales linearly with BH mass. The thermal channel for energy injection is implemented identically to that in the purely thermal AGN feedback model, but with a spin-dependent radiative efficiency for the thin disc, while separate terms are used to represent the effects of winds in the thick and slim disc states. The amounts of energy available for injection through the thermal and kinetic jet channels are determined based on the BH's instantaneous accretion rate, expressed as a fraction of the Eddington rate, and its spin. In both the thermal and hybrid AGN feedback models, the BH accretion rate is capped at 100 times the Eddington rate, allowing for super-Eddington accretion events.

\subsubsection{Calibration}
\label{subsection: calibration}

As described in detail in \citet{2026COLIBREcalibration}, the SN and AGN energy feedback in the \colibre{} simulations has been calibrated to reproduce the observed $z = 0$ \gsmf{} from GAMA DR4 \citep{2022MNRAS.513..439D} and the observed $z \approx 0$ galaxy stellar mass -- size relation reported by \citet{2022MNRAS.509.3751H}. For the \colibre{} models with thermal AGN feedback, the calibration involved adjusting the strength and functional form of SN and AGN energy feedback. At m7 resolution, this was achieved by training Gaussian process emulators on $\sim 200$ simulations in a  ($50$~cMpc)$^{3}$ volume with varied SN and AGN feedback parameter values. The trained emulators were then fit to observational data in the stellar mass range $10^9 < M_*/\mathrm{M_\odot} < 10^{11.3}$ to infer the best-fitting parameter values. At m6 and m5 resolutions, the calibration was performed through small manual adjustments of the subgrid parameters for SN and AGN feedback, relative to their best-fitting values determined using the emulators at m7 resolution. Additionally, and independently from the calibration of all other feedback parameters, the coupling efficiency of AGN feedback was tuned to match the $z=0$ observed BH mass -- stellar mass relation of massive galaxies for which dynamical measurements of BH masses are possible.

For the \colibre{} models incorporating hybrid AGN feedback, AGN feedback parameters were manually tuned at each resolution to match the same calibration data as for the thermal models, as well as the $z=0.2$ AGN bolometric luminosity function \citep[see][]{2026MNRAS.547ag324H}. Among the SN feedback parameters, only the normalization of the $\Delta T_{\rm SN}(\rho_{\rm gas}) \propto \rho_{\rm gas}^{2/3}$ scaling was adjusted, while all other SN feedback parameters retained the same values as in the \colibre{} models with thermal AGN feedback at the corresponding resolution. 
\subsection{Halo finding and calculation of galaxy properties}
\label{subsection: halo_finding_and_galaxy_properties}

We identify subhaloes using HBT-HERONS\footnote{\url{https://hbt-herons.strw.leidenuniv.nl/}} \citep{2025MNRAS.543.1339F}, which is an updated version of the Hierarchical Bound Tracing algorithm (HBT+) originally presented by \citet{2018MNRAS.474..604H}. HBT-HERONS identifies structures by applying an iterative unbinding algorithm to particles within spatial FoF groups, starting at the highest-redshift simulation output. Once self-bound objects are identified, their particles are tracked across subsequent simulation outputs to trace the evolution of these objects forward in time. This tracking improves the identification of satellite subhaloes, as particles remain associated with their subhaloes even after the subhalo is accreted into a larger host halo. Each candidate substructure undergoes additional checks for self-boundness and phase-space consistency to determine its status -- whether it remains resolved, has merged with another substructure, or has been disrupted.

We further process the output of HBT-HERONS using the Spherical Overdensity and Aperture Processor (SOAP; \citealt{McGibbon_2025}) to obtain a comprehensive list of subhalo and galaxy properties in various apertures. This list includes, in particular, galaxy stellar masses and star formation rates (SFRs) used in this work. For a given subhalo, 
\begin{itemize}
    \item We use the stellar mass computed by summing up the masses of individual stellar particles that are both gravitationally bound to the subhalo and located within 3D spherical apertures of radius 50~pkpc. The aperture is centred on the position of the particle with the largest binding energy within the subhalo. We chose the 50~pkpc aperture because \citet{2022MNRAS.511.2544D} found it best reproduces masses inferred from virtual observations of the \textsc{eagle} simulations. We discuss the impact of the aperture choice on the \gsmf{} in Appendix \ref{appendix: apertures}, finding a noticeable effect only for stellar masses $M_* \gtrsim 10^{11}~\mathrm{M_\odot}$.
    
    \item The SFR is calculated by summing the instantaneous SFRs of gas particles that are both bound to the subhalo and within a 50~pkpc 3D spherical aperture\footnote{We verified that the results of this work are largely insensitive to whether galaxy SFRs are computed using all star-forming gas particles within a 50~kpc aperture or only the gravitationally bound particles within that same distance.}. Additionally, in Appendix \ref{appendix_sfr}, we explore the impact of using 10- and 100-Myr averaged SFRs instead of instantaneous SFRs, finding generally minor differences at all redshifts.

    \item The specific star formation rate (sSFR) is calculated by dividing the subhalo's instantaneous SFR by its stellar mass, both measured at the same time and within the same apertures.
\end{itemize}
We use SOAP outputs at integer redshifts from $z = 20$ to $z = 12$; at integer and half-integer redshifts from $z = 12$ to $z = 10$; at integer, half-, and quarter-integer redshifts from $z = 10$ to $z = 1$; and at lower redshifts with a step of $\Delta z = 0.05$, except at $z = 0.15$, where we instead use $z = 0.14$ due to the structure of the \colibre{} output redshifts.

\subsection{Assumptions in the analysis}
\label{subsection:assumptions}

\begin{figure*}
    \centering
    \includegraphics[width=0.99\textwidth]{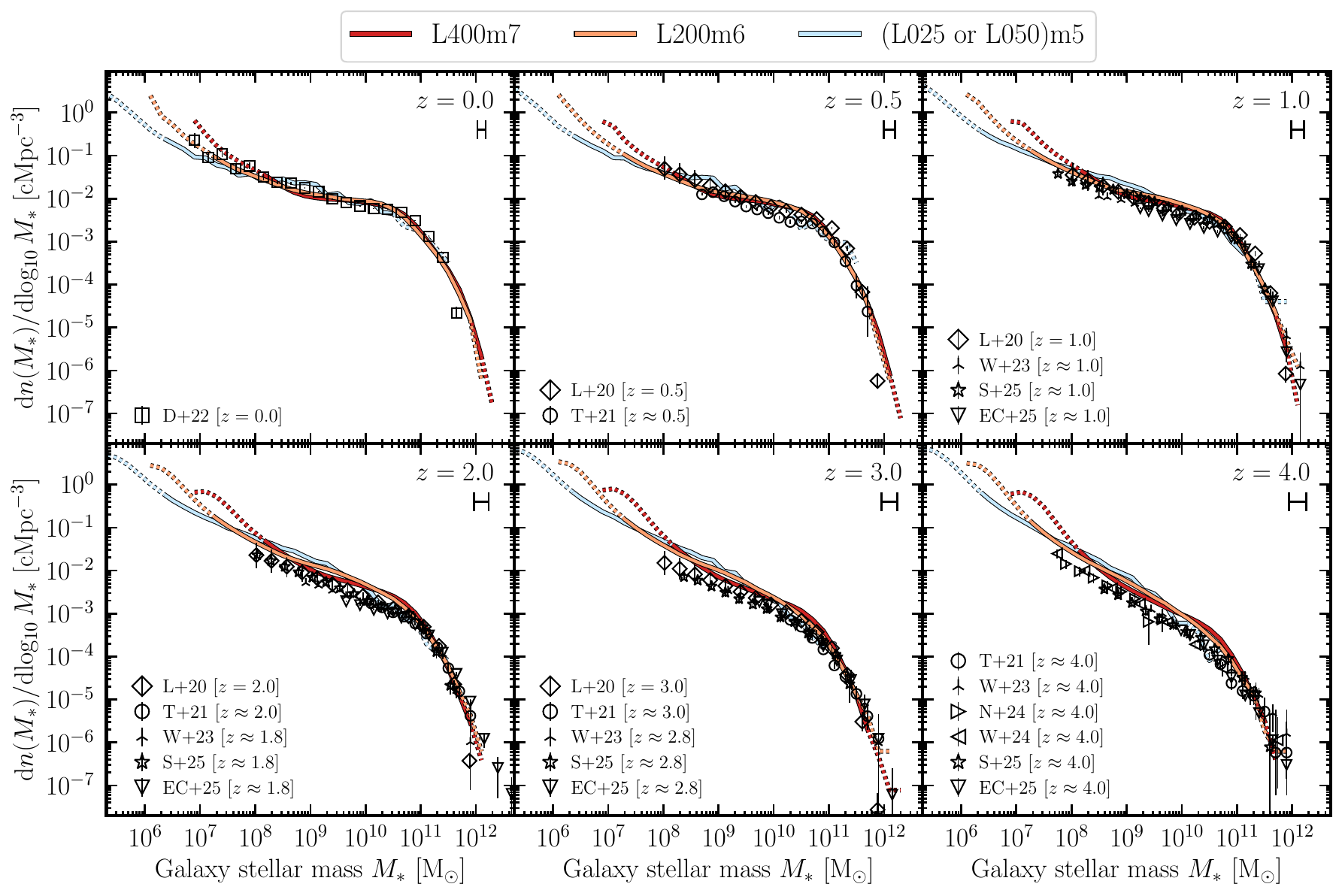}
    \caption{Evolution of the galaxy stellar mass function (\gsmf) from $z = 0$ (\textit{top left}) to $z = 4$ (\textit{bottom right}) in the \colibre{} m7, m6, and m5 simulations using the largest cosmological volumes available (see Table \ref{table: simulations}), with solid \msevencolor, \msixcolor, and \mfivecolor{} curves representing each resolution, respectively. To highlight resolution and volume limitations, the solid lines become dotted at stellar masses where galaxies are poorly resolved ($M_*<10$ times the mean initial baryonic particle mass) or where the number of galaxies per $0.2$~dex mass bin is less than $10$. A lognormal scatter with a standard deviation given by equation~(\ref{eq: random_scatter}) is added to galaxy stellar masses to account for Eddington bias. For reference, the (full extent of the) horizontal black error bar in each panel indicates the typical systematic uncertainty in stellar mass measurements, as given by equation~(\ref{eq: sys_error}). For comparison, we include a compilation of observational data from \citet{2020ApJ...893..111L}, \citet{2021ApJ...922...29S}, \citet{2021MNRAS.505..540T}, \citet{2022MNRAS.513..439D}, \citet{2023A&A...677A.184W}, \citet{2024MNRAS.533.1808W}, \citet{2024ApJ...961..207N}, \citet{2025A&A...695A..20S}, and \citet{2025arXiv250417867E} (black symbols). \colibre{} is consistent with the selected observational data, given the systematic uncertainties on observationally inferred stellar masses, and achieves very good convergence with resolution at all explored redshifts.}
    \label{fig:gsmf_evolution_lowz}
\end{figure*}

Throughout this work, we assume a \citet{2003PASP..115..763C} IMF. All observational data based on \citet{1955ApJ...121..161S} or \citet{2001MNRAS.322..231K} IMFs that we include for comparison with the simulations are converted to the \citet{2003PASP..115..763C} IMF using the conversion factors for SFRs and stellar masses provided by \citet{2014ARA&A..52..415M}. Additionally, we account for differences in the assumed value of the dimensionless Hubble parameter $h$ between cosmologies used in the comparison data, $h_{\rm obs}$, and in \colibre, $h=0.681$. This is done by applying the scaling factor $(h_{\rm obs}/0.681)^{\alpha}$ to the reported observational volumes, stellar masses, and SFRs, where the exponent $\alpha$ is chosen according to each quantity's dependence on cosmological distances. 

Random errors in observationally inferred stellar masses, arising from the application of SED-fitting algorithms to broad-band photometric data and uncertainties in SPS modelling, are estimated to be $\approx 0.1-0.2$~dex at $z \lesssim 1$ \citep[e.g.][]{2009ApJS..185..253G,2011MNRAS.418.1587T,2014ApJS..214...15S,2015ApJ...808..101M,2020MNRAS.495..905R,10.1093/mnras/stac1667,lee2025resolvedstellarmassestimation}, and $\approx 0.2-0.3$~dex or higher at $z \gtrsim 1$ \citep[e.g.][]{2009ApJ...699..486C,2023ApJ...942L..27S,2023ApJ...944..141P,2025ApJ...978L..42C}. These random errors affect the shape of the observed \gsmf{} through the \cite{1913MNRAS..73..359E} bias, particularly at the high-mass end. Due to the steep slope of the \gsmf{} at high $M_*$, there are significantly more low-mass galaxies than high-mass ones, leading to a greater upward scatter of random errors in stellar mass of low-mass objects compared to the downward scatter of high-mass objects. As we will explore in Appendix \ref{appendix: effect of eddington bias}, this effect has an impact on any statistic related to galaxy stellar mass, including the \gsmf, the \sfms, the galaxy quenched fraction, and the number density of massive quiescent galaxies. 

To account for this Eddington bias, we add a lognormal scatter to galaxy stellar masses in the simulations, with a mean of zero and a redshift-dependent standard deviation of $\log_{10} M_*$ following the parametrization of \citet{2019MNRAS.488.3143B},

\begin{equation}
\label{eq: random_scatter}
    \sigma_{\rm random}(z) = \min(\sigma_{\mathrm{random},0} + \sigma_{\mathrm{random},z} \, z, \sigma_\mathrm{random,cap})
    \, \mathrm{dex} \, .
\end{equation}
Based on the values of random errors reported in the literature, we choose $\sigma_{\mathrm{random},0} = 0.1$, $\sigma_{\mathrm{random},z} = 0.1$, and $\sigma_\mathrm{random,cap}=0.3$. Unless stated otherwise, this scatter is included in all subsequent figures presenting statistics that depend on $M_*$. In particular, we note the added scatter also affects galaxy sSFR values, as $M_*$ appears in the denominator of the sSFR definition.

In addition to random errors, SED-fitting codes introduce \mbox{\textit{systematic}} uncertainties in the inferred stellar masses, as the assumptions underlying SED modelling -- most notably the choice of SPS model, SFH parameterization, and dust attenuation -- affect \textit{all} galaxies within a given sample. By applying different SED-fitting pipelines with varying assumptions to the same data, systematic errors have been estimated to be $\approx 0.1-0.3$ dex for $0 \lesssim z\lesssim 5$ \citep{2009ApJ...701.1839M,2019ApJ...877..140L,2020MNRAS.492.5592K,2020MNRAS.495..905R,10.1093/mnras/stac1667,lee2025resolvedstellarmassestimation,2025A&A...695A..20S}. At higher redshifts, where photometric data are more limited, these estimates become progressively more uncertain \citep[e.g.][]{2023ApJ...949L..18P, 2024ApJ...961...73N,2025ApJ...978...89H,2025ApJ...988L..35W}. However, \citet{2025ApJ...978L..42C} showed that even when relying solely on \textit{JWST}/NIRCam photometry, systematic errors remain within $\approx 0.4$~dex up to $z \approx 10$; though these estimates neglect the effect of IMF choice, which can introduce additional systematic errors \citep[e.g.][]{10.1093/mnras/stae651}.

Unlike the Eddington bias, which primarily affects the \gsmf{} by \textit{increasing} the number density of massive galaxies while having a negligible impact at low $M_*$, systematic errors can shift the \gsmf{} in \textit{either direction} — leftward or rightward. Therefore, we do not add a systematic error to the simulated galaxy stellar masses but instead show it for reference in the relevant plots (see below), assuming the parametrization,

\begin{equation}
\label{eq: sys_error}
    \sigma_{\rm sys}(z) = \min(\sigma_{\mathrm{sys},0} + \sigma_{\mathrm{sys},z} \, \sqrt{z}, \sigma_{\mathrm{sys,cap}})
    \,  \mathrm{dex}  \, ,
\end{equation}
where $\sigma_{\mathrm{sys},0} = 0.15$, $\sigma_{\mathrm{sys},z}=0.08$, and $\sigma_{\mathrm{sys,cap}}=0.4$ based on the values reported in the literature. Finally, we note that systematic errors can vary with galaxy stellar mass by up to $\approx 0.15$ dex relative to the average offset (see, e.g., fig. 3 in \citet{2019ApJ...877..140L} and fig. F1 in \citealt{2025A&A...695A..20S}), an effect that we do not incorporate into equation~(\ref{eq: sys_error}).

\begin{figure*}
    \centering
    \includegraphics[width=0.99\textwidth]{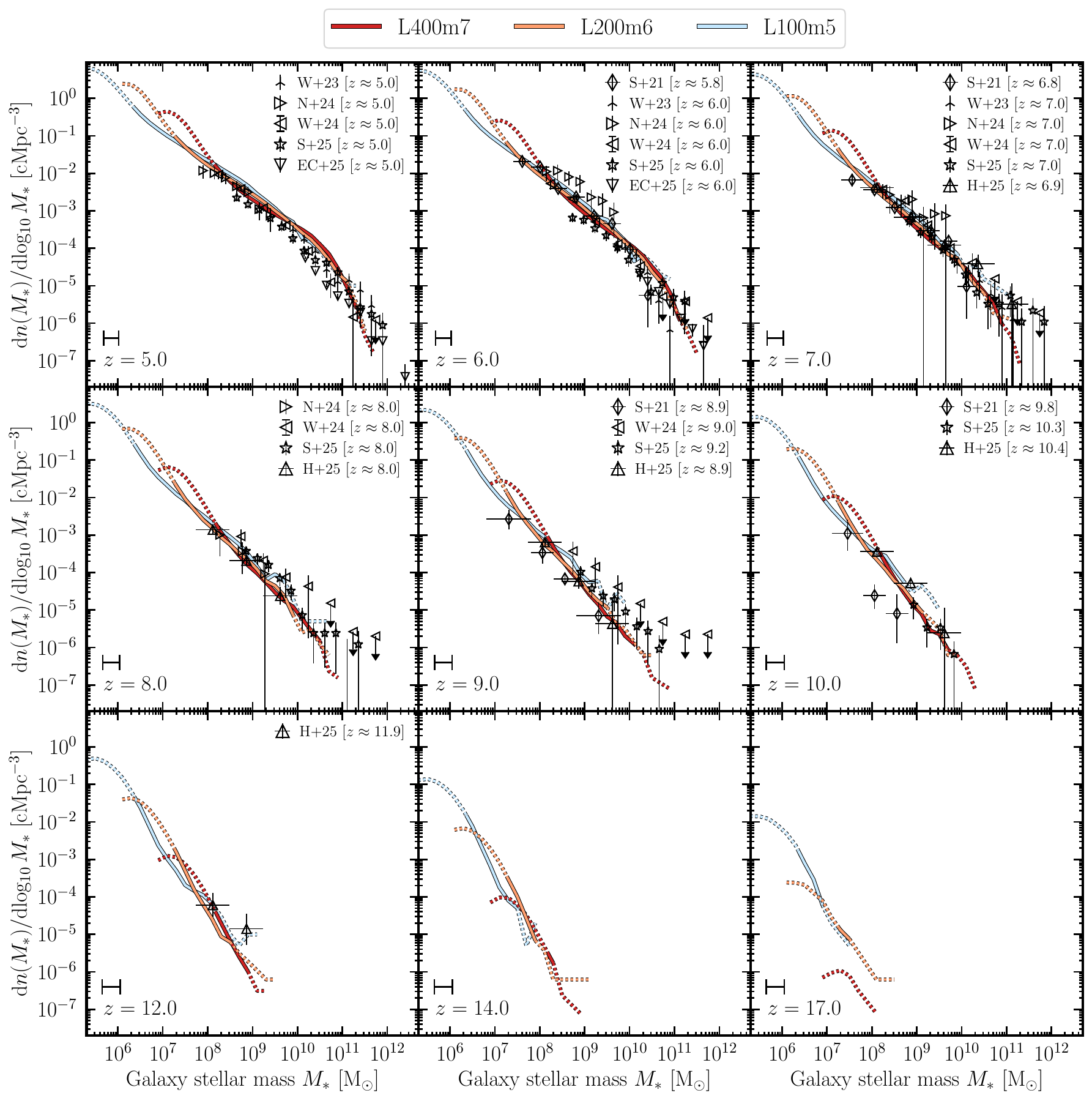}
    \caption{As Fig.~\ref{fig:gsmf_evolution_lowz}, but for higher redshifts, ranging from $z = 5$ (\textit{top left}) to $z = 17$ (\textit{bottom right}). Black symbols represent observational data from \citet{2021ApJ...922...29S}, \citet{2023A&A...677A.184W}, \citet{2024MNRAS.533.1808W}, \citet{2024ApJ...961..207N}, \citet{2025A&A...695A..20S}, \citet{2025ApJ...978...89H}, and \citet{2025arXiv250417867E}. \colibre{} again shows very good convergence with resolution and is consistent with the observational data.}
    \label{fig:gsmf_evolution_highz}
\end{figure*}

\section{Results}
\label{section: results}

\subsection{Evolution of the galaxy stellar mass function}
\label{subsection: evolution_of_gsmf}

Fig.~\ref{fig:gsmf_evolution_lowz} shows the evolution of the \gsmf{} in the \colibre{} simulations from $z=0$ to $z=4$, and Fig.~\ref{fig:gsmf_evolution_highz} extends the evolution to higher redshifts, from $z=5$ to $z=17$\footnote{The choice of $z=17$ is motivated by the fact that this is the highest integer redshift for which our lowest-resolution simulation (m7) contains a sufficient number of subhaloes with non-zero stellar masses ($\approx 60$), so that the \gsmf{} can still be estimated.}. The \gsmf{} is computed using fixed mass bins of $0.2$~dex. The \colibre{} models with thermal AGN feedback at m7, m6, and m5 resolutions are depicted by the solid \msevencolor, \msixcolor, and \mfivecolor{} curves, respectively. To highlight limitations due to resolution and volume, the curves transition from solid to dotted at stellar masses below $10$ times the mean initial baryonic particle mass (i.e. where galaxies contain fewer than $\sim 10$ stellar particles) or where fewer than 10 galaxies are present per bin. We emphasize that the chosen resolution limit is shown only to guide the eye and does not necessarily correspond to the true resolution limit. The binned \gsmf{} values for the L200m6 simulation at $0<z<17$ are provided in Appendix \ref{appendix: binned_gsmf}, and those for other simulations will be made available on the \colibre{} website upon publication\footnote{\href{https://colibre-simulations.org/}{https://colibre-simulations.org/}}.

As a comparison, we present the following observed \gsmfs{} from the literature:

\begin{itemize} \item The \gsmf{} from GAMA DR4 at $z=0$ \citep{2022MNRAS.513..439D}, to which the \colibre{} simulations were calibrated. GAMA is a highly complete multi-wavelength survey with over $200,000$ spectroscopic redshifts. Stellar masses were inferred using UV-to-IR SED fitting with the Bayesian code \textsc{prospect} \citep{2020MNRAS.495..905R}, assuming \citet{2003MNRAS.344.1000B} stellar population models. The analysis employed parametric SFHs modelled as a skewed-Normal distribution combined with a prescription for time-evolving metallicity \citep[for more details, see][]{2020MNRAS.498.5581B}. We use the GAMA \gsmf{} derived from the data across four primary GAMA regions (in total covering  $\approx 230$ deg$^2$), corrected to evolve to $z=0$ and re-normalized to match the larger SDSS area of $\approx 5,012$ deg$^2$.

\item The $0 < z < 4$ \gsmf{} from the DEVILS survey \citep{2018MNRAS.480..768D}, based on the D10-COSMOS field ($\approx 1.5$ deg$^{2}$), as reported by \citet{2021MNRAS.505..540T}. Stellar masses were derived by applying the \textsc{prospect} code to $\approx 494,000$ galaxies using the same data processing pipeline as for the GAMA survey.

\item The $0.2 < z < 7.5$ \gsmf{} from the COSMOS2020 photometric redshift catalogue \citep{2023A&A...677A.184W}, based on $\sim 10^6$ objects observed over an effective area of $\approx 1.27$ deg$^2$. Stellar masses were estimated via UV-to-IR SED fitting using the \textsc{LePhare} code \citep{2011ascl.soft08009A}, assuming stellar population models of \citet{2003MNRAS.344.1000B} with fixed (i.e. non-evolving) metallicities and either exponentially declining or delayed SFHs. Photometric fluxes were measured across $\sim 30$ bands using the FARMER pipeline \citep{2023ApJS..269...20W}.

\item The $0.2 < z < 3$ \gsmf{} from \citet{2020ApJ...893..111L} based on $58,461$ galaxies in the 3D-\textit{HST} survey (covering an effective area of $\sim 900$ arcmin$^2$) and $48,443$ galaxies in the COSMOS field (an effective area of $\approx 1.38$ deg$^2$). The authors employed the \textsc{prospector} SED-fitting code \citep{2017ApJ...837..170L} to model rest-frame UV-to-IR galaxy photometry. Stellar masses were estimated using the Flexible Stellar Population Synthesis model \citep{2009ApJ...699..486C}, utilizing non-parametric SFHs and treating metallicity as a free parameter. All analysis was performed within a Bayesian framework.

\item The $6 < z < 10$ GSMF from \citet{2021ApJ...922...29S} based on data from the GOODS, HUDF/XDF, and five CANDELS fields, covering a combined effective area of $\approx 731$ arcmin$^{2}$. The authors constructed a sample of $\sim 800$ Lyman-break galaxies at $z \gtrsim 6$, including six galaxies at $z \sim 10$ and making use of deep IR data from \textit{Spitzer}/IRAC at 3.6 and 4.5 $\mu$m. Stellar masses were estimated using \citet{2003MNRAS.344.1000B} stellar population models, assuming a $0.2 \, \rm Z_\odot$ metallicity and a constant SFH. SED fitting was performed using the \textsc{FAST} code \citep{2009ApJ...700..221K}. Photometric redshifts were derived with \textsc{eazy} \citep{2008ApJ...686.1503B}.

\item The $4 < z < 9$ \gsmf{} from \citet{2024MNRAS.533.1808W} based on data from the \textit{JWST}/NIRCam imaging programs CEERS, PRIMER, and JADES, encompassing an effective area of $\sim 500$ arcmin$^2$. The authors compiled a sample of over $30,000$ galaxies. Stellar masses and photometric redshifts were estimated using SED fitting with the Bayesian framework \textsc{bagpipes} \citep{2018MNRAS.480.4379C}. The authors adopted the \textsc{bpass-v2.2.1} stellar population models \citep{2018MNRAS.479...75S} with delayed-$\tau$ SFHs and treated metallicity as a free parameter.

\item The \textit{JWST} \gsmf{} from \citet{2024ApJ...961..207N} based on a sample of $3,300$ galaxies spanning $z \approx 3.5$ to $8.5$. These galaxies were selected from \textit{JWST} imaging in the HUDF ($\approx 12 \, \mathrm{arcmin}^2$) and the UKIDSS Ultra Deep Survey field ($\approx 7 \, \mathrm{arcmin}^2$), complemented by auxiliary \textit{HST}/WFC3 and WFC3/ACS data in the same fields. Stellar masses were estimated using \citet{2003MNRAS.344.1000B} stellar population models with exponentially declining SFHs and metallicities of $Z= 0.2 \, \rm Z_\odot$ and $1 \, \rm Z_\odot$, as well as using \citet{1999ApJS..123....3L} models with constant SFHs and $Z= 0.05 \, \rm Z_\odot$ and $0.4 \, \rm Z_\odot$. Photometric redshifts were determined through SED fitting using \textsc{LePhare}.

\item The \gsmf{} from the COSMOS-Web \textit{JWST} photometric survey \citep{2025A&A...695A..20S} based on imaging with four NIRCam filters. The data span a redshift range of $0.2 < z < 12$, containing $289,844$ galaxies over an effective area of $0.53$~deg$^2$, with  $\approx 0.2$~deg$^{2}$ also covered by one MIRI filter. Stellar masses were estimated using \citet{2003MNRAS.344.1000B} stellar population models with fixed metallicities, assuming delayed and exponential SFHs. SED fitting was performed using \textsc{LePhare}, applied to photometry in the UV-to-IR spectral range. The authors incorporated pre-\textit{JWST} multi-band imaging data in the COSMOS field alongside the \textit{JWST} data.

\item The \gsmf{} from \citet{2025ApJ...978...89H} based on \textit{JWST}/NIRCam imaging of $1,120$ galaxies at $6.5 < z < 13.5$ from multiple \textit{JWST} fields, including CEERS, JADES, and PEARLS (covering a total area of $\approx 187$ arcmin$^2$). Stellar masses were estimated using \textsc{bagpipes}, adopting the \citet{2003MNRAS.344.1000B} stellar population models with lognormal SFHs. The allowed metallicity values ranged from $0.005 \, \rm Z_{\odot}$ to $5 \, \rm Z_{\odot}$. Photometric redshifts were determined using \textsc{eazy}.

\item The \gsmf{} from \citet{2025arXiv250417867E} at $0.2 < z < 6.5$ derived using the Cosmic Dawn Survey pre-launch catalogues \citep{2025A&A...695A.229E}, with an effective area of $\approx 10.13~\textrm{deg}^2$ and $2,091,740$ galaxies in the final sample after all selection criteria were applied. Multi-wavelength data were provided by \textit{Spitzer}/IRAC, the Hyper Suprime-Cam instrument on the Subaru Telescope, and the MegaCam instrument on the Canada–France–Hawaii Telescope. Photometric redshifts were inferred using \textsc{LePhare}. Stellar masses were also derived with \textsc{LePhare}, using the same configuration and stellar templates as in COSMOS2020 \citep{2023A&A...677A.184W}.

\begin{figure*}
    \centering
    \includegraphics[width=0.99\textwidth]{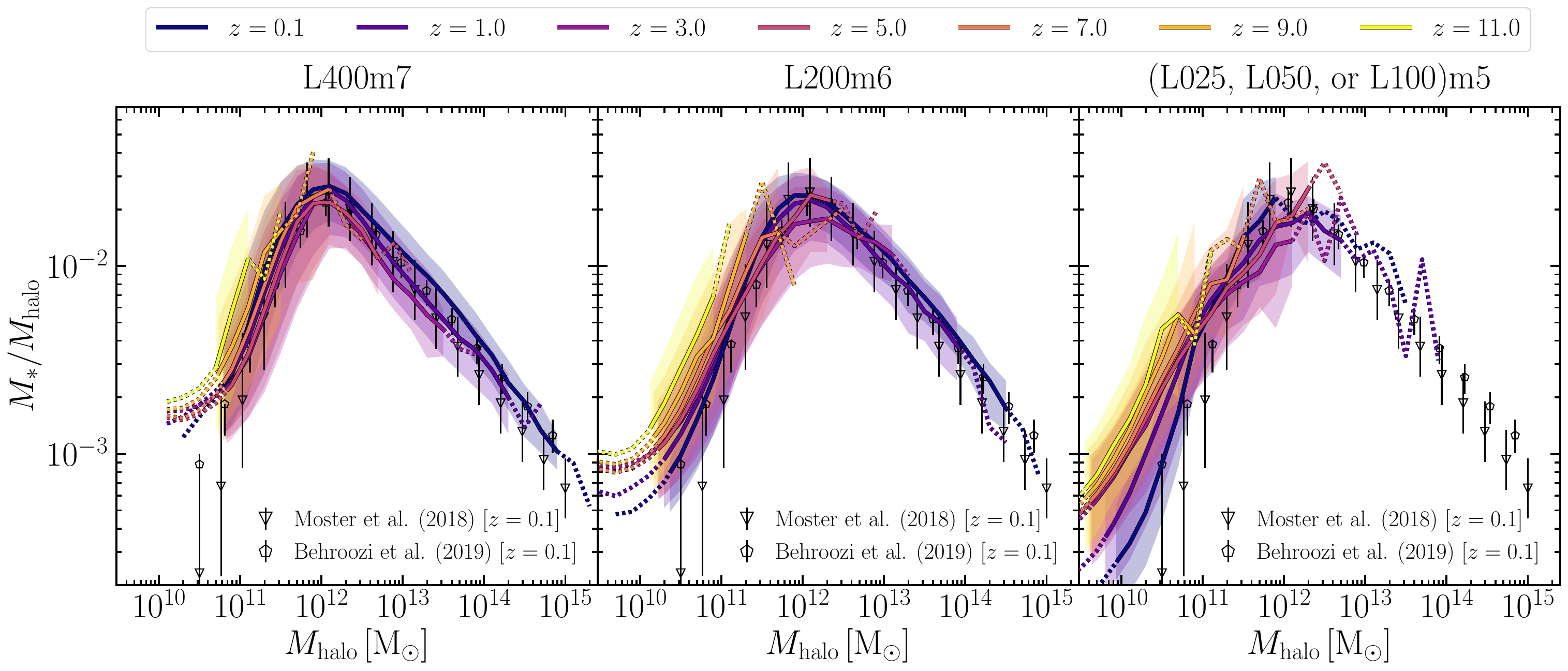}
    \caption{Evolution of the stellar-to-halo mass relation (\shmr) for central subhaloes in the \colibre{} simulations at m7 (\textit{left}), m6 (\textit{middle}), and m5 (\textit{right}) resolutions. Solid curves represent the median stellar-to-halo mass ratios from the simulations, while the shaded regions show the 16$^{\rm th}$ to 84$^{\rm th}$ percentile scatter. Different colours indicate different redshifts, ranging from $z=0.1$ (dark blue) to $z=11$ (yellow). The solid lines switch to dotted at halo masses corresponding to galaxies with $M_*<10$ times the mean initial baryonic particle mass, or where fewer than 10 objects fall within a given halo mass bin. For comparison, the semi-empirical \shmrs{} from \citet{2018MNRAS.477.1822M} and \citet{2019MNRAS.488.3143B} at $z=0.1$ are shown by black symbols. Here we do not add a lognormal scatter to galaxy stellar masses from the simulations because the Eddington bias is absent in the comparison data. The \colibre{} \shmr{} is consistent with the semi-empirical comparison data at $M_{\rm halo} \gtrsim 10^{11}~\mathrm{M_\odot}$, where it shows excellent convergence with resolution, peaks at a halo mass $M_{\rm halo} \sim 10^{12}~\mathrm{M_\odot}$, and exhibits only weak evolution with redshift.}
    \label{fig:\shmr{}_evolution}
\end{figure*}
\end{itemize}

We note that, except for the \gsmf{} from \citet{2020ApJ...893..111L}, none of the comparison data listed above have been corrected to remove the Eddington bias, which is included in our simulated \gsmfs{} (as explained in $\S$\ref{subsection:assumptions}).

It is evident from Figs.~\ref{fig:gsmf_evolution_lowz} and \ref{fig:gsmf_evolution_highz} that the \colibre{} simulations are in good agreement with the observationally inferred \gsmf{} at all redshifts where observational data are available. At $z = 0$, the excellent agreement of the \colibre{} m7 model is expected, as it was calibrated using emulators to match the \gsmf{} from \citet{2022MNRAS.513..439D}. Although the m6 and m5 models were calibrated manually, they also show very good agreement, with deviations within $\approx 0.1$~dex and no systematic offset from the observations.

For all \colibre{} resolutions, deviations from the data increase with redshift, peaking at $z \approx 2-4$, where the simulations predict a \gsmf{} with number densities that are on average $\approx 0.3$~dex higher than observed at $M_* \lesssim 10^{10.5}~\mathrm{M_\odot}$. The smallest discrepancies (around $0.2$~dex) are found for the m7 and m6 models when compared to the measurements of \citet{2020ApJ...893..111L}, whereas the largest deviations -- up to $\approx 0.4$~dex -- occur in the m5 model relative to \citet{2023A&A...677A.184W} and \citet{2025A&A...695A..20S}, as well as in the m7 and m6 models relative to \citet{2025arXiv250417867E}. While these minor discrepancies may point to limitations of the simulations, we also note that multiple studies have shown that, depending on the assumed SFH and SPS models, inferred stellar masses at $z \approx 2-4$ may be systematically offset by $0.1-0.3$~dex relative to the true values \citep[e.g.][]{2020MNRAS.492.5592K,2025A&A...695A..20S}. With this in mind, reducing the stellar masses in the simulations by $0.2$~dex at $M_* \lesssim 10^{10.5}~\mathrm{M_\odot}$ would bring the simulations into much closer agreement with the data, within $\approx 0.2$~dex.

Above $z = 4$, the differences between \colibre{} and observations become slightly smaller, on average falling within $\approx 0.25$~dex for all resolutions and remaining at that level up to $z = 10$. Notably, from $z = 5$ to $10$, \colibre{} is consistent with (albeit on average somewhat on the lower side of) the \gsmf{} reported by \citet{2025A&A...695A..20S}, which is based on the largest contiguous \textit{JWST} imaging survey to date ($\approx 0.53$~deg$^{2}$; \citealt{2023ApJ...954...31C}). Furthermore, \colibre{} reproduces the $7 < z < 10$ \gsmf{} from \citet{2025ApJ...978...89H}, also based on \textit{JWST} data, as well as the $6 < z < 9$ \gsmf{} from \citet{2021ApJ...922...29S}, which relies on pre-\textit{JWST} observations. At the highest redshift probed by \citet{2021ApJ...922...29S}, $z \approx 10$, \colibre{} overestimates their \gsmf{} by up to $\approx 1$~dex at $M_* \gtrsim 10^8~\mathrm{M_\odot}$, supporting the scenario of a higher abundance of massive galaxies at early times, in line with \textit{JWST} observations.

One notable exception to the good agreement between \colibre{} and observations is the discrepancy with the \textit{JWST} measurements from \citet{2024ApJ...961..207N} at $z = 6$. At this redshift, the \gsmfs{} from both \citet{2025A&A...695A..20S} and \colibre{} undershoot the \gsmf{} reported by \citet{2024ApJ...961..207N} by $\approx 0.5$~dex. \citet{2025A&A...695A..20S} suggest three possible explanations for the discrepancy in normalization: cosmic variance (the effective area of the \textit{JWST} observations by \citealt{2024ApJ...961..207N} is only $\approx 20$ arcmin$^{2}$), photometric uncertainties, and/or systematic effects in the SED fitting. Furthermore, at $7 < z < 9$, the \colibre{} \gsmf{} falls more than $0.5$~dex below the measurements of \citet{2024MNRAS.533.1808W} for $M_* \gtrsim 10^{10}~\mathrm{M_\odot}$. However, \citet{2024MNRAS.533.1808W} caution that the redshifts and stellar masses of the objects contributing to the \gsmf{} at these high $M_*$ and $z$ are poorly constrained, and that the resulting \gsmf{} values -- nearly all of which are upper limits -- may be significantly overestimated.

At the highest redshifts shown in Fig.~\ref{fig:gsmf_evolution_highz}, from $z = 12$ to $17$, \colibre{} predicts a gradual evolution of the \gsmf{} without sharp drop-offs, with the normalization changing by $\approx 2$~dex over this redshift interval. At $z=12$, we compare to the only available observational dataset, \citet{2025ApJ...978...89H}, who provide \gsmf{} estimates based on \textit{JWST}/NIRCam imaging at a median redshift of $z\approx 11.94$ and in two stellar mass bins: $M_* \sim 10^{8.125}$ and $10^{8.875}~\mathrm{M_\odot}$, each containing 5 objects (see their table 3). We find that \colibre{} is consistent with the lower mass bin within $1\sigma$ at all three resolutions but underpredicts the value in the higher mass bin by up to $\approx 2\sigma$, depending on the resolution. Note, however, that we are not accounting for potential systematic uncertainties. A shift in $M_*$ by $0.3$~dex, which is a relatively conservative estimate for the magnitude of the systematic error on stellar mass at $z=12$, would bring the \colibre{} predictions at all three resolutions into agreement with the second bin within $1\sigma$. Alternatively, increasing the assumed stellar mass uncertainty for simulated galaxies from our $z\geq 2$ default (and likely conservative for
$z=12$) value of $0.3$~dex to $0.5$~dex would also resolve the discrepancy (see Fig.~\ref{fig:eddington_bias_plot_multile_z} in Appendix~\ref{appendix: effect of eddington bias}).

Finally, we note that at $z = 17$, the combination of different volumes and resolutions of \colibre{} enables us to predict a \gsmf{} that spans more than two orders of magnitude in stellar mass: $10^{5.5} < M_*/\mathrm{M_\odot} < 10^8$. 

\begin{figure}
    \centering
    \includegraphics[width=0.49\textwidth]{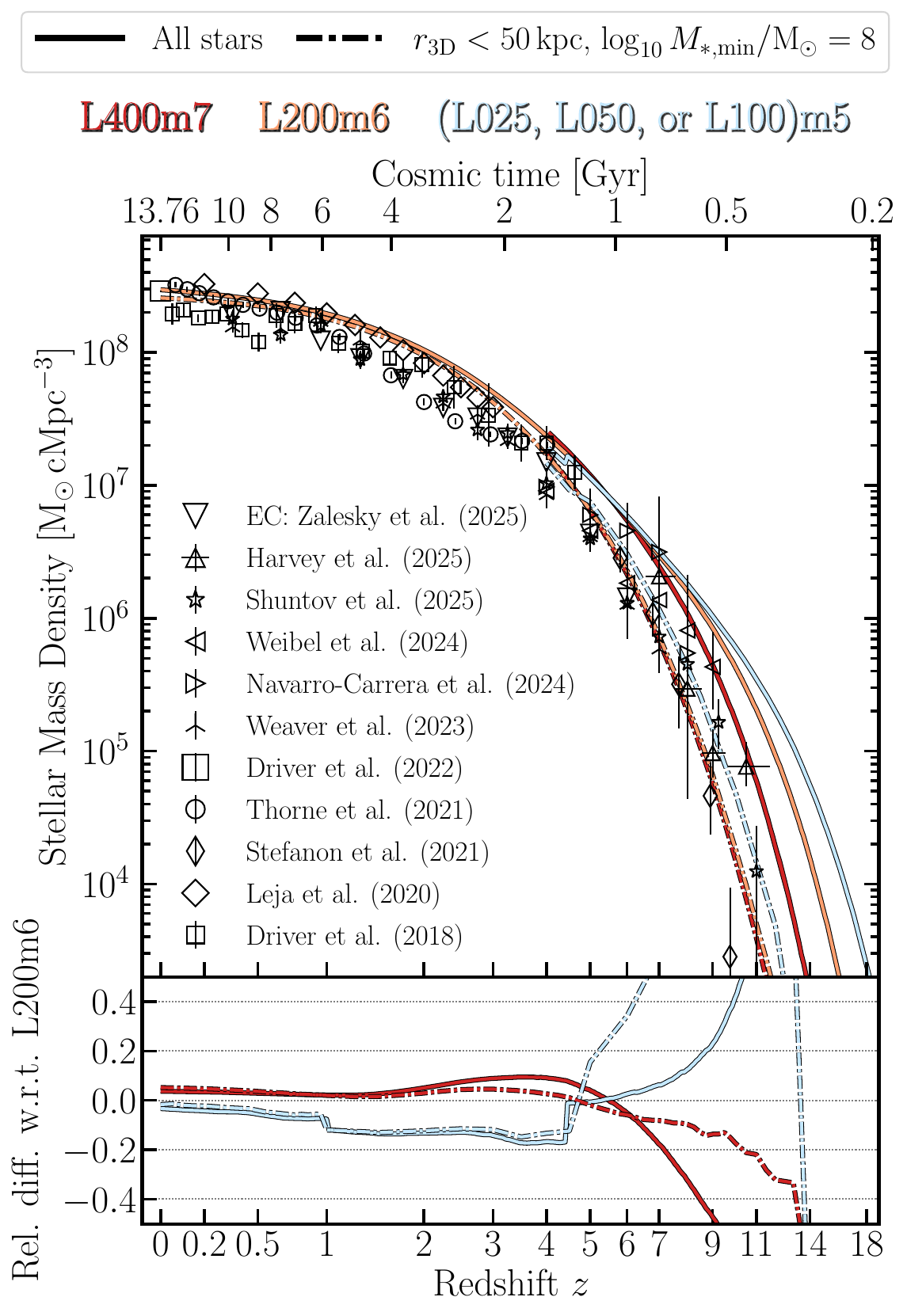}
    \caption{\textit{Top panel:} Evolution of the cosmic stellar mass density (CSMD) in the \colibre{} m7, m6, and m5 simulations. Different resolutions are indicated by different colours. We show the total CSMD, including all stellar mass within the simulated volumes (solid lines), and the CSMD contributed only by galaxies with stellar masses $M_* \geq 10^8 \, \mathrm{M_\odot}$ (dash-dotted lines), where the stellar mass is measured within 3D spherical apertures of radius 50~pkpc. While the total CSMD reflects the `raw' simulation values, the CSMD shown by the dash-dotted lines incorporates common observational cuts and can be directly compared to observations. For clarity, the solid and dash-dotted lines for the m7 and m5 simulations are truncated at $z < 4$ due to their overlap with the m6 lines. For comparison, a compilation of recent observational measurements is shown as black symbols. No lognormal scatter is added to the simulated galaxy stellar masses, as the comparison data account for Eddington bias when estimating the CSMD. \textit{Bottom panel:} Relative difference in the CSMD between simulations at different resolutions, shown for both versions of CSMD and using the m6 simulation as the baseline. The CSMD in \colibre{} agrees well with the observational data at all redshifts and shows very good convergence between resolutions for $z < 6$. Selecting only galaxies with $M_* \geq 10^8~\mathrm{M_\odot}$ reduces the CSMD by more than $0.5$~dex at $z \gtrsim 6$, while having only a minor impact below $z = 2$.}
    \label{fig:CSMD_evolution}
\end{figure}

\subsection{Stellar-to-halo mass relation}

Having discussed the evolution of the \colibre{} \gsmf, we turn to the stellar-to-halo mass relation (\shmr), which is displayed in Fig.~\ref{fig:\shmr{}_evolution}. Specifically, we show the evolution of \shmrs{} for central subhaloes at m7 (left panel), m6 (middle panel), and m5 (right panel) resolutions. The solid curves represent the median stellar-to-halo mass ratio for each $0.2$-dex halo mass bin and the shaded regions mark the 16$^{\rm th}$ to 84$^{\rm th}$ percentile scatter. The colours indicate different redshifts, smoothly transitioning from yellow at $z=11$ to dark blue at $z=0.1$. The solid curves turn to dotted at the low- and high-mass ends, corresponding to haloes hosting galaxies with stellar masses $M_*<10$ times the mean initial baryonic particle mass (i.e. containing fewer than $\sim 10$ stellar particles) and halo mass bins containing fewer than 10 objects, respectively.

For comparison, we take \shmrs{} from the semi-empirical models \textsc{UniverseMachine} \citep{2019MNRAS.488.3143B} and \textsc{emerge} \citep{2018MNRAS.477.1822M} for central galaxies at $z=0.1$. Both \citet{2019MNRAS.488.3143B} and \cite{2018MNRAS.477.1822M} used the peak halo mass defined within a sphere whose mean density is $\Delta$ times greater than the critical density of the Universe, with $\Delta$ being the redshift-dependent value from \citet{1998ApJ...495...80B} derived by fitting the results of \citet{1996MNRAS.282..263E}. In our analysis, we adopt the same \citet{1998ApJ...495...80B} halo mass definition, $M_{\rm halo}$, but use the actual halo masses at each redshift instead of their peak values. As we study the \shmr{} of \textit{central} subhaloes, this distinction should introduce negligible differences to the \shmr.

We find that at all redshifts, the \colibre{} \shmrs{} are well converged among the three resolutions for halo masses of $M_{\rm halo} \gtrsim 10^{11}~\mathrm{M_\odot}$. At lower $M_{\rm halo}$, convergence degrades due to resolution limitations of the lower-resolution simulations, with the \shmrs{} predicted by the m7 model flattening and approaching $\sim 10^{-3}$ from above as $M_{\rm halo}$ decreases to $\sim 10^{10}~\mathrm{M_\odot}$. This `floor' of the median relation is expected, as the lowest non-zero stellar mass a subhalo can have at m7 resolution is comparable to the mass of a single stellar particle ($\sim 10^7~\mathrm{M_\odot}$), which, at $M_{\rm halo} \sim 10^{10}~\mathrm{M_\odot}$, results in a stellar-to-halo mass ratio of $\sim 10^{-3}$. At even lower halo masses \mbox{($M_{\rm halo} \lesssim 10^{9}~\mathrm{M_\odot}$),} the median stellar-to-halo mass ratio drops to zero as the majority of these low-mass subhaloes have not formed any stellar particles (not shown).

At $z = 0.1$, \colibre{} agrees with the \shmrs{} reported by \citet{2018MNRAS.477.1822M} and \citet{2019MNRAS.488.3143B} for halo masses $M_{\rm halo} \gtrsim 10^{11.5}~\mathrm{M_\odot}$, matching both the shape and normalization of the \shmr. For $M_{\rm halo} \lesssim 10^{11.5}~\mathrm{M_\odot}$, the simulations remain in good agreement with \citet{2019MNRAS.488.3143B} but are only broadly consistent with \citet{2018MNRAS.477.1822M}, which predict a slightly steeper decline of the \shmr{} at lower halo masses. We note, however, that both \citet{2018MNRAS.477.1822M} and \citet{2019MNRAS.488.3143B} derive their halo masses from DMO simulations, whereas \colibre{} includes baryonic effects throughout. For haloes in the mass range $10^{11.5} \lesssim M_{\rm halo} \lesssim 10^{13} \, \mathrm{M_\odot}$ ($10^{10} \lesssim M_{\rm halo} \lesssim 10^{11.5}~\mathrm{M_\odot}$), halo masses may increase by $\approx 15$ ($30$) per cent when measured from a DMO simulation \citep[e.g.][]{2015MNRAS.451.1247S}, thereby likely bringing \colibre{} into closer agreement with \citet{2018MNRAS.477.1822M} for $M_{\rm halo} \lesssim 10^{11.5}~\mathrm{M_\odot}$. Another difference between \colibre{} and the two semi-empirical models lies in the choice of the halo finder. While \colibre{} employs HBT-HERONS, both \citet{2018MNRAS.477.1822M} and \citet{2019MNRAS.488.3143B} use \textsc{rockstar} \citep{2013ApJ...762..109B}. \citet{2025MNRAS.543.1339F} found that differences in halo-finding algorithms can introduce variations in the number density of central subhaloes of up to $\approx 10$ per cent, which may, in turn, impact the \shmr.

The \colibre{} simulations predict only weak evolution\footnote{We verified that the \colibre{} \shmr{} remains weakly dependent on redshift also when defining halo mass as $M_{\rm 200c}$ -- the mass enclosed within a sphere where the average density is 200 times the critical density of the Universe.} of the \shmr{} with redshift, with the normalization varying by only $\approx 0.5~\mathrm{dex}$ between $z=0.1$ and $11$, while the overall shape remains largely unchanged. On average, the normalization increases weakly with redshift. At all resolutions and times, the peak of the \shmr{} never exceeds a value of $0.04$. The most notable redshift-dependent variation is a monotonic shift of the high-mass end of the \shmr{} towards higher $M_{\rm halo}$ at lower redshifts, reflecting the fact that more massive haloes on average require more time to assemble. Importantly, the \colibre{} models at different resolutions show good convergence not only at $z = 0.1$ but also at all higher redshifts. 

We note that the peak of the $z = 0.1$ \shmr{} in \colibre{} is $M_* / M_{\rm halo}\approx 0.025$. Dividing this value by $\Omega_{\rm b,0}/\Omega_{\rm m,0}$ yields a baryon conversion efficiency into stars of $\approx 16$ per cent. The logarithmic slope of the \colibre{} \shmr{} before (after) the peak is approximately $1$ ($-0.5$). These values are comparable to those used by \citet{2025MNRAS.539.2409D}, who presented a simple theoretical model that reproduces both the high-$z$ \gsmf{} and high-$z$ UV luminosity functions measured by \textit{JWST}. Their model combines an HMF with a redshift-independent galaxy formation efficiency (i.e. a non-evolving \shmr) and a fixed IMF, and assumes a mass-to-light ratio that evolves with redshift driven by the increasingly younger stellar population at higher redshifts. The fact that the \colibre{} simulations also predict a (nearly) redshift-independent galaxy formation efficiency and are consistent with the high-$z$ \gsmf{} based on \textit{JWST} data further supports the findings of the authors. At the same time, \colibre{} appears to be in tension with \citet{2025A&A...695A..20S}, who argue that galaxy formation efficiencies as high as $\approx 0.5$ may be necessary to match the observed \gsmf{} in the $z\gtrsim 6$ Universe. As shown in Fig.~\ref{fig:gsmf_evolution_highz}, \colibre{} is broadly consistent with the \gsmf{} from \citet{2025A&A...695A..20S}, yet it predicts a nearly constant galaxy formation efficiency of $\approx 16$ per cent at the peak of \shmr.

\subsection{Cosmic stellar mass density}

The top panel of Fig.~\ref{fig:CSMD_evolution} illustrates the evolution of the cosmic stellar mass density (CSMD) as a function of redshift (bottom $x$-axis) and cosmic time (top $x$-axis). The \colibre{} m7, m6, and m5 simulations are shown in \msevencolor{}, \msixcolor{}, and \mfivecolor{} colours, respectively. The solid lines indicate the CSMD computed using the mass of all stellar particles that are present in the simulated volume. The dash-dotted lines represent the CSMD derived from only subhaloes with stellar masses equal to or exceeding $10^8~\mathrm{M_\odot}$. This selection mimics a common observational method for estimating the CSMD by integrating a (double) \citet{1976ApJ...203..297S} function down to a stellar mass limit of $\sim 10^8~\mathrm{M_\odot}$. The subhalo stellar masses are calculated by summing the masses of stellar particles that are gravitationally bound and located within 50~pkpc 3D apertures.

For clarity, the solid and dash-dotted lines for the m5 and m7 simulations are truncated below $z=4$, where they overlap with the CSMD from the m6 simulation. The bottom panel shows the relative difference in CSMD between the m7 and m5 simulations compared to the m6 simulation. The relative difference for a simulation at resolution m$X$, where $X$ represents 5, 6, or 7, is calculated as $\Delta \mathrm{CSMD}_{\mathrm{m}X} \equiv \mathrm{CSMD}_{\mathrm{m}X}/\mathrm{CSMD}_{\mathrm{m6}} - 1$, where ${\mathrm{CSMD}_{\mathrm{m}X}}$ is the CSMD from the m$X$ simulation.  We compute these differences for two cases: using all stellar mass in the simulated volume (solid lines) and applying the $M_* \geq 10^8~\mathrm{M_\odot}$ selection criterion (dash-dotted lines). Because the L100m5 and L50m5 simulations have not yet progressed to $z=0$ at the time of writing (they are currently at $z=4.6$ and $z=0.9$, respectively; see Table~\ref{table: simulations}), we show the CSMD at m5 resolution from the $100^3$~cMpc$^3$ volume for $z > 4.6$, from $50^3$~cMpc$^3$ between $z = 0.9$ and $z = 4.6$, and from the $25^3$~cMpc$^3$ volume at lower redshifts. In Appendix \ref{appendix:convergence}, we demonstrate that at $z<4$, the CSMDs in these three volumes converge very well, while at higher redshifts the $100^3$~cMpc$^3$ volume yields a slightly higher CSMD.

For comparison, we include CSMD estimates from the same ten studies shown in Figs.~\ref{fig:gsmf_evolution_lowz} and \ref{fig:gsmf_evolution_highz}, as well as additional measurements from \citet{2018MNRAS.475.2891D}. Specifically, we show the $z=0$ CSMD value from \citet{2022MNRAS.513..439D}, obtained by integrating the GAMA DR4 \gsmf{} down to $M_* = 10^{6.75}~\mathrm{M_\odot}$. This estimate accounts for re-normalization of the GAMA fields to a larger region selected from the SDSS DR16, along with a correction for CSMD evolution between the GAMA sample’s median observed redshift ($z\approx 0.079$) and $z=0$. We also include CSMD estimates from \citet{2020ApJ...893..111L}, \citet{2021ApJ...922...29S}, \citet{2021MNRAS.505..540T}, \citet{2023A&A...677A.184W}, \citet{2024MNRAS.533.1808W}, \citet{2025A&A...695A..20S}, \citet{2024ApJ...961..207N}, \citet{2025ApJ...978...89H}, and \citet{2025arXiv250417867E}, all derived by integrating their respective best-fitting Schechter functions down to $M_* = 10^8~\mathrm{M_\odot}$. Finally, we include the $0 < z < 5$ CSMD estimates from \citet{2018MNRAS.475.2891D}, based on a combined dataset of $\sim10^6$ galaxies from GAMA DR2, G10-COSMOS, and the 3D-\textit{HST} survey. In their analysis, the authors fit a spline to their \gsmf{} measurements and integrated it down to $M_* \sim 10^7~\mathrm{M_\odot}$, relying increasingly on extrapolation at higher redshifts.

We first compare the solid and dash-dotted lines of the same colour in the top panel, finding that applying a stellar mass threshold of $10^8~\mathrm{M_\odot}$ to subhaloes contributing to the CSMD has a noticeable impact at $z \gtrsim 2$. This threshold becomes increasingly significant at higher redshift, reducing the CSMD by about $0.5$ and $1$~dex at $z \sim 6$ and $z \sim 9$, respectively. These results are consistent with the findings of \citet{2015MNRAS.450.4486F}, who performed a similar analysis using the \textsc{eagle} simulations. Furthermore, while the $M_* = 10^8~\mathrm{M_\odot}$ threshold accounts for most of the difference between the solid and dash-dotted curves, at $z\lesssim 2$ there is also a minor contribution from the aperture requirement\footnote{The impact of this and other restrictions is discussed in detail in Appendix \ref{appendix: particle_selection}.}. At $z=0$, \colibre{} predicts a total CSMD of $\approx 3 \times 10^{8}~\mathrm{M_\odot \, cMpc^{-3}}$, with $\approx 90$~per cent bound to subhaloes of $M_* \geq 10^8~\mathrm{M_\odot}$ and within $50$~pkpc of their centres. 

Next, by comparing the observational data with the dash-dotted lines, which best mimic the observational selection criteria, we find that the CSMD in \colibre{} is generally in good agreement with the observed CSMD across all redshifts. Specifically, the \colibre{} CSMD at $z=0$ best matches the measurements from \citet{2022MNRAS.513..439D}; from $z=0$ to $z=1$, it is in good agreement with \citet{2021MNRAS.505..540T}; between $z=1$ and $4$, \colibre{} remains consistent with the CSMD from \citet{2018MNRAS.475.2891D} and \citet{2020ApJ...893..111L}; and at $4<z<12$, it agrees reasonably well with the CSMD reported by \citet{2025A&A...695A..20S}. At the same time, \colibre{} slightly overestimates the CSMD reported by \citet{2023A&A...677A.184W}, \citet{2025A&A...695A..20S}, and \citet{2025arXiv250417867E} at $1.5 < z < 4$, and underestimates the measurements from \citet{2024ApJ...961..207N} at $z > 6$ and from \citet{2024MNRAS.533.1808W} at $z > 7$. These discrepancies are consistent with those seen in the \gsmf{} (see Figs.~\ref{fig:gsmf_evolution_lowz} and \ref{fig:gsmf_evolution_highz}, and \S\ref{subsection: evolution_of_gsmf} for discussion). The \colibre{} CSMD is consistent with the observations from \citet{2021ApJ...922...29S} and \citet{2025ApJ...978...89H} over the redshift range $6 < z < 9$, but at $z \approx 10$ it falls $\approx 0.5$~dex below the measurement from \citet{2025ApJ...978...89H}, while exceeding that of \citet{2021ApJ...922...29S} by $\approx 1$~dex.

Finally, in the lower panel, we observe that the relative differences in the CSMD between the m5 and m6 models, and between the m6 and m7 models, remain within $\approx 20$~per cent at $0 < z < 5$, increase to within $\approx 40$~per cent at $5 < z < 6$, and rise to beyond $40$~per cent at higher redshifts. As we clarify in Appendix \ref{appendix:convergence}, the sharp rise in the relative differences at $z > 7$ is caused by resolution effects, while the increase in the difference between the m6 and m5 resolutions at $5 < z < 7$ is due to the switch from the L050m5 simulation to L100m5 at $z = 5$, with the larger volume containing a massive protocluster at $z \sim 5-6$ that is atypical for a $(100~\mathrm{cMpc})^3$ volume and hence yields a higher CSMD compared to L200m6.

\subsection{Cosmic star formation rate density}
\label{subsection:CSFRD}

\begin{figure}
    \centering
    \includegraphics[width=0.49\textwidth]{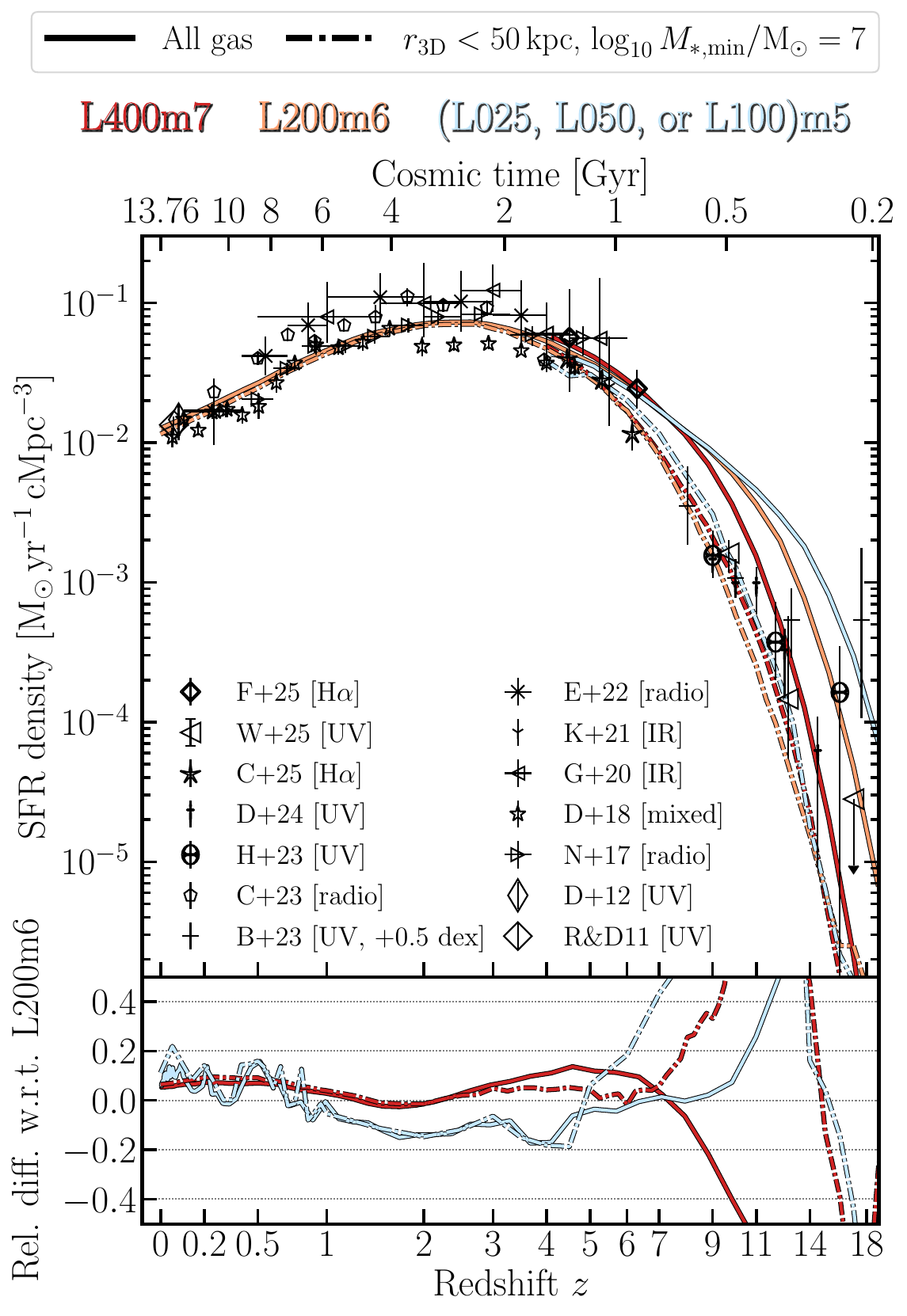}
    \caption{Evolution of the cosmic star formation rate density (CSFRD). The figure follows the same layout as Fig.~\ref{fig:CSMD_evolution}. We show the total CSFRD, contributed by all star-forming gas particles within the simulated volumes (solid lines), and the CSFRD from galaxies with stellar masses $M_* \geq 10^7~\mathrm{M_\odot}$ (dash-dotted lines). Black symbols represent a compilation of observational data from \citet{2011MNRAS.413.2570R,2012MNRAS.427.3244D,2017A&A...602A...5N,2018MNRAS.475.2891D,2020A&A...643A...8G,2021A&A...649A.152K,2022ApJ...927..204E,2023MNRAS.523.6082C,2023MNRAS.523.1009B,2023ApJS..265....5H,2024MNRAS.533.3222D,2025A&A...694A.178C,fu2025,2025arXiv250706292W}. The CSFRD in \colibre{} exhibits a plausible shape and normalization, showing good agreement with the data at all redshifts and very good numerical convergence at $z < 6$.}
    \label{fig:CSFRD_evolution}
\end{figure}

The top panel of Fig.~\ref{fig:CSFRD_evolution} shows the evolution of the cosmic star formation rate density (CSFRD). The solid curves show the SFR from all star-forming gas particles in the simulation, while the dash-dotted curves indicate the SFR from gas particles that are bound to and located within 50~pkpc of subhaloes containing at least $10^7~\mathrm{M_\odot}$ of stellar mass. This mass threshold is chosen to facilitate a more consistent comparison with \textit{JWST} observations at high redshift. As in Fig.~\ref{fig:CSMD_evolution}, we highlight the m6 simulation, while the CSFRD curves for the m7 and m5 simulations are truncated below $z=4$ for visual clarity. The bottom panel displays the differences in CSFRD relative to the m6 simulation. As in Fig.~\ref{fig:CSMD_evolution}, we show the predictions of the \colibre{} m5 model from the $100^3$~cMpc$^3$ volume for $z > 4.6$, from the $50^3$~cMpc$^3$ volume for $0.9 < z < 4.6$, and from the $25^3$~cMpc$^3$ volume at lower redshifts.

For comparison, we include a compilation of observational estimates of the CSFRD based on various SFR indicators. At $z \approx 0$, we show FUV-based estimates from \citet{2012MNRAS.427.3244D}, derived from the GAMA DR1 data, and from \citet{2011MNRAS.413.2570R}, which are based on a matched catalogue between the \textit{GALEX} Medium Imaging Survey and SDSS DR7 data. At higher redshifts, we present SED-based measurements from \citet{2018MNRAS.475.2891D}, derived from a combined dataset of the GAMA, G10-COSMOS, and 3D-\textit{HST} surveys, yielding a total of $582,314$ galaxies over $0<z<5$. We complement our comparison dataset with three radio continuum-based CSFRDs: measurements from the VLA COSMOS survey, derived from 3-GHz observations of $5,915$ objects over $0.3<z<5$ \citep{2017A&A...602A...5N}; VLA observations of 554 radio-selected galaxies in the GOODS-N field, covering $0.1<z<3$ \citep{2022ApJ...927..204E}; and 150-MHz measurements from the LOFAR Two Metre Sky Survey, which include $55,581$ radio sources spanning $0.1<z\lesssim 4$ \citep{2023MNRAS.523.6082C}. At $z>0.5$, we show CSFRD estimates from \citet{2020A&A...643A...8G}, based on 56 sources within the ALPINE multi-wavelength survey (ECDFS and COSMOS fields) detected in sub-mm continuum by ALMA, as well as two high-redshift ($z\approx 4.5$ and $\approx 5.5$) values of the CSFRD also measured by ALMA \citep{2021A&A...649A.152K}. We further include H~$\upalpha$-based measurements at $4<z<7$ from \citet{2025A&A...694A.178C} and \citet{fu2025}. Both studies utilized \textit{JWST}/NIRCam grism observations and derived CSFRD values by integrating their SFR functions down to $\approx 0.25~\mathrm{M_\odot \, yr^{-1}}$, corresponding to an absolute UV magnitude limit of $M_{\rm UV} = -17$. Finally, at very high redshifts ($8<z<18$), we show the CSFRD based on \textit{JWST}-measured UV luminosity function (UVLF) reported by \citet{2023MNRAS.523.1009B}, \citet{2023ApJS..265....5H}, \citet{2024MNRAS.533.3222D}, and \citet{2025arXiv250706292W}. While the latter three studies estimated the CSFRD by integrating their measured UVLF down to $-17$~mag, \citet{2023MNRAS.523.1009B} adopted a fainter limit of $-19$~mag. To account for this difference, we shift the CSFRD values reported in \citet{2023MNRAS.523.1009B} by $0.5$~dex upwards. 

\begin{figure*}
    \centering
    \includegraphics[width=0.99\textwidth]{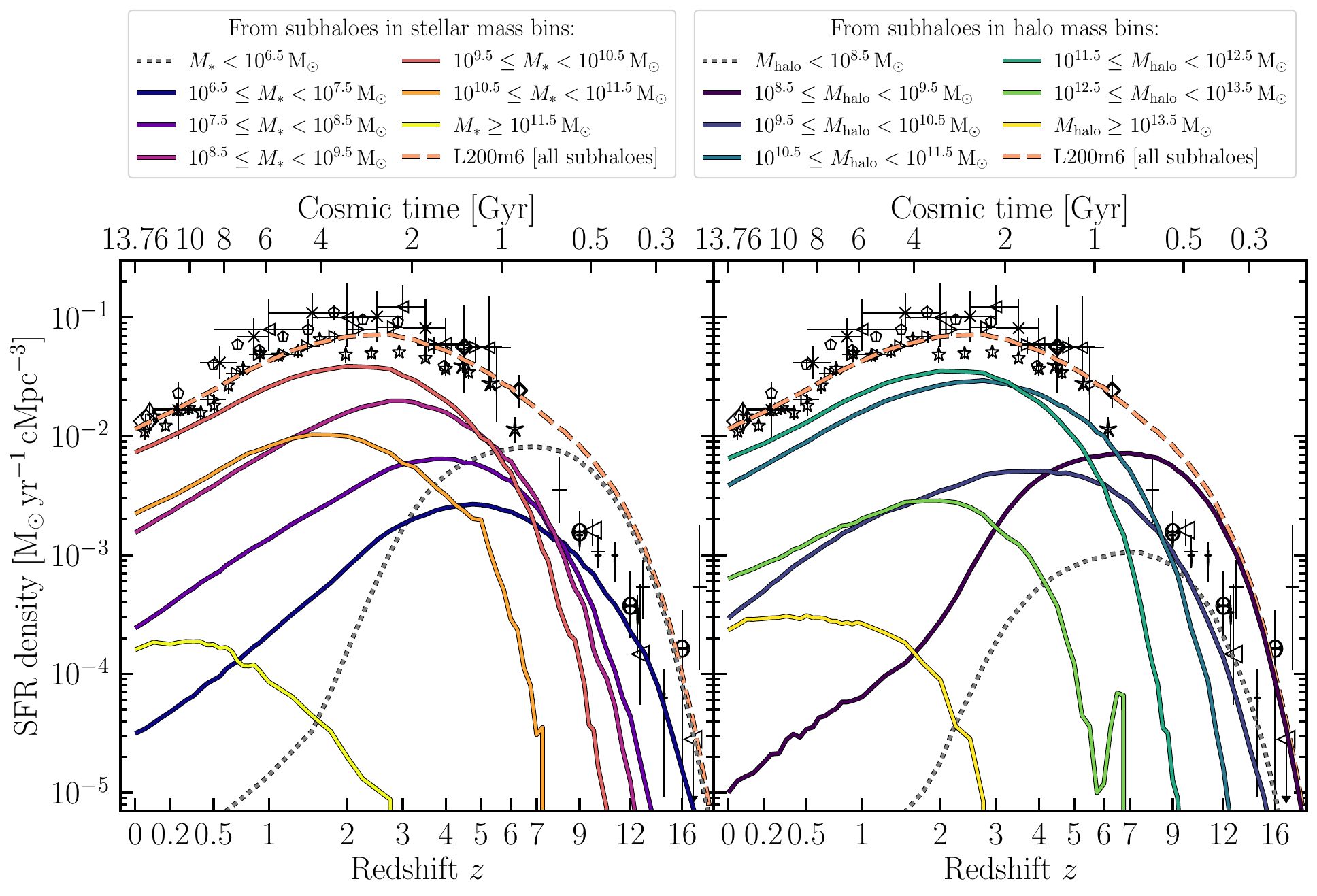}
    \caption{Cosmic star formation rate density (CSFRD) as a function of redshift in the \colibre{} L200m6 simulation, split into $1$-dex stellar mass bins (\textit{left panel}) and $1$-dex halo mass bins (\textit{right panel}). Differently coloured solid lines represent the contributions from each bin, except for the lowest mass bin, which is shown as a grey dotted line to highlight resolution limitations. The lower (upper) edge of the lowest (highest) stellar and halo mass bins is unbounded to include all remaining objects in the simulation at the low-mass (high-mass) end. The halo mass is the \citet{1998ApJ...495...80B} mass for central subhaloes and the total gravitationally bound mass for satellites. Subhalo SFRs are obtained by summing the instantaneous SFRs of gas particles that are bound to the subhalo and within a 50~pkpc 3D spherical aperture. The thick dashed \msixcolor{} line represents the CSFRD calculated by summing the SFRs of all subhaloes within the simulation volume. To guide the eye, the black symbols show the same observational measurements as in Fig.~\ref{fig:CSFRD_evolution}. At $0 < z < 3$, subhaloes with masses of $M_{\rm halo}\sim 10^{12}~\mathrm{M_\odot}$, which host galaxies with stellar masses of $\sim 10^{10}~\mathrm{M_\odot}$, dominate the CSFRD. At higher redshifts ($z > 3$), the primary contribution to the CSFRD shifts progressively towards subhaloes of lower mass, while the most massive objects ($M_{\rm halo} \geq 10^{13.5}~\mathrm{M_\odot}$) contribute negligibly at all $z$.}
    \label{fig:SFH_per_mass}
\end{figure*}

The \colibre{} simulations exhibit good numerical convergence for both particle selection criteria. At $0 < z < 6$, the relative differences between the simulations remain within $\approx 20$~per cent. At $z > 6$, the convergence of the CSFRD from $M \geq 10^{7}~\mathrm{M_\odot}$ galaxies becomes worse, driven by differences in the simulated volumes used at each resolution (see Appendix \ref{appendix:convergence}). At even higher redshifts ($z \gtrsim 10$), the predicted total CSFRD begins to strongly diverge, with higher-resolution simulations yielding systematically higher SFRs than their lower-resolution counterparts. As clarified later in Fig.~\ref{fig:SFH_per_mass}, this behaviour is driven by resolution effects. At $z \gtrsim 10$, the CSFRD is dominated by SFRs from low-mass subhaloes ($M_{\rm halo} \lesssim 10^{9.5}~\mathrm{M_\odot}$). Higher-resolution simulations 
resolve more of these low-mass objects and begin to track them at earlier times, which together lead to a higher CSFRD.

Turning to the comparison with observational data, at $z \approx 0$, \colibre{} predicts a CSFRD of $\approx 0.012 - 0.013~\mathrm{M_\odot \, yr^{-1}~cMpc^{-3}}$, which is consistent with the $z \sim 0$ measurements from \citet{2012MNRAS.427.3244D}, \citet{2011MNRAS.413.2570R}, and \citet{2018MNRAS.475.2891D}. From $z = 0$ to $2$, the CSFRD in the simulations increases by about $0.7$~dex, closely following the measurements from   \citet{2017A&A...602A...5N} and \citet{2018MNRAS.475.2891D}. The simulated CSFRD peaks at $\approx 0.075~\mathrm{M_\odot \, yr^{-1} \, cMpc^{-3}}$ between $z = 2$ and $3$, in reasonable agreement with \citet{2017A&A...602A...5N} and \citet{2018MNRAS.475.2891D}. However, it falls short by $\approx 0.3$~dex compared to the measurements of \citet{2020A&A...643A...8G}, \citet{2022ApJ...927..204E}, and \citet{2023MNRAS.523.6082C}. The factor of $\approx 2$ discrepancy at $z \approx 2$ between the CSFRD computed from intrinsic SFRs in simulations and that derived from direct SFR indicators in observations, with the former being lower, has been seen in nearly all modern galaxy formation simulations in representative volumes that predict a realistic $z=0$ galaxy population, including \textsc{eagle}, \textsc{IllustrisTNG}, and \textsc{Simba} \citep{2015MNRAS.450.4486F, 2018MNRAS.473.4077P, 2019MNRAS.486.2827D}. Furthermore, the same discrepancy has long been seen in observational data themselves: the SFH derived from the observationally inferred CSMD is typically a factor of $\approx 2$ lower around cosmic noon compared to the SFH constructed from observationally inferred SFRs \citep[e.g.][]{2008MNRAS.385..687W, 2016ApJ...820..114Y, 2019MNRAS.490.5359W, 2025A&A...695A..20S}. The ubiquity of this discrepancy was clearly demonstrated by \citet{2019MNRAS.488.3143B}, who calibrated the semi-empirical model \textsc{UniverseMachine} by fitting it to the observed CSFRD and CSMD, among other constraints. Although by design the \textsc{UniverseMachine} is highly flexible, the authors still had to introduce a redshift-dependent correction factor to the model's SFRs to satisfy the CSMD and CSFRD constraints simultaneously, which increased the 'observed' SFRs relative to the true SFRs predicted by the \textsc{UniverseMachine} by $\approx 0.3$~dex at $z=2$. The source of this discrepancy remains unclear, but it is often attributed to systematic offsets in observational data, arising from uncertainties in the calibration of SFR indicators and assumptions regarding the dust attenuation curve, the stellar IMF, and/or gas metallicity. 

Moving to even higher redshifts, we compare the \colibre{} CSFRD from subhaloes with $M_* \geq 10^7~\mathrm{M_\odot}$ (dash-dotted curves) to H $\upalpha$- and UV-based \textit{JWST} measurements at $z > 4$ from \citet{2023MNRAS.523.1009B}, \citet{2023ApJS..265....5H}, \citet{2024MNRAS.533.3222D}, \citet{2025A&A...694A.178C},  \citet{fu2025}, and \citet{2025arXiv250706292W}. At $4 < z < 8$, a stellar mass\footnote{While in Fig.~\ref{fig:CSFRD_evolution} we show the effect of applying the $M_* \geq 10^7~\mathrm{M_\odot}$ cut, in Appendix \ref{appendix: particle_selection} we additionally demonstrate the impact of a stricter $M_* \geq 10^8~\mathrm{M_\odot}$ threshold. Compared to $M_* \geq 10^7~\mathrm{M_\odot}$, adopting $M_* \geq 10^8~\mathrm{M_\odot}$ further reduces the CSFRD predicted by the simulations by $\approx 0.03$~dex at $z = 4$, increasing to about $0.15$~dex at $z = 8$.} of $M_* \sim 10^7-10^8~\mathrm{M_\odot}$ roughly corresponds to a rest-frame UV magnitude of $M_{\rm UV} = -17$ \citep[e.g.][]{2016ApJ...825....5S,2021ApJ...922...29S}, to which all considered \textit{JWST} data are normalized. At higher redshifts ($z > 8$), the $M_{\rm UV}-M_*$ relation becomes increasingly uncertain; however, both theoretical models and observations suggest that galaxies with $M_* \sim 10^7~\mathrm{M_\odot}$ should correspond to $-18 < M_{\rm UV}  <- 16$ \citep[e.g.][]{2023ApJ...942L..27S,2025MNRAS.536..988F}. \colibre{} agrees with the five said datasets within $\approx 1.5\sigma$ over $4 < z < 12$, while at $12 < z < 18$ it undershoots \citet{2023MNRAS.523.1009B} but remains roughly within the error bars -- albeit on the lower side -- of the measurements from \citet{2024MNRAS.533.3222D}, \citet{2023ApJS..265....5H}, and \citet{2025arXiv250706292W}. We caution, however, that at these highest redshifts, the observationally inferred CSFRD relies heavily on extrapolation down to $M_{\rm UV} = -17$, and that the CSFRD in the simulations is almost entirely dominated by subhaloes that have formed at most only a handful of stellar particles, indicating resolution limitations of the \colibre{} simulations. 

Finally, we note that although the CSFRD from $M_*\geq 10^{7}~\mathrm{M_\odot}$ subhaloes predicted by the m5 model becomes negligible above $z=14$ due to the lack of $M_*\geq 10^{7}~\mathrm{M_\odot}$ objects in the $100^3$~cMpc$^3$ volume at these very high redshifts, its total CSFRD (indicated by the solid \mfivecolor{} curve) at $z=14$ remains at $\sim 10^{-3}~\mathrm{M_\odot \, yr^{-1} \, cMpc^{-3}}$, decreasing by an order of magnitude by $z=18$. At $8 \lesssim z \lesssim 14$, this total CSFRD exceeds \textit{all} \textit{JWST} measurements shown in Fig.~\ref{fig:CSFRD_evolution} by $\approx 0.5$ to $\approx 1.5$~dex, which are based on integrating the UV luminosity function down to $M_{\rm UV} = -17$. This result implies that, due to the contribution from objects too faint to be currently observed -- with $M_{\rm UV} \gtrsim -17$ or, equivalently, $M_*\lesssim 10^{7}~\mathrm{M_\odot}$ -- the true CSFRD in the Universe at high redshifts may be significantly higher than current observational estimates suggest.

Fig.~\ref{fig:SFH_per_mass} decomposes the \colibre{} CSFRD into contributions from subhaloes in different mass bins (with an analogous decomposition of the CSMD presented in Appendix \ref{appendix: SMD_per_mass}). Here for clarity, we display only the m6 simulation. The CSFRD is calculated by summing the instantaneous SFRs of individual gas particles that are located within 50~pkpc of subhaloes and gravitationally bound.

In the left panel, the CSFRD is split into contributions from $1$-dex stellar mass bins with centres (in log space) at $M_{\rm *} = 10^{6}$, $10^{7}$, $10^{8}$, $10^{9}$, $10^{10}$, $10^{11}$, and $10^{12} \, \mathrm{M_\odot}$. As an exception, the lower edge of the lowest-mass bin and the upper edge of the highest-mass bin are left unbounded to include SFR contributions from subhaloes with $M_* < 10^{5.5}~\mathrm{M_\odot}$ (including those with zero stellar mass) and $M_* > 10^{12.5}~\mathrm{M_\odot}$, respectively. At each redshift, only subhaloes with stellar masses within the specified bin range contribute to the total SFR density of that bin. The corresponding CSFRDs from these subhaloes are shown as solid lines, with colours indicating different mass bins. To emphasize resolution limitations, the CSFRD from the lowest mass bin is represented by the dotted line in grey. The total CSFRD from all subhaloes in the simulated volume is depicted by the dashed \msixcolor{} line. The comparison data (black symbols) are the same as in Fig.~\ref{fig:CSFRD_evolution}. In the right panel of Fig.~\ref{fig:SFH_per_mass}, we perform the same CSFRD decomposition as in the left panel, but based on halo mass instead of stellar mass. The (logarithmic) centres of the 1-dex halo mass bins are $M_{\rm halo} = 10^{8}$, $10^{9}$, $10^{10}$, $10^{11}$, $10^{12}$, $10^{13}$, and $10^{14}~\mathrm{M_\odot}$. As with the stellar mass decomposition, the lower edge of the lowest bin and the upper edge of the highest bin are unbounded. The halo mass is calculated using the definition from \citet{1998ApJ...495...80B} for central subhaloes and as the total gravitationally bound mass for satellites. 

At $0 < z \lesssim 3$, the CSFRD is dominated by subhaloes located around the peak of the \shmr{}, with $M_{\rm halo} \sim 10^{12}~\mathrm{M_\odot}$ and $M_{\rm *} \sim 10^{10}~\mathrm{M_\odot}$. The second-largest contribution at these redshifts comes from subhaloes in the $M_{\rm halo} \sim 10^{11}~\mathrm{M_\odot}$ bin. Together, these haloes dominate the CSFRD up to $z \approx 6$, above which less massive subhaloes dominate.

In terms of stellar mass, the CSFRD at $z \gtrsim 6$ is dominated by the lowest stellar mass bin, reflecting the resolution limit of the m6 simulation. The galaxies in this bin either contain only a few stellar particles (with a mass of $m_{\rm *} \sim 10^{6}~\mathrm{M_\odot}$) or have not yet formed any stars. Focussing again on the right panel, we find that the majority of these galaxies must reside in subhaloes with masses of $10^{8.5} \lesssim M_{\rm halo}/\mathrm{M_\odot} \lesssim 10^{9.5}$, as these subhaloes dominate the $z \gtrsim 6$ CSFRD\footnote{The spike at $6.5 < z < 6.75$ in the CSFRD from the $10^{12.5} \leq M_{\rm halo}/\mathrm{M_\odot} < 10^{13.5}$ bin is caused by a single starburst galaxy with $M_{\rm halo} \approx 10^{12.52}~\mathrm{M_\odot}$ ($\approx 10^{12.56}~\mathrm{M_\odot}$) at  $z = 6.75$ ($z=6.5$) whose SFR drops from $\approx 5 \times 10^2~\mathrm{M_\odot\, yr^{-1}}$ at $6.5 < z <6.75$ to $\approx 10^2~\mathrm{M_\odot\,yr^{-1}}$ by $z = 6$.}.

The most massive objects ($M_{\rm halo} \gtrsim 10^{13.5}~\mathrm{M_\odot}$, $M_{\rm *} \gtrsim 10^{11.5}~\mathrm{M_\odot}$) contribute negligibly to the CSFRD at all redshifts due to their rarity. At $z \approx 0$, their contribution to the total CSFRD is only a few per cent, which gets even smaller at higher redshifts. We emphasize, however, that the mass used to split the CSFRD into bins is the actual mass at each redshift. This implies that the lower-mass progenitors of the $z\approx 0$ objects with $M_{\rm halo} \sim 10^{14}~\mathrm{M_\odot}$ contribute to progressively lower-mass bins at higher redshifts.

\begin{figure*}
    \centering
    \includegraphics[width=0.999\textwidth]{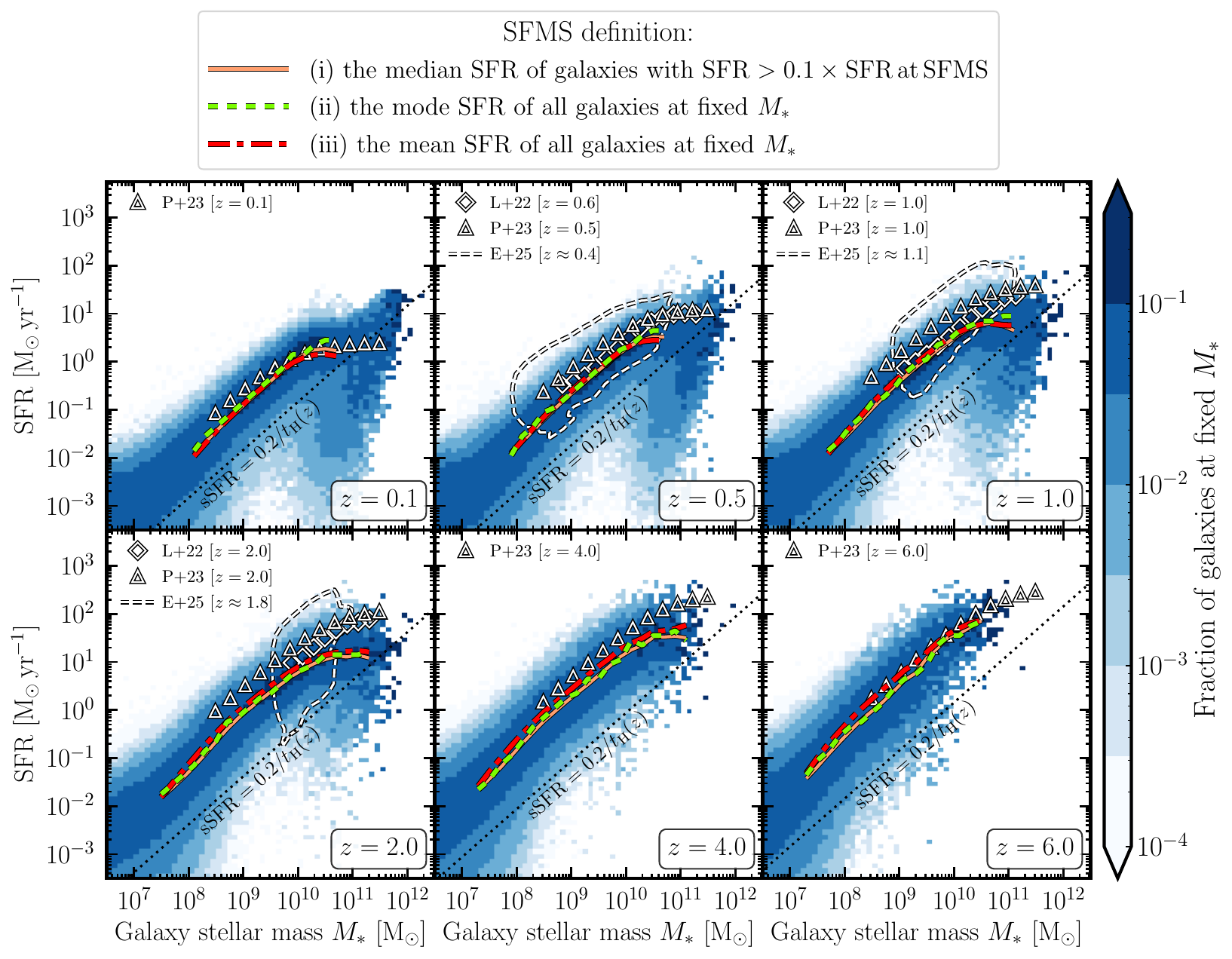}
    \caption{Star-forming main sequence (\sfms) in the L200m6 simulation at $z = 0.1$, $0.5$, $1$, $2$, $4$, and $6$ (different panels). The colour scale shows the fraction of galaxies per pixel with a given $M_*$ and SFR, normalized by the total number of galaxies across all pixels at the same $M_*$. The solid \msixcolor{} lines represent the \colibre{} \sfms, defined as the median SFR of star-forming galaxies ($\mathrm{SFR} > 0.1\, \times$ the SFR at the \sfms{} determined iteratively). The short-dashed light-green and red dash-dotted curves show alternative \sfms{} definitions based on the mode and mean of the SFR distribution at fixed $M_*$, respectively. For reference, the diagonal thin dotted line shows $\mathrm{sSFR} = 0.2 / t_{\rm H}(z)$, which is commonly used as an sSFR threshold for defining star-forming galaxies. Furthermore, also for reference, white diamonds and triangles indicate the observed \sfmss{} from \citet{2022ApJ...936..165L} and \citet{2023MNRAS.519.1526P}, respectively, while the white dashed contours mark the regions enclosing 95~per cent of the sample of star-forming galaxies from \citet{2025arXiv250315314E}. All three definitions of the \sfms{} in the simulation yield very similar results at all redshifts.}
    \label{fig:sfms_hist}
\end{figure*}

\subsection{Evolution of the star-forming main sequence}
\label{subsection:sfr}

An alternative but equally important diagnostic of stellar mass evolution, alongside the \gsmf, is the relation between galaxy SFR and its stellar mass. The SFR of star-forming galaxies is known to follow a remarkably tight sequence in the stellar mass -- SFR plane, known as the star-forming main sequence (\sfms{}), with a scatter of $0.2 - 0.3$~dex \citep[e.g.][]{2004MNRAS.351.1151B,2012ApJ...754L..29W}. This strong correlation between the SFR and stellar mass has been observed to hold from $z=0$ up to $z \approx 6$, with the SFR at a fixed stellar mass declining at lower redshifts as $\mathrm{SFR}\propto (1+z)^\alpha$, where $2<\alpha<4$ \citep[e.g.][]{2014ApJS..214...15S,2023MNRAS.519.1526P}.

Modern galaxy formation models, including \textsc{eagle}, \textsc{IllustrisTNG}, and \textsc{Simba}, have successfully reproduced the qualitative features of the observed \sfms{} over $0 < z < 5$. However, at $1<z<4$, the simulated \sfms{} typically falls short in normalization by $0.2-0.5$ dex compared to the observed relation \citep[e.g.][]{2015ARA&A..53...51S,2019MNRAS.485.4817D,2019MNRAS.486.2827D,2020MNRAS.492.5592K}. Growing evidence suggests that the longstanding offset between simulated and observed \sfms{} arises from methodological inconsistencies in how SFRs are measured in simulations versus observations, as already briefly discussed in $\S$\ref{subsection:CSFRD} in the context of the CSFRD. While simulations predict SFRs directly, observational SFR estimates rely on a series of assumptions in stellar population modelling -- including the assumed dust attenuation curve, gas metallicity, and SFHs -- all of which may introduce significant systematic errors \citep[e.g.][]{2020MNRAS.498.5581B,2023ApJ...944..141P}.

\begin{figure*}
    \centering
    \includegraphics[width=0.99\textwidth]{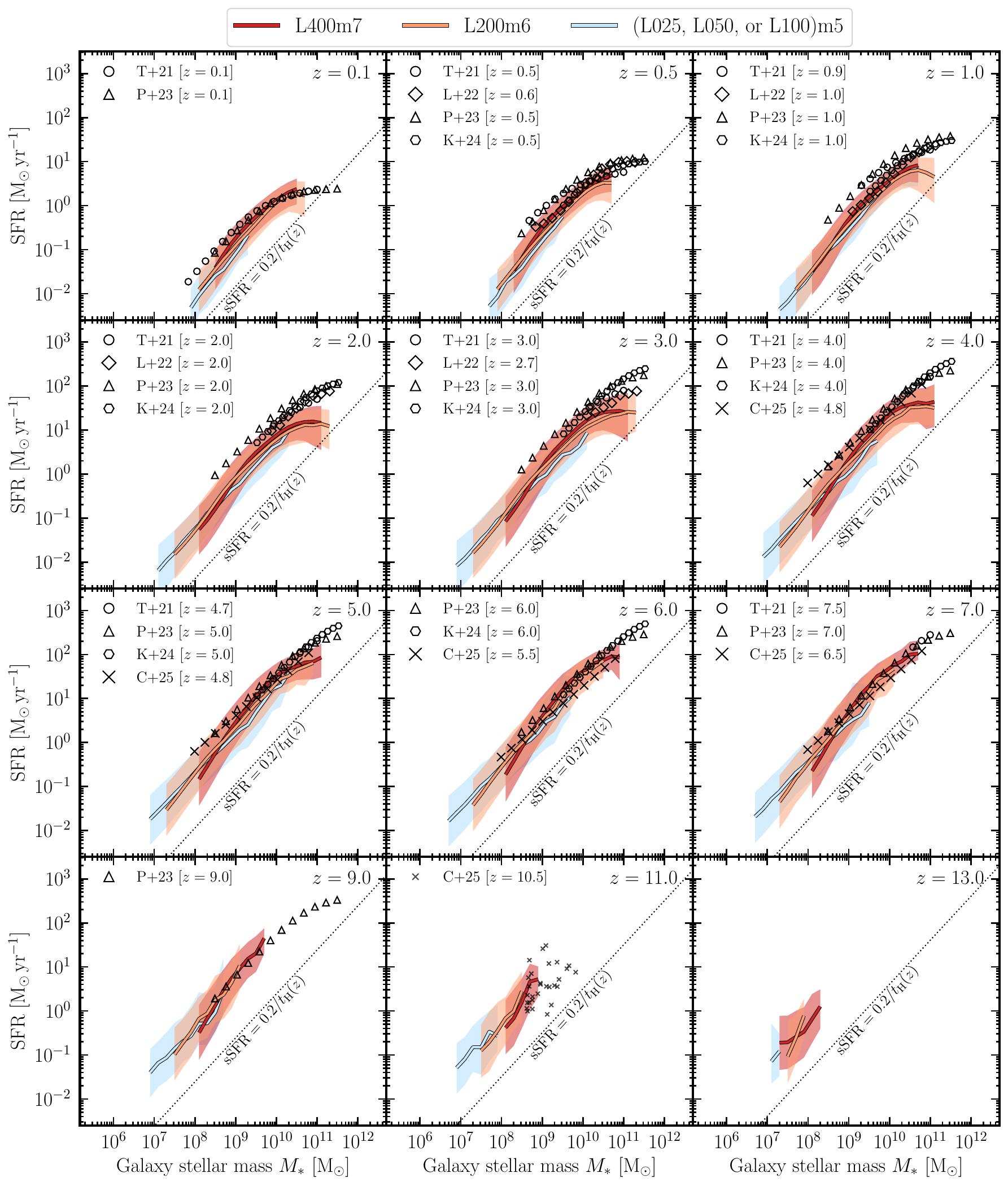}
    \caption{Evolution of the star-forming main sequence (\sfms) from $z = 0.1$ (\textit{top left}) to $z = 13$ (\textit{bottom right}) in the m7, m6, and m5 \colibre{} simulations, shown in \msevencolor, \msixcolor, and \mfivecolor{}, respectively. The thick solid lines represent the \colibre{} \sfms, defined as the median SFR of star-forming galaxies, identified iteratively by $\mathrm{SFR} > 0.1 \, \times$ the SFR at the \sfms. The shaded regions indicate the 16$^{\rm th}$ to 84$^{\rm th}$ percentiles of the SFR of star-forming galaxies ($\mathrm{SFR} > 0.1 \, \times$ the SFR at the \sfms) in each mass bin. A compilation of observed \sfms{} data from \citet{2021MNRAS.505..540T}, 
     \citet{2022ApJ...936..165L}, \citet{2023MNRAS.519.1526P}, \citet{2024A&A...691A.164K}, and \citet{2025ApJ...979..193C} are shown as black symbols. For reference, the diagonal thin dotted line indicates another commonly used criterion for defining star-forming galaxies: $\mathrm{sSFR} \geq 0.2 / t_{\rm H}(z)$. \colibre{} shows good agreement with observational data at low ($z \lesssim 0.5$) and high ($z \gtrsim 5$) redshifts. At intermediate redshifts ($0.5 < z < 4$), the simulations underestimate the observed \sfms{} in all datasets at $M_* \gtrsim 10^{10.5}~\mathrm{M_\odot}$, but remain broadly consistent with \citet{2022ApJ...936..165L} and \citet{2024A&A...691A.164K} at lower stellar masses.}
    \label{fig:sfr_evolution}
\end{figure*}

Bearing these caveats in mind, we now turn to the \sfms{} in \colibre. Because there is no unique way to define the \sfms{} \citep[e.g.][]{2022ApJ...936..165L}, we begin by comparing several definitions of the \sfms{} in the \colibre{} m6 simulation over a limited redshift range ($0.1 < z < 6$; Fig.~\ref{fig:sfms_hist}). After selecting a fiducial definition, we use it to compute the \sfms{} in the \colibre{} simulations at all resolutions and compare these predictions with a range of observational data within an extended redshift range ($0.1 < z < 13$; Fig.~\ref{fig:sfr_evolution}).

Fig.~\ref{fig:sfms_hist} shows the evolution of the \sfms{} in the L200m6 simulation at redshifts $z = 0.1$, $0.5$, $1$, $2$, $4$, and $6$ (different panels). The colour scale of the pixels indicates the fraction of galaxies per $0.1$ dex bin with a given stellar mass and SFR, normalized by the total number of galaxies across all SFR bins at the same $M_*$. We show the \colibre{} \sfms{} defined in three different ways:
\begin{enumerate}
\item \textit{solid \msixcolor{} lines:} the median SFR of active galaxies, where `active' refers to galaxies whose SFR lies within 1~dex of (or higher than) the \sfms{}, which is determined iteratively;
\item \textit{short-dashed light-green lines:} the mode of the SFR distribution of all galaxies (both star-forming and quiescent) at fixed $M_*$;
\item \textit{dash-dotted red lines:} the mean of the SFR distribution of all galaxies (both star-forming and quiescent) at fixed $M_*$;
\end{enumerate}
where in all three cases, the \sfms{} is computed in $0.2$-dex (as opposed to $0.1$-dex) stellar mass bins. For reference, the diagonal thin dotted line in each panel corresponds to $\mathrm{sSFR} = 0.2 / t_{\rm H}(z)$, providing an alternative criterion for selecting active galaxies: $\mathrm{sSFR} \geq 0.2 / t_{\rm H}(z)$, where $t_{\rm H}(z)$ is the Hubble time at redshift $z$. The \sfmss{} are displayed only for mass bins that contain at least 30 galaxies and for which the fraction of quenched galaxies (defined\footnote{We do not define the quenched fraction based on a distance cut from the \sfms{} at fixed $M_*$ because the \sfms{} begins to bend downward at $M_* \gtrsim 10^{10}~\mathrm{M_\odot}$, which would bias the quenched fractions at the high-mass end. However, we verified that defining the quenched fraction as the fraction of galaxies with $\mathrm{SFR} < 0.1 \times$ the SFR at the \textit{linearly extrapolated} \sfms{} (fit to the simulation data within $10^9 < M_* / \mathrm{M_\odot} < 10^{10}$) yields results very similar to the definition based on $\mathrm{sSFR} < 0.2 / t_{\rm H}(z)$.} as $\mathrm{sSFR} < 0.2 / t_{\rm H}(z)$, i.e., independently of the \sfms{}) is less than $50$~per cent, ensuring that the first two \sfms{} definitions always correspond to the star-forming galaxy population. 

In Fig.~\ref{fig:sfms_hist}, we see that \colibre{} predicts a realistic \sfms{}, with the SFR increasing roughly linearly with galaxy stellar mass up to a critical value of $M_* \sim 10^{10}-10^{11}~\mathrm{M_\odot}$, above which the \sfms{} begins to flatten. This critical mass increases monotonically with redshift. A second important observation from Fig.~\ref{fig:sfms_hist} is that, at all redshifts ($0 < z < 6$), the first and second \sfms{} definitions produce nearly identical \sfmss{}, while the third definition yields an \sfms{} with a marginally higher normalization at $z \geq 2$, but is otherwise consistent with the first two definitions. Repeating the same test using the m5 simulation (not shown), we found similar trends. However, we noticed that due to the much smaller number of galaxies per bin available in the m5 volume compared to m6, the second definition -- based on the peak of the galaxy number density at fixed $M_*$ -- yields a slightly noisier \sfms, whereas the third definition -- based on the mean SFR -- becomes more sensitive to outliers, biasing the \sfms{} towards higher SFR values. Based on these findings, we adopt the first definition (the median SFR of galaxies with $\mathrm{SFR} > 0.1\, \times$ the SFR at the \sfms) as our fiducial choice and use it for comparisons with observational data in Fig.~\ref{fig:sfr_evolution}, though we note that using either of the other two definitions would have little to no effect on our conclusions.

Fig.~\ref{fig:sfms_hist} also shows the $0.2 < z < 3$ measurements from the Euclid Q1 data release \citep{2025arXiv250315314E}, which is based on the Euclid Deep Fields (an effective area of $\approx 63$~deg$^2$) and includes over $4 \times 10^6$ objects above the 95 per cent stellar mass completeness limit up to $z = 3$. Specifically, from this sample, we show the distribution of star-forming galaxies in the SFR -- $M_*$ plane in three redshift bins (median values of $z=0.37$, $1.15$, and $1.83$), indicated by white dashed contours enclosing 95 per cent of the star-forming galaxies (i.e. $2\sigma$ of the star-forming distribution). The distribution of galaxies in the L200m6 simulation (i.e. pixels coloured in different shades of blue) largely covers the $2\sigma$ contours from \citet{2025arXiv250315314E} at all three redshifts. Finally, Fig.~\ref{fig:sfms_hist} includes the observed \sfmss{} from \citet[][]{2022ApJ...936..165L} and \citet{2023MNRAS.519.1526P}, indicated by white diamonds and triangles, respectively. These two \sfmss{} are shown for reference only, and we defer discussion of how well \colibre{} reproduces these two (and other) observed relations to the next figure.

Fig.~\ref{fig:sfr_evolution} shows the evolution of the \colibre{} \sfms{} from $z = 0.1$ to $z = 13$. The thick solid lines in \msevencolor, \msixcolor, and \mfivecolor{} denote the \sfms{} in the \colibre{} m7, m6, and m5 simulations, respectively. The shaded regions in the corresponding colours indicate the 16$^{\rm th}$ to 84$^{\rm th}$ percentiles of the SFR distribution of galaxies whose SFR is within $1$~dex of the \sfms{} or higher. 

For comparison, we include the \sfmss{} from several recent observational studies. We show the $0 < z < 9$ \sfms{} best-fitting relation derived from the DEVILS survey \citep{2021MNRAS.505..540T}; the $0.2 < z < 3$ \sfms{} from \citet{2022ApJ...936..165L}, defined as the ‘ridge line’ in the SFR -- $M_*$ plane and based on the 3D-\textit{HST} and COSMOS catalogues, with photometry and SED fitting methods identical to those in \citet{2020ApJ...893..111L}; the \sfms{} best-fitting relation from \citet[][their equation 15]{2023MNRAS.519.1526P} based on the authors’ comprehensive compilation of 27 recent studies spanning $0 < z < 6$, which we further extrapolated to $z = 9$; and the best-fitting formula for the \sfms{} from \citet{2024A&A...691A.164K}, derived through the stacking of FIR imaging of $\sim 100,000$ K-band-selected galaxies observed with \textit{Herschel} and James Clerk Maxwell Telescope in the UKIDSS Ultra Deep Survey and COSMOS fields over $0.5 < z <6$. Finally, we show the $4.5 < z < 12$ \sfms{} from \citet{2025ApJ...979..193C}, based on CEERS data obtained with \textit{JWST}, supplemented by \textit{HST}/ACS and \textit{HST}/WFC3 data. Their sample includes $1,863$ galaxies, with stellar masses and SFRs derived through SED fitting using \textsc{bagpipes} and \textsc{bpass} stellar evolution models. At $z=11$, we plot their individual object measurements due to the relatively small number of galaxies. 

The \sfms{} in the m5, m6, and m7 \colibre{} models qualitatively matches the observational data at all examined redshifts. At $z \geq 5$, \colibre{} reproduces both the normalization and slope of the \sfms{} reported by \citet{2021MNRAS.505..540T}, \citet{2023MNRAS.519.1526P}, and \citet{2025ApJ...979..193C}. For $1 < z < 4$, the \colibre{} \sfms{} is $\approx 0.2-0.6$~dex below all observed \sfmss{} at the high-mass end ($M_* \gtrsim 10^{10.5}~\mathrm{M_\odot}$), while still broadly matching the \sfms{} of \citet{2022ApJ...936..165L} and, to a somewhat lesser degree, \citet{2024A&A...691A.164K} at $M_* \lesssim 10^{10.5}~\mathrm{M_\odot}$. Below $z = 1$, the \colibre{} \sfms{} once again becomes broadly consistent with \citet{2021MNRAS.505..540T} and \citet{2023MNRAS.519.1526P}. While \colibre{} shows good convergence with resolution at $z \gtrsim 5$, at lower redshifts the higher-resolution models tend to predict systematically lower SFRs at fixed stellar mass, leading to larger deviations from the observations. This lower normalization of the \sfms{} in the higher-resolution simulations appears to correlate with a slightly higher normalization of their \gsmfs{} at similar or lower $M_*$ and corresponding redshifts (see Fig.~\ref{fig:gsmf_evolution_lowz}), which may point to small but systematic differences in stellar mass growth between different resolution levels.

The fact that \colibre{} falls short of the \sfms{} from \citet{2021MNRAS.505..540T} and \citet{2023MNRAS.519.1526P} by $0.2-0.6$~dex near cosmic noon warrants further discussion. As previously mentioned, a discrepancy of this magnitude between theoretical models and observations at $1 < z < 4$ is common in galaxy formation studies, and is believed to stem largely from methodological inconsistencies between theoretical predictions and observations. One way to mitigate these discrepancies is therefore to mock-observe simulated galaxies rather than relying on the internal SFRs and stellar masses predicted by the simulations. This approach involves generating realistic SEDs based on the distribution of stellar particle ages and metallicities, which are then processed through the same observational pipelines used for real data to infer SFRs and stellar masses, ensuring a more consistent comparison. By applying such a method to the \textsc{eagle} simulations, \citet{2020MNRAS.492.5592K} showed that the common tension between observed and simulated \sfms{} relations is largely alleviated.

Alternatively, the tension can be reduced by improving observational techniques to minimize inevitable systematic errors in the inferred stellar masses and SFRs, which would yield values closer to the true ones, thereby enabling a more consistent comparison with the intrinsic values from simulations \citep[see, e.g., fig. 1 in][]{2021MNRAS.508..219N}. Among the most promising improvements are the use of Bayesian SED fitting codes that model galaxy SFHs in a non-parametric way. Unlike traditional parametric models, which assume fixed functional forms for SFHs (e.g. a constant or exponentially declining SFR), non-parametric methods offer greater flexibility, enabling a more accurate inference of the complex and diverse shapes of the SFHs found in real galaxies. A notable example of the non-parametric approach is the work of \citet[][shown as black diamonds in Fig. \ref{fig:sfr_evolution}]{2022ApJ...936..165L}, who analyzed data from the COSMOS and 3D-\textit{HST} surveys over the redshift range $0.2 < z < 3$ using the Bayesian SED-fitting code \textsc{prospector} with non-parametric SFHs. They found systematically higher stellar masses and lower SFRs compared to studies relying on parametric SFHs combined with SFR indicators such as IR, IR+UV, H $\upalpha$, or radio. The authors attributed the SFR discrepancy to dust heating by older stellar populations, as inferred by \textsc{prospector}, while arguing that studies reporting higher SFRs have (mis)attributed dust heating to younger stars, leading to overestimated SFRs. In contrast, the increase in stellar masses is ascribed to the contribution of an additional component of old stars captured by non-parametric SFHs \citep[see the discussion in][]{2020ApJ...893..111L}. Overall, \citet{2022ApJ...936..165L} showed that their observationally inferred \sfms{} aligns closely with predictions from galaxy formation simulations, including \textsc{Eagle} \citep{2015MNRAS.446..521S}, \textsc{IllustrisTNG} \citep{2018MNRAS.473.4077P}, and \textsc{Simba} \citep{2019MNRAS.486.2827D}, thereby resolving (or at least strongly mitigating) the longstanding discrepancies between observed and simulated SFRs. 

To sum up, in Fig.~\ref{fig:sfr_evolution} we find that \colibre{} reproduces the observed \sfms{} reasonably well at $0 < z < 1$ and $z > 4$, whereas at intermediate redshifts the discrepancy between \colibre{} and observations is $\approx 0.2-0.6$~dex. While we cannot rule out the possibility that this discrepancy is due to intrinsic limitations of the simulations, and that accounting for systematic errors in observations might even exacerbate it, the growing evidence in the literature suggests that the discrepancy between simulations and observations around cosmic noon is dominated by systematic errors on the observational side, which can likely be alleviated by moving simulated galaxies into observational space and/or by improving observational techniques to reduce these systematic errors.

\begin{figure*}
    \centering
    \includegraphics[width=0.99\textwidth]{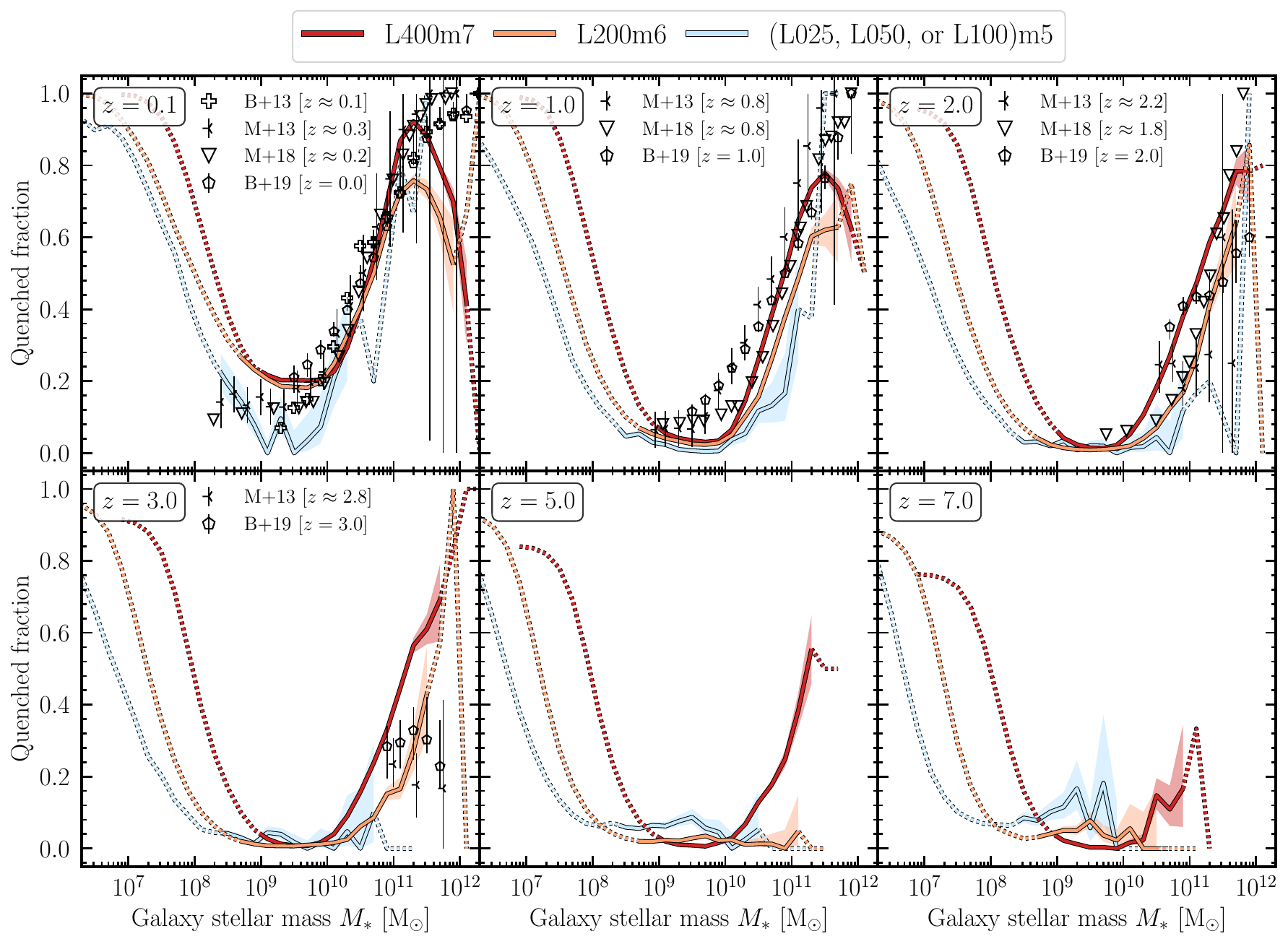}
    \caption{Evolution of the galaxy quenched fraction from $z=0.1$ to $z=7$ (different panels). Thick solid lines show the quenched fraction in the simulations, where galaxies are considered quenched if their $\mathrm{sSFR} < 0.2 / t_{\rm H}(z)$. Shaded regions indicate the uncertainty in the predicted quenched fractions, calculated as the Clopper–Pearson interval at the 68~per cent confidence level. At low stellar masses ($M_* \lesssim 10^{9}~\mathrm{M_\odot}$), the solid lines become dotted to indicate that the increase in quenched fraction may be an artefact of limited numerical resolution. The lines also become dotted at high $M_*$ where the number of objects per bin drops below 10. Black symbols represent a compilation of observational data from \citet{2013ApJ...777...18M}, \citet{2013MNRAS.434..209B}, \citet{2018MNRAS.477.1822M}, and \citet{2019MNRAS.488.3143B}. Overall, \colibre{} shows good agreement with the data at all redshifts. However, at $z \sim 0$, the most massive galaxies ($M_* \gtrsim 10^{11.5}~\mathrm{M_\odot}$) appear more star-forming than observed.}
    \label{fig:qf_evolution}
\end{figure*}

\begin{figure*}
    \centering
    \includegraphics[width=0.99\textwidth]{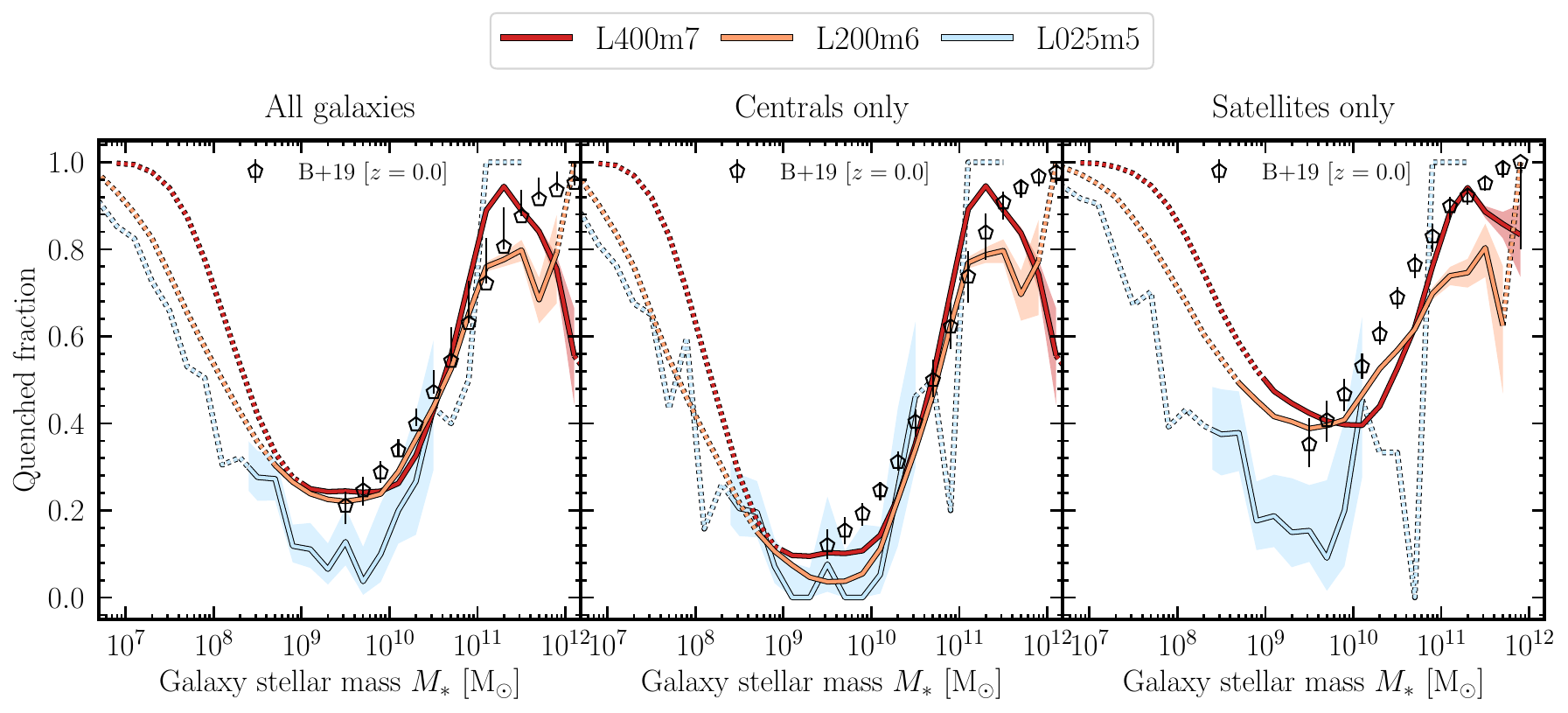}
    \caption{Quenched fractions at $z=0$ computed for all galaxies (\textit{left panel}), central galaxies (\textit{middle panel}), and satellite galaxies (\textit{right panel}). Black symbols show data from \citet{2019MNRAS.488.3143B}. \colibre{} predicts that, at fixed stellar mass, satellite galaxies are on average more quenched than centrals, in agreement with \citet{2019MNRAS.488.3143B}. The m5 model predicts systematically lower satellite quenched fractions compared to m6 and m7, due to its relatively small cosmological volume, which lacks massive host haloes (see Appendix \ref{appendix:convergence} for details).}
    \label{fig:qf_central_sat_difference}
\end{figure*}

\subsection{Evolution of the galaxy quenched fraction}
\label{subsection: evolution_of_qf}

Another valuable statistic derived from SFRs is the fraction of quenched galaxies at a given stellar mass. These are galaxies that are forming stars at such a low rate that their stellar mass remains nearly unchanged over a Hubble time. Accurately reproducing quenching is important, as quenched galaxies account for more than \mbox{$\approx 40$ per cent} of the observed CSMD at $z\approx 0$ \citep[e.g.][]{2023A&A...677A.184W}. Observations indicate that the fraction of quenched galaxies increases with cosmic time at fixed stellar mass, and with stellar mass at fixed redshift \citep[see][for a recent review on cosmic quenching]{2024A&A...687A..68D}. The primary mechanisms driving galaxy quenching include (i) AGN feedback in massive galaxies ($M_* \gtrsim 10^{10}~\mathrm{M_\odot}$), (ii) environmental effects such as tidal and ram pressure stripping in satellite galaxies of massive haloes, and (iii) stellar feedback in low-mass galaxies.

We show the fraction of quenched galaxies as a function of stellar mass in Fig.~\ref{fig:qf_evolution}. Each panel corresponds to a different redshift, ranging from $z = 0.1$ (top left) to $z = 7$ (bottom right). The quenched fraction in the simulations (shown as solid lines) is defined as the fraction of galaxies within a given stellar mass bin with sSFRs below $0.2 / t_{\rm H}(z)$ (black dotted lines in Fig.~\ref{fig:sfms_hist}). Such thresholds -- which compare galaxy sSFRs to a fraction of either the Hubble time or the age of the Universe at a given redshift -- are commonly used in the literature to distinguish between quiescent and star-forming galaxy populations \citep[e.g.][]{2008ApJ...688..770F, 2014ApJ...788...72G, 2025MNRAS.539..557B, 2024MNRAS.534..325C, 2025MNRAS.544.4482R}. At low stellar masses ($M_* \lesssim 10^{9}~\mathrm{M_\odot}$), the solid lines become dotted to emphasize that the rise in quenched fractions may be driven by limited numerical resolution (see below). The solid lines also become dotted at the high-mass end, indicating that there are fewer than 10 objects per stellar mass bin. The shaded regions indicate the uncertainty in the quenched fractions, computed as the Clopper–Pearson interval at the 68 per cent confidence level.

For comparison, we include the $0 < z < 4$ quenched fractions inferred from the semi-empirical model \textsc{emerge} \citep{2018MNRAS.477.1822M}, where quiescent galaxies are defined using $\mathrm{sSFR} < 0.3 / t_{\rm H}(z)$, and the $0 < z < 5$ quenched fractions inferred from the semi-empirical model \textsc{UniverseMachine} \citep{2019MNRAS.488.3143B}, where quiescent galaxies satisfy $\mathrm{sSFR} < 10^{-11}~\mathrm{yr}^{-1}$. Additionally, we include the $z\approx 0.1$ quenched fractions from GAMA reported by \citet{2013MNRAS.434..209B}, which were re-calculated for the $\mathrm{sSFR} < 10^{-11}~\mathrm{yr}^{-1}$ threshold by \citet{2019MNRAS.488.3143B}. Finally, we incorporate the $0.2 < z < 4$  quenched fractions from the COSMOS/UltraVISTA Survey \citep{2013ApJ...777...18M}, where a quenched galaxy population is defined using UVJ colour selection.

\colibre{} matches the $z \approx 0$ observational data reasonably well, with the exception of the dip at $M_* \gtrsim 10^{11.5}~\mathrm{M_\odot}$ and the increase at $M_* \lesssim 10^{9}~\mathrm{M_\odot}$. The \colibre{} quenched fraction for galaxies with $M_* \sim 10^9-10^{10}~\mathrm{M_\odot}$ stays at $\approx 20$ per cent in the m6 and m7 models, and at $\approx 10$ per cent in the m5 model. These values increase steeply for higher stellar masses, reaching \mbox{$\approx 75-90$ per cent} by $M_* \sim 10^{11.5}~\mathrm{M_\odot}$, in agreement with observations. For even higher stellar masses ($M_* \gtrsim 10^{11.5}~\mathrm{M_\odot}$), which correspond to brightest cluster galaxies (BCGs) and are only sampled in the volumes of the m7 and m6 simulations, the quenched fraction drops from $\approx 75-90$ per cent to $\approx 40-60$ per cent. This drop is consistent with \citet{2013ApJ...777...18M} but is not seen in the other datasets we compare to. 

To better understand the shape of the quenched fraction -- $M_*$ relation at the high-mass end and the importance of AGN feedback in driving it, we re-ran the \colibre{} m7 simulation in a ($50~\mathrm{cMpc}$)$^{3}$ cosmological volume without SMBHs. This test confirmed that the rise in the quenched fraction from $M_* \sim 10^{10}$ to $\sim 10^{11.5}~\mathrm{M_\odot}$ is driven by AGN feedback in massive haloes, while the decrease at $M_* \gtrsim 10^{11.5}~\mathrm{M_\odot}$ is likely an indication of cooling flows in the BCGs, which the \colibre{} AGN feedback model cannot fully counteract, resulting in overly high SFRs in these objects. However, the dip at $M_* \gtrsim 10^{11.5}~\mathrm{M_\odot}$ disappears when we increase the assumed lognormal scatter on the stellar mass from $\approx 0.1$~dex (as given by equation \ref{eq: random_scatter} at $z\approx 0$) to $\gtrsim 0.3$~dex (see Fig. \ref{fig:eddington_bias_plot} in Appendix \ref{appendix: effect of eddington bias}). Similarly, the dip also disappears if the SFR is computed within an aperture of $\lesssim 10$~pkpc (comparable to the physical size at $z=0.1$ of a fibre with a diameter of a few arcseconds in surveys such as SDSS or GAMA), rather than the default $50$~pkpc aperture used for all quantities in this work (not shown).

Below $M_* \sim 10^9~\mathrm{M_\odot}$, the increase in galaxy quenched fraction with decreasing stellar mass is likely driven (or at least exacerbated) by numerical effects due to insufficient sampling of the galaxy SFR by star-forming gas particles. As stellar mass decreases, the (star-forming) gas mass also declines, which worsens SFR sampling. Eventually (around a stellar mass of $M_* \sim 10^8 - 10^9~\mathrm{M_\odot}$ depending on resolution and redshift), the sampling becomes so sparse that galaxies frequently lack any star-forming gas particles, which artificially boosts the quenched fraction at lower stellar masses. To indicate this limitation of the simulations, in Fig.~\ref{fig:qf_evolution}, we switch the solid lines to dotted at stellar masses of $1 \times 10^9~\mathrm{M_\odot}$, $0.5 \times 10^9~\mathrm{M_\odot}$, and $0.25 \times 10^9~\mathrm{M_\odot}$ for the m7, m6, and m5 resolutions, respectively (i.e. $\propto m_{\rm gas}^{1/3}$), ignoring for simplicity the dependence on redshift. We observe that the increase in quenched fraction at low $M_*$ becomes less severe at higher resolutions but remains prominent even in the m5 simulation. A similar trend in quenched fraction was found in the \textsc{eagle} simulations by \citet{2015MNRAS.450.4486F}, although the differences between resolutions were somewhat larger.

At higher redshifts ($z\gtrsim 1$), \colibre{} remains in reasonably good agreement with the observational data for $M_* \gtrsim 10^9~\mathrm{M_\odot}$, with the quenched fraction steadily decreasing with increasing redshift at all stellar masses. However, caution should be taken when drawing more detailed conclusions, as differences in the criteria for defining quiescent galaxy populations between the observations and \colibre{} may influence the results. For galaxies with $M_* \sim 10^9-10^{10}~\mathrm{M_\odot}$, \colibre{} predicts a nearly zero quenched fraction by $z = 1$, which remains negligible also at higher redshifts. For $M_* \sim 10^{11}~\mathrm{M_\odot}$, the quenched fraction decreases from $\approx 70-80$~per cent at $z = 0.1$ to $20-40$~per cent by $z = 2$, and further drops to $5-30$~per cent by $z = 5$. 

Although the $z = 0.1$ quenched fractions show good convergence with resolution for $10^{10}\lesssim M_*/\mathrm{M_\odot}\lesssim 10^{11}$, at $z \geq 1$ the higher-resolution simulations predict systematically lower quenched fractions (i.e. weaker quenching) than m7 in this mass range. The explanation for this discrepancy is likely twofold. First, the gas surrounding SMBHs in the higher-resolution simulations is, on average, denser than in the m7 simulation, which may increase radiative cooling losses, thereby reducing the effectiveness of AGN feedback on the host galaxy. Second, according to the \colibre{} prescription for thermal AGN feedback, BHs inject energy into neighbouring gas particles such that the gas temperature increases by $\Delta T_{\rm AGN}$ upon injection. Because the gas particle mass is a factor of 8 (64) lower in the m6 (m5) simulation compared to m7, the energy required to heat a gas particle by the same $\Delta T_{\rm AGN}$ is also 8 (64) times lower in m6 (m5) than in m7. In other words, less energy is injected per AGN event in higher-resolution simulations, making the AGN feedback less bursty and potentially reducing its efficiency in expelling gas and quenching the galaxy. This may also explain why, at $z=0.1$, the dip in the galaxy quenched fraction at $M_* \gtrsim 10^{11.5}~\mathrm{M_\odot}$ is slightly more pronounced at m6 resolution than at m7.

Lastly, at $z = 7$, the \colibre{} m7 model predicts a quenched fraction of about 20~per cent at $M_* \sim 10^{11}~\mathrm{M_\odot}$, while the cosmological volumes of the m6 and m5 simulations are not large enough to make robust predictions at such high stellar mass and redshift.

Fig.~\ref{fig:qf_central_sat_difference} shows the $z=0$ quenched fractions for all galaxies (left panel), central galaxies (middle panel), and satellite galaxies (right panel). For comparison, we include observational data from \citet{2019MNRAS.488.3143B}, which also provides quenched fractions separately for centrals, satellites, and the overall galaxy population.

Satellites in \colibre{} exhibit systematically higher quenched fractions than centrals at fixed stellar mass, a trend consistent at all resolutions. For galaxies with $M_* \sim 10^9 - 10^{10}~\mathrm{M_\odot}$, the quenched fraction of centrals ranges between $0$ and $10$~per cent, whereas satellites show values of $\approx 15-40$ per cent, reflecting the important role of environmental effects. The impact of environmental effects on the quenched fraction diminishes for more massive satellites ($M_* \gtrsim 10^{10}~\mathrm{M_\odot}$). This can be explained by the deeper gravitational potential wells of more massive satellites, which allow them to better retain their gas reservoirs within the host environment (as seen in some numerical simulations and observations; e.g., \citealt{2024MNRAS.528.4891G}, though see also \citealt{2012MNRAS.424..232W,2023MNRAS.518.4782K}), and by the fact that they are, on average, accreted later and therefore spend less time within their host haloes. Instead, quenching in these massive satellites should primarily be driven by internal processes (in particular, AGN feedback), in a manner similar to central galaxies. As a result, the quenched fractions of massive satellites become increasingly similar to those of central galaxies at comparable stellar masses. A detailed study of environmental effects on the galaxy quenched fraction and other galaxy properties will be presented in Bah\'{e} et al. (in preparation). 

For central galaxies, the quenched fraction shows good convergence in the mass range $10^{10} \lesssim M_*/\mathrm{M_\odot} \lesssim 10^{11}$ and reasonable convergence at the low ($M_* \lesssim 10^{10}~\mathrm{M_\odot}$) and high ($M_* \gtrsim 10^{11}~\mathrm{M_\odot}$) mass ends, where the higher-resolution simulations tend to predict somewhat lower values. For satellite galaxies, the convergence is worse: in the $10^{9} \lesssim M_*/\mathrm{M_\odot} \lesssim 10^{10}$ range, the quenched fraction in the m5 model is a factor of $\approx 2$ lower than in the m6 and m7 counterparts. The cause for this discrepancy is the relatively small cosmological volume used in the m5 simulation, which biases the predicted satellite quenched fraction towards lower values due to the lack of massive haloes ($M_{\rm halo} \gtrsim 10^{13.5}~\mathrm{M_\odot}$). In Fig. \ref{fig:numerical_convergence_qf} of Appendix \ref{appendix:convergence}, we show that the bias is significantly reduced when all three resolutions are compared within the same cosmological volume $(25^3~\mathrm{cMpc}^3$).

\subsection{The abundance of massive quiescent galaxies at high redshift}

We next quantify the abundance of quenched galaxies as a function of redshift and assess whether \colibre{} predicts a number density of massive quiescent objects at high redshifts consistent with recent \textit{JWST} constraints.

\begin{figure}
    \centering
    \includegraphics[width=0.49\textwidth]{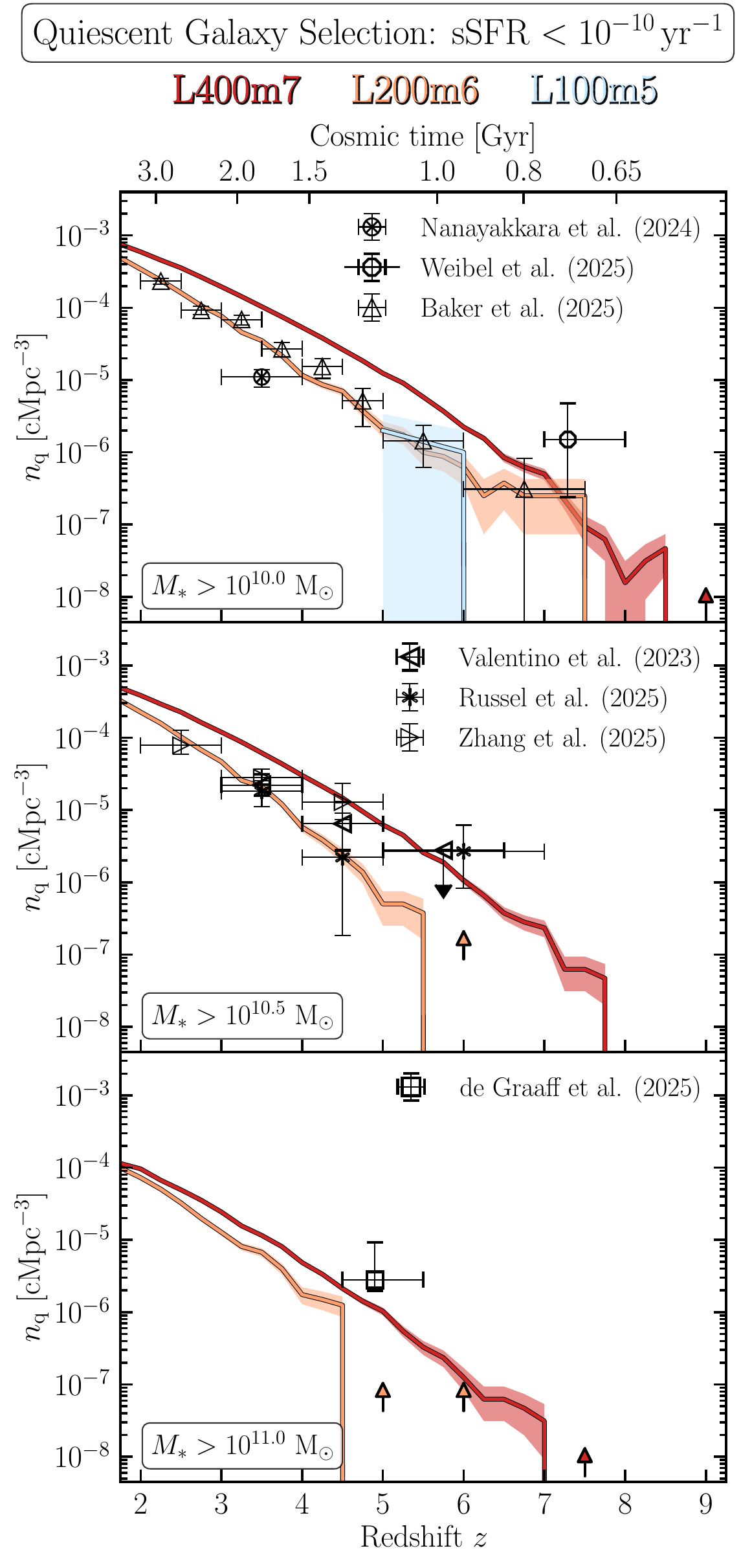}
     \vspace{-0.6cm}
    \caption{Evolution of the comoving number density of massive quiescent galaxies in the \colibre{} L400m7, L200m6, and L100m5 simulations (coloured curves). Quiescent galaxies are defined as those with $\mathrm{sSFR} < 10^{-10}~\mathrm{yr}^{-1}$, mimicking observational selection criteria at high redshift. We show the number densities of quiescent galaxies with stellar masses $M_* > 10^{10}~\mathrm{M_\odot}$ (\textit{top panel}), $M_* > 10^{10.5}~\mathrm{M_\odot}$ (\textit{middle panel}), and $M_* > 10^{11}~\mathrm{M_\odot}$ (\textit{bottom panel}). The shaded regions indicate the Poisson uncertainties on the simulation predictions. The upward-pointing arrows denote non-zero number density values in the regime of small-number statistics at high redshift and are coloured consistently with the corresponding curves (see text for details). For comparison, we include \textit{JWST} data with similar $M_*$ and sSFR thresholds from \citet{2023ApJ...947...20V, 2025ApJ...983...11W, 2024NatSR..14.3724N,2025MNRAS.544.4482R}, \citet{2025NatAs...9..280D}, \citet{2025A&A...702A.270B}, and \citet{2025arXiv250808577Z}. All \textit{JWST} data up to $z \approx 7$ are consistent with at least one of the \colibre{} simulations, although the m6 model systematically predicts a lower number density of quiescent galaxies than m7 at fixed redshift.}
    \label{fig:n_passive_vs_z}
\end{figure}

Fig.~\ref{fig:n_passive_vs_z} shows the redshift evolution of the number density of quiescent galaxies at $1.75<z < 9.25$. Unlike in Figs. \ref{fig:qf_evolution} and \ref{fig:qf_central_sat_difference}, here we define quenched galaxies as those with $\mathrm{sSFR} < 10^{-10}~\mathrm{yr}^{-1}$, to facilitate a fair comparison with the observational data (discussed later). We present the number densities of quiescent galaxies in \colibre{} for several stellar mass bins commonly used in observational studies: $M_* > 10^{10}~\mathrm{M_\odot}$ (top panel), $M_* > 10^{10.5}~\mathrm{M_\odot}$ (middle panel), and $M_* > 10^{11}~\mathrm{M_\odot}$ (bottom panel). The shaded regions indicate Poisson uncertainties in the simulation predictions. We show the predictions from the largest \colibre{} simulations at each resolution: L400m7, L200m6, and L100m5 (colours). Unlike in previous figures, we do not switch from the L100m5 simulation to L050m5 below $z = 4.6$, as the $50^{3}~\mathrm{cMpc}^{3}$ volume is too small for a meaningful comparison (see Fig.~\ref{fig:n_passive_vs_z_boxsize} in Appendix~\ref{appendix:convergence}).

Following $\S$\ref{subsection:assumptions}, the number densities predicted by the simulations are computed using galaxy stellar masses with a lognormal scatter (equation~\ref{eq: random_scatter}) to account for Eddington bias. The effect of Eddington bias on the number density of quiescent galaxies is discussed in Appendix~\ref{appendix: effect of eddington bias}, where we show that adding the scatter improves the agreement with observations (see Fig.~\ref{fig:eddington_bias_n_passive}). At very high redshift, the number of quiescent objects may fluctuate between zero and a few because of small-number statistics. In such cases, we use upward-pointing arrows (instead of plotting a continuous curve) to indicate redshift bins where the number density of quiescent galaxies is greater than zero but both adjacent bins are zero. The arrows are coloured consistently with the curves corresponding to the different resolutions, with their tips marking the corresponding non-zero number density values.

For comparison, in Fig.~\ref{fig:n_passive_vs_z} we include \textit{JWST} data from \citet{2023ApJ...947...20V}, \citet{2024NatSR..14.3724N}, \citet{2025MNRAS.544.4482R},  \citet{2025ApJ...983...11W}, \citet{2025NatAs...9..280D}, \citet{2025A&A...702A.270B}, and \citet{2025arXiv250808577Z}. Briefly, 
\begin{itemize}
    \item \citet{2024NatSR..14.3724N} reported 12 quiescent galaxy candidates with spectroscopically confirmed redshifts in the redshift range $z=3-4$. All galaxies have stellar masses $M_* > 10^{10}~\mathrm{M_\odot}$ and were classified as quiescent using the UVJ colour selection criterion. Additionally, the authors showed that all but one galaxy have $\text{sSFR} < 10^{-10}~\mathrm{yr}^{-1}$ over the past 10 Myr. 
    
    \item \citet{2025ApJ...983...11W} reported the discovery of a massive quiescent galaxy at redshift $z = 7.29$ within the $\sim 300$ arcmin$^2$ survey area covered by PRIMER and CEERS. Using \textit{JWST}/NIRCam and MIRI photometry, along with \textit{JWST}/NIRSpec PRISM spectroscopy from the RUBIES program, they estimated the galaxy’s stellar mass to be $M_*\approx 10^{10.2}~\mathrm{M_\odot}$, with an sSFR of $< 10^{-10}~\mathrm{yr}^{-1}$ over the past 50 Myr. 
    
    \item \citet{2023ApJ...947...20V} found $\approx 80$ quiescent galaxy candidates with photometric redshifts of $3 < z < 5$ based on NIRCam imaging in 11 \textit{JWST} fields, covering a total effective area of $\approx 145$ arcmin$^2$. From this study, we plot the `strict' galaxy subsample with stellar masses $M_* > 10^{10.6}~\mathrm{M_\odot}$ that were classified as quiescent based on rest-frame UVJ colours. 
    
    \item \citet{2025MNRAS.544.4482R} identified $\sim 100$ quiescent galaxies at redshifts $3 < z < 7$ in the CEERS, NEP, and JADES fields (covering a total area of $\approx 144.45$ arcmin$^2$), using \textit{JWST}/NIRCam imaging data. The galaxies were classified as quiescent if their sSFR satisfied $\mathrm{sSFR} < 0.2 / t_{\rm age}(z)$, where $t_{\rm age}(z)$ is the age of the Universe at redshift $z$. At $z=5$, this threshold corresponds to an $\text{sSFR} \approx 1.7 \times 10^{-10}~\mathrm{yr}^{-1}$. In Fig.~\ref{fig:n_passive_vs_z}, we plot the `robust' subsample of galaxies from this study with stellar masses $M_* > 10^{10.6}~\mathrm{M_\odot}$ ($12$ galaxies in total).
    
    \item  \citet{2025NatAs...9..280D} identified one quiescent galaxy with $M_* \sim 10^{11}~\mathrm{M_\odot}$ at $z = 4.9$ within the RUBIES survey, observed using both \textit{JWST}/NIRCam and \textit{JWST}/NIRSpec. An effective surface area used in the study was $\sim 100$ arcmin$^2$. The authors defined the galaxy as quiescent based on its spectrum and determined that the galaxy has an sSFR over the past $100$~Myr of just $\sim 10^{-11}~\mathrm{yr}^{-1}$. Since the authors do not report an uncertainty on their measurement, we estimate it assuming a Poisson distribution with an unknown mean $\lambda$ and an observed count of $k=1$. Adopting a flat prior on $\lambda$ and using Bayesian inference, the posterior is a gamma distribution with shape parameter $\alpha=2$. We take the $16^{\rm th}$ and $84^{\rm th}$ percentiles of this distribution as an estimate of the uncertainty. 
    
    \item \citet{2025A&A...702A.270B} analyzed a sample of 745 massive quiescent galaxies over the redshift range $2 < z < 7$, using \textit{JWST}/NIRCam imaging from a compilation of public \textit{JWST} fields covering a combined area of over $800$ arcmin$^2$. Photometric redshifts were derived using \textsc{eazy}, and quiescent galaxies were identified based on an sSFR threshold of $\mathrm{sSFR} < 0.2 / t_{\rm age}(z)$. For our comparison, we consider only the subset of their sample with stellar masses $M_* > 10^{10}~\mathrm{M_\odot}$.
    
    \item \citet{2025arXiv250808577Z} presented a sample of 20 massive ($M_* > 10^{10.3}~\mathrm{M_\odot}$) quiescent galaxies at $2 < z < 5$  with \textit{JWST}/NIRSpec PRISM spectra from the RUBIES survey, with a total surface area of $\approx 150$ arcmin$^2$. We plot their subsample of galaxies satisfying $\mathrm{sSFR} < 10^{-10}~\mathrm{yr}^{-1}$ and $M_* > 10^{10.5}~\mathrm{M_\odot}$.
\end{itemize}

We find that the number density of massive quiescent galaxies, $n_{\rm q}$, predicted by the \colibre{} simulations increases steeply with decreasing redshift, rising by more than four orders of magnitude between $z = 9$ and $z = 2$ in all three stellar mass bins. Furthermore, comparing the different panels at fixed redshift, we see that $n_{\rm q}$ decreases as the stellar mass threshold increases from $M_* = 10^{10}~\mathrm{M_\odot}$ to $10^{11}~\mathrm{M_\odot}$. Thanks to the relatively large volumes of the m7 and m6 simulations, quiescent galaxies with $M_*> 10^{10}~\mathrm{M_\odot}$ are sampled out to $z = 9$ and $z = 7.5$ in the m7 and m6 simulations, respectively. For the most massive quiescent galaxies ($M_* > 10^{11}~\mathrm{M_\odot}$), the m7 (m6) simulation predicts a non-zero number density at $z \leq 7.5$ ($z \leq 6$). The predictions from the m5 simulation are present only over a very narrow redshift interval, $5 < z < 6$, and only for the lowest stellar mass bin ($M_* > 10^{10}~\mathrm{M_\odot}$). This is expected: at $z \lesssim 5$ the simulation is not yet available, while at higher redshifts the $(100~\mathrm{cMpc})^{3}$ volume of the m5 simulation is too small to yield robust predictions, particularly in the higher stellar mass bins.

When comparing the m6 and m7 simulations at fixed redshift and for the same stellar mass bin, we find that the predicted values of $n_{\rm q}$ are systematically higher in the m7 simulation, with differences ranging from $\approx 0.2$ to $\approx 1$~dex\footnote{Unlike the m6 and m7 models, the m5 and m6 simulations formally converge for the lowest stellar mass bin in the narrow redshift range where the predictions of the L100m5 simulation are available ($5 < z < 6$). However, given the tendency of the $100^3~\mathrm{cMpc}^3$ volume to produce higher $n_{\rm q}$ than larger volumes at these redshifts (see Fig.~\ref{fig:n_passive_vs_z_boxsize}), the predictions of the m5 model in a larger volume would likely lie slightly below those of L200m6.}. As shown in Appendix \ref{appendix:convergence}, these discrepancies cannot be due to the eight times larger volume of the m7 simulation. Instead, they arise from weaker, more gradual quenching at high redshifts in the higher-resolution \colibre{} simulations -- a trend already observed in the evolution of the galaxy quenched fraction (see Fig.~\ref{fig:qf_evolution}). As discussed in $\S$\ref{subsection: evolution_of_qf}, a likely explanation for this behaviour is stronger radiative cooling losses at higher resolution, combined with less bursty AGN feedback, leading to overall weaker AGN feedback and consequently weaker quenching of massive galaxies.

All \textit{JWST} measurements are reproduced by at least one of the \colibre{} simulations. In the $M_* > 10^{10}~\mathrm{M_\odot}$ mass bin, the m6 and m7 simulations both predict $n_{\rm q} \sim 10^{-6.5}~\mathrm{cMpc}^{-3}$ at $z \approx 7$, which lies within $\approx 1\sigma$ of the measurements by \citet{2025ApJ...983...11W} and \citet{2025A&A...702A.270B}. The m6 simulation is also consistent with \citet{2025A&A...702A.270B} at all lower redshifts probed by the observations ($2 < z < 6$), as well as with the measurement of \citet{2024NatSR..14.3724N} at $z \approx 3.5$. The m5 model, where available, is likewise consistent with \citet{2025A&A...702A.270B}. In contrast, the m7 model overpredicts the observationally inferred number densities from \citet{2025A&A...702A.270B} and \citet{2024NatSR..14.3724N} by $\approx 0.25$~dex to $\approx 1$~dex at these redshifts. In the $M_* > 10^{10.5}~\mathrm{M_\odot}$ bin, the m7 and m6 simulations predict quiescent galaxy number densities of $n_{\rm q} \sim 10^{-6}~\mathrm{cMpc}^{-3}$ at $z \approx 6$ and $z \approx 5$, respectively, rising in both cases to $\sim 10^{-3.5}~\mathrm{cMpc}^{-3}$ by $z \approx 2$. The high-redshift measurements reported by \citet{2023ApJ...947...20V} and \citet{2025MNRAS.544.4482R} at $z \approx 6$, as well as by \citet{2025arXiv250808577Z} at $z \approx 4.5$, are reproduced by the m7 simulation, while the m6 simulation shows good agreement with the lower-redshift bins ($z \lesssim 4$) reported in these studies. For the highest mass bin, the m7 (m6) simulation predicts a number density of $\approx 1.5 \times 10^{-6}~\mathrm{cMpc}^{-3}$ ($\approx 10^{-7}~\mathrm{cMpc}^{-3}$) at $z \approx 4.9$, placing it within $\approx 1.5\sigma$ ($\approx 3\sigma$) of the single galaxy-based estimate by \citet{2025NatAs...9..280D}.

We note that earlier galaxy formation simulations and semi-analytic models that predict reasonably realistic low-redshift galaxy populations typically strongly underpredict the number density of massive quiescent galaxies observed by \textit{JWST} at $z \gtrsim 5$ (see, e.g., fig. 6 in \citealt{2025ApJ...983...11W} and the section `Theoretical predictions' in \citealt{2025NatAs...9..280D}). The improved performance of \colibre{} in this metric may be attributed to several factors. First, unlike most earlier galaxy simulations, \colibre{} does not implement an effective equation of state (EoS) to model the ISM. The absence of an EoS results in denser and cooler gas within the ISM, which can enable more rapid BH accretion. Consequently, the more efficiently growing SMBHs may trigger stronger AGN feedback, facilitating the quenching of massive galaxies at earlier times. Second, \colibre{} does not impose a cap on SMBH accretion at the Eddington rate. This approach also allows faster SMBH growth and potentially stronger quenching, particularly at high redshifts where super-Eddington accretion events are expected to be more common \citep[e.g.][]{2025MNRAS.537.2559H}. A detailed investigation of the relevance of these factors will be conducted in future work. 

Furthermore, as explained in $\S$\ref{subsection:assumptions}, galaxy stellar masses in this work include a lognormal scatter with a standard deviation given by equation (\ref{eq: random_scatter}) to account for Eddington bias, which is present in observational data due to non-negligible uncertainties in the inferred stellar masses. As we show in Fig.~\ref{fig:eddington_bias_n_passive} in Appendix~\ref{appendix: effect of eddington bias}, the number density of massive quiescent galaxies is particularly sensitive to this scatter, which on average boosts the value of $n_{\rm q}$ at fixed redshift and stellar mass bin. This effect may not have been fully accounted for in some previous works comparing simulations with observations, potentially biasing simulation-based estimates of $n_{\rm q}$ towards lower values.

Lastly, we note that many earlier galaxy formation simulations were conducted in cosmological volumes of $\sim 100^{3}~\mathrm{cMpc}^{3}$, which strongly limits the non-zero number density of quenched galaxies that can be predicted at high redshift. For example, \citet{2023MNRAS.525.5520L} showed that the highest redshift at which a non-zero abundance of massive quenched systems is found in the \textsc{eagle} simulations increases from $z \approx 5$ in the original \textsc{eagle} volume of $100^{3}~\mathrm{cMpc}^{3}$ to $z \approx 8$ in the \textsc{Flares} simulations, which are zoom-in simulations selected from a parent volume of $3.2^{3}~\mathrm{Gpc}^{3}$ and employing the same \textsc{eagle} galaxy formation model.

We emphasize that the number density of quiescent galaxies presented in this work is computed using an $\mathrm{sSFR} < 10^{-10}~\mathrm{yr}^{-1}$ threshold to define quiescent galaxies. In all but one of the observational studies that we show for comparison in Fig.~\ref{fig:n_passive_vs_z}, galaxies were selected using (or satisfy) a similar sSFR threshold (see precise definitions above) to maximize consistency in the comparison and minimize potential bias in our conclusions. Furthermore, very recently, our results were confirmed by \citet{2025arXiv251216208C}, who conducted a detailed study of the properties of massive quenched galaxies in \colibre{} at high redshift, finding generally good agreement with \textit{JWST} data while adopting $\mathrm{sSFR} < 0.2/t_{\rm age}(z)$ to define quiescent systems.

\subsection{Comparison of models with thermal and hybrid AGN feedback}
\label{subsection:thermal_vs_hybrid}

\begin{figure*}
    \centering
    \includegraphics[width=0.99\textwidth]{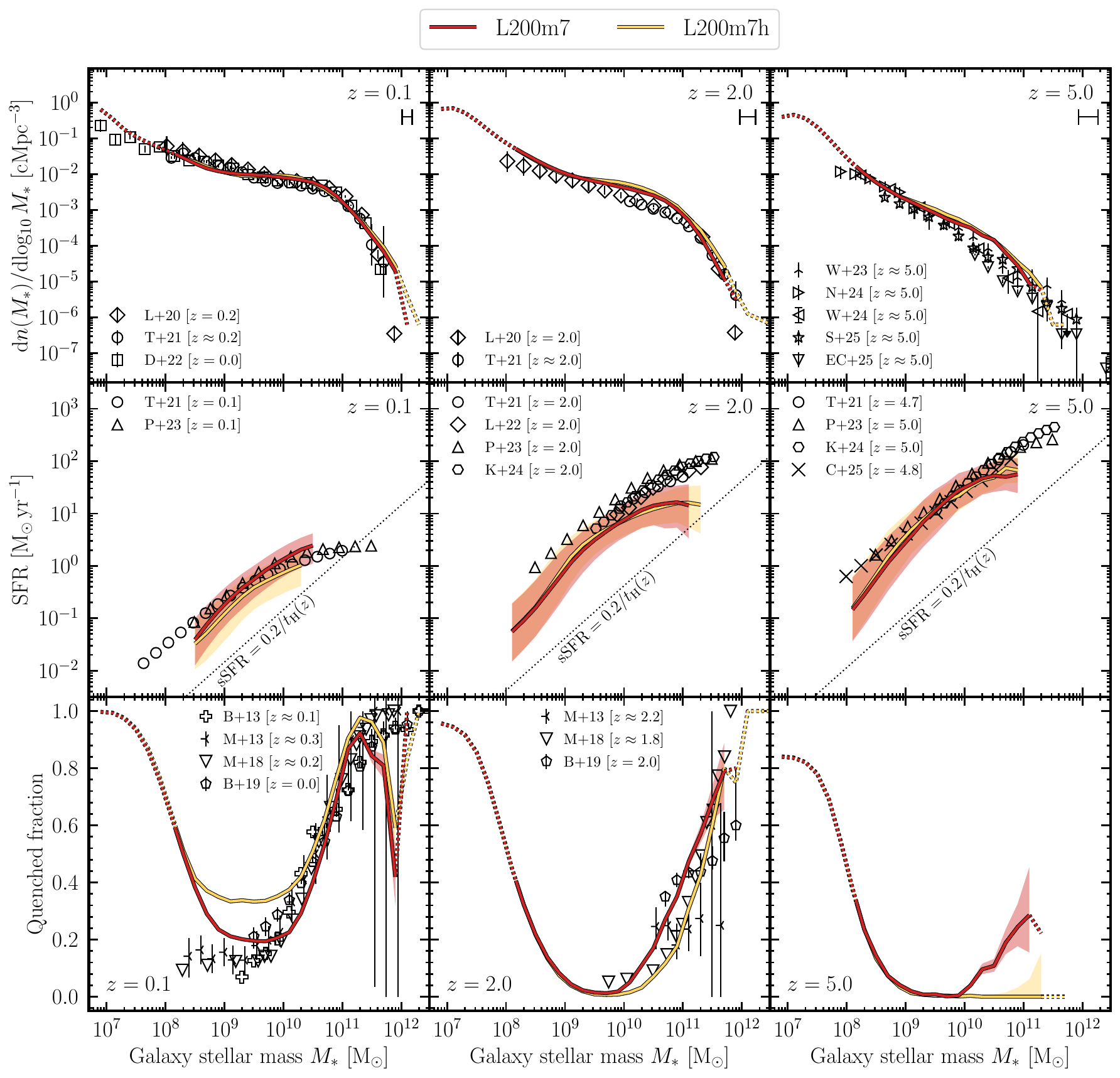}
    \caption{Comparison of the \colibre{} m7 simulations in a ($200~\mathrm{cMpc}$)$^3$ volume with thermal (\msevencolor) and hybrid (\msevenhybridcolor) AGN feedback at $z=0.1$ (\textit{left}), $z=2$ (\textit{middle}), and $z=5$ (\textit{right}). \textit{Top row:} galaxy stellar mass function (\gsmf). \textit{Middle row:} star-forming main sequence (\sfms), defined as the median SFR of galaxies with $\mathrm{SFR} > 0.1 \, \times$ the SFR at the \sfms. \textit{Bottom row:} galaxy quenched fraction, where quenched galaxies are defined as $\mathrm{sSFR} < 0.2 / t_{\mathrm{H}}(z)$. The shaded regions in the middle row indicate the 16$^{\rm th}$ to 84$^{\rm th}$ percentiles of the SFR distribution of galaxies with $\mathrm{SFR} > 0.1 \, \times$ the SFR at the \sfms, while those in the bottom row show the uncertainty in the quenched fractions, calculated using the Clopper–Pearson interval at the 68 per cent confidence level. For reference, the horizontal black error bar in the top panels marks the systematic uncertainty in stellar mass measurements given by equation~(\ref{eq: sys_error}), while the diagonal thin dotted lines in the middle panels indicate $\mathrm{sSFR} = 0.2 / t_{\rm H}(z)$. The observational data indicated by black symbols are the same as in Figs. \ref{fig:gsmf_evolution_lowz}, \ref{fig:gsmf_evolution_highz}, \ref{fig:sfr_evolution}, and \ref{fig:qf_evolution}. While the hybrid and thermal models yield nearly identical \gsmfs{} and \sfmss, the hybrid model shows weaker quenching at $z = 2$ and $z = 5$, but stronger quenching at $z = 0.1$ compared to the thermal model.}
    \label{fig:thermal_vs_hybrid_agn}
\end{figure*}

\begin{figure*}
    \centering
    \includegraphics[width=0.49\linewidth]{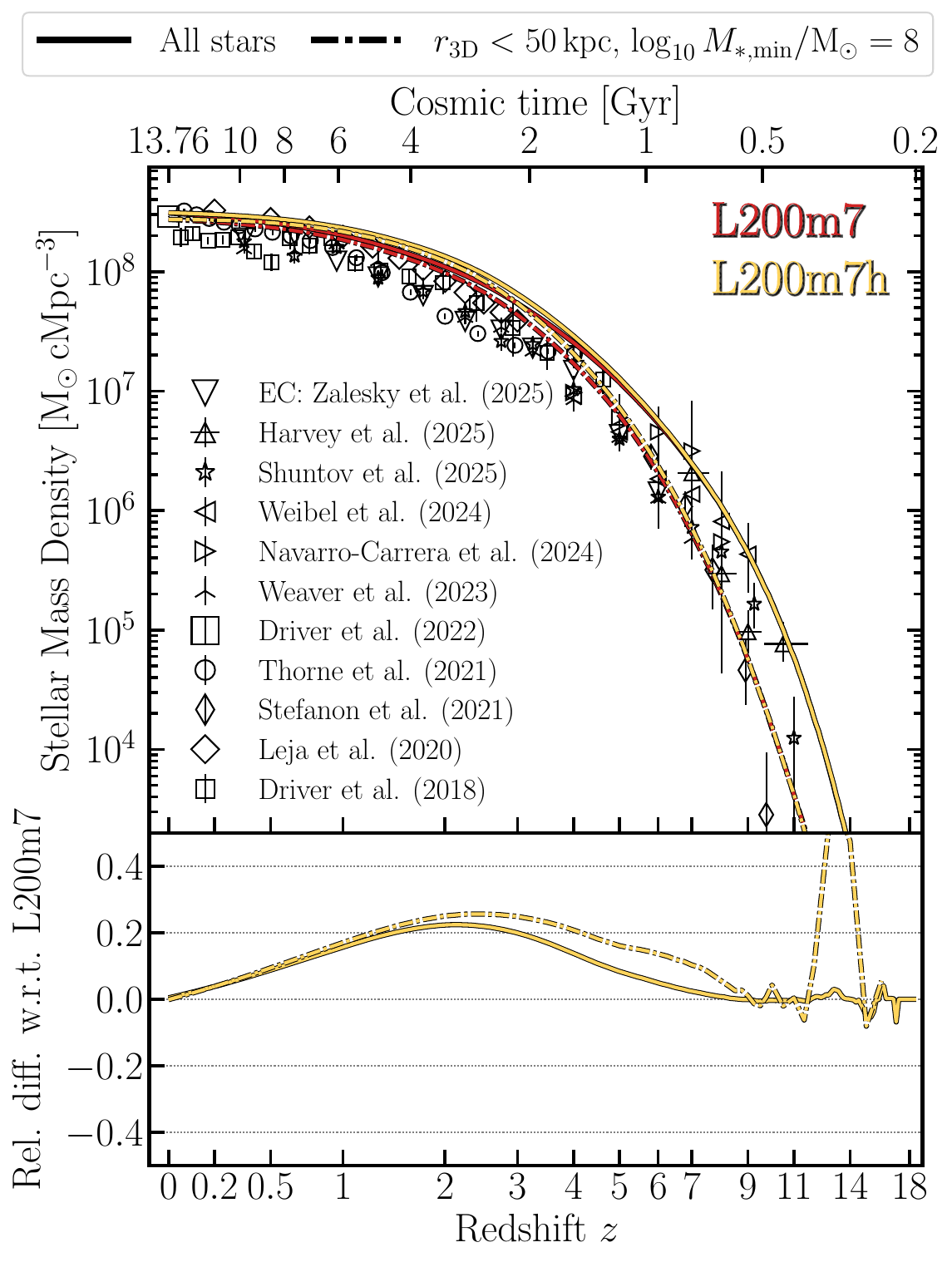}
    \includegraphics[width=0.49\linewidth]{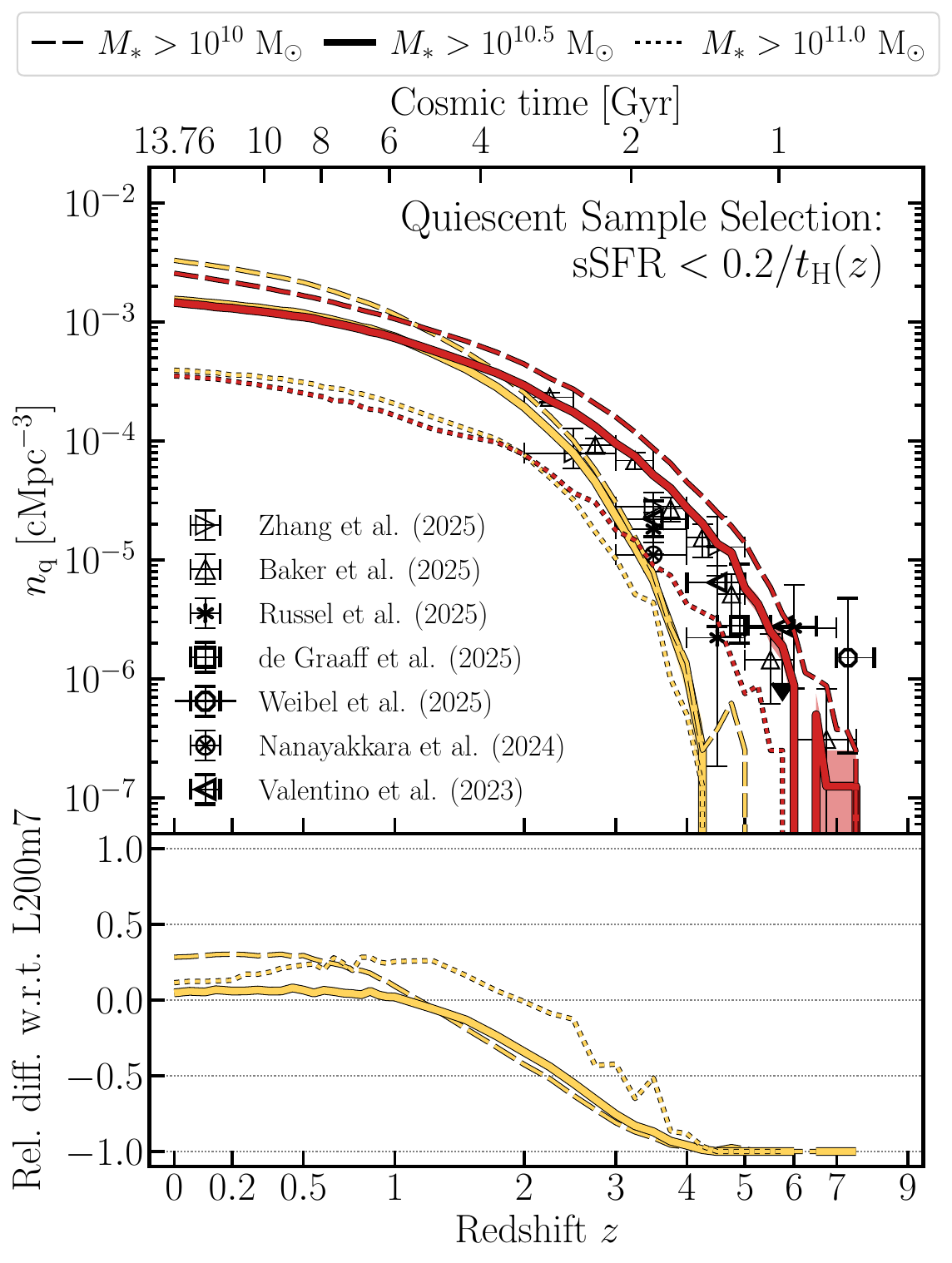}
    \caption{Evolution of the cosmic stellar mass density (CSMD, \textit{left}) and the number density of massive quiescent galaxies, $n_{\rm q}$, (\textit{right}) in the \colibre{} L200m7 simulations with thermal (\msevencolor) and hybrid (\msevenhybridcolor) AGN feedback. The top panels show simulation predictions (curves) compared to observational data (black symbols), while the bottom panels display the relative differences between the two simulations, defined as in Fig. \ref{fig:CSMD_evolution}. Different line styles indicate the different selection criteria used to compute the CSMD and $n_{\rm q}$ in the simulations (see main text for details). At $z \gtrsim 1$, the hybrid model predicts a higher CSMD than the thermal model, owing predominantly to weaker quenching at early cosmic times. However, by $z = 0$, the CSMDs of the two models converge, as quenching becomes stronger in the hybrid model than in the thermal model at $z \lesssim 1$.}
    \label{fig:thermal_vs_hybrid_agn_evolution}
\end{figure*}

Thus far, we have discussed exclusively the \colibre{} simulations with thermal AGN feedback. However, the \colibre{} suite also includes models with hybrid (thermal + kinetic jet) AGN feedback. In this section, we make use of these additional simulations to investigate how differences in AGN feedback prescriptions impact the key galaxy properties examined in this study: stellar masses, SFRs, and quenched fractions. We emphasize that both the thermal and hybrid models were calibrated to reproduce the $z=0$ \gsmf{} from \citet{2022MNRAS.513..439D} and the $z \approx 0$ stellar mass -- size relation from \citet{2022MNRAS.509.3751H} in the stellar mass range $10^9 < M_* / \mathrm{M}_\odot < 10^{11.3}$, as well as the $z \approx 0$ masses of SMBHs in massive galaxies. This implies that the properties of galaxies other than their $z=0$ stellar masses, half-mass radii, and SMBH masses should not necessarily be expected to match, especially at higher redshifts and outside the calibrated stellar mass range, where no observational constraints were applied to tune the simulations. Additionally, although the primary difference between the thermal and hybrid models lies in the AGN feedback prescription, we note that one of the SN feedback parameters also differs slightly. Namely, at all resolutions, the normalization of the gas density-dependent heating temperature $\Delta T_{\rm SN}(\rho_{\rm gas}) \propto \rho_{\rm gas}^{2/3}$ is increased in the hybrid model, resulting in lower heating temperatures at fixed gas density and therefore weaker SN feedback. This adjustment was introduced to counterbalance the stronger AGN feedback in the hybrid model in intermediate-mass galaxies (see \citealt{2026MNRAS.547ag324H} for details).

For clarity, we present a comparison of the \colibre{} thermal and hybrid models at a single resolution level: m7. We selected this resolution\footnote{We have verified that the general trends discussed below for m7 resolution also hold at m6 resolution.} to maximize galaxy number statistics at the high-mass end, as the largest \colibre{} volumes are available for m7 resolution. Specifically, the largest m7 volume in which the hybrid model was run is $(200~\mathrm{cMpc})^3$, whereas the thermal model is available in $(200~\mathrm{cMpc})^3$ and $(400~\mathrm{cMpc})^3$ volumes. Therefore, to ensure a consistent comparison, we use the $(200~\mathrm{cMpc})^3$ volume for both models.

The top, middle, and bottom row of Fig.~\ref{fig:thermal_vs_hybrid_agn} show, respectively, the \gsmf, the \sfms, and the galaxy quenched fraction, all in the m7 \colibre{} simulations using thermal (\msevencolor) and hybrid (\msevenhybridcolor) AGN feedback. The three columns, from left to right, correspond to redshifts $z = 0.1$, $z = 2$, and $z = 5$. The solid lines and shaded regions in the second and third rows are defined in the same way as in Figs. \ref{fig:sfr_evolution}, and \ref{fig:qf_evolution}, respectively. The redshifts $z=0.1$, $2$, and $5$ have been chosen to highlight various evolutionary phases of the galaxy populations:

\begin{itemize}
    \item At $z = 5$, the \gsmfs{} in the thermal and hybrid models are nearly indistinguishable, with the hybrid \gsmf{} being marginally higher and both \gsmfs{} reproducing the observational data. The \sfmss{} are also nearly identical. However, the quenched fractions differ: the thermal model predicts a monotonic increase in the quenched fraction with stellar mass, from $\approx 0$ per cent at $M_* \sim 10^{10}~\mathrm{M_\odot}$ to $\approx 30$ per cent at $M_* \sim 10^{11}~\mathrm{M_\odot}$, whereas the hybrid model predicts a quenched fraction of zero in the same stellar mass range.

    \item At $z=2$, the \sfmss{} in the hybrid and thermal models remain indistinguishable. Quenching continues to be stronger in the thermal model, although the differences are no longer as pronounced: in the hybrid model, the quenched fraction increases with stellar mass from $\approx 0$ per cent at $M_* \sim 10^{10}~\mathrm{M_\odot}$ to $\approx 30$ per cent at $M_* \sim 10^{11}~\mathrm{M_\odot}$, whereas in the thermal model, it rises from $\approx 0$ to $\approx 40$ per cent over the same mass range. At even higher stellar masses ($M_* \gtrsim 10^{11.5}~\mathrm{M_\odot}$), the quenched fractions in the two models converge, while continuing to rise, reaching $\approx 80$ per cent by $M_* \approx 5\times 10^{11}~\mathrm{M_\odot}$. At $z=2$, non-negligible differences also emerge in the \gsmf, with the hybrid model predicting values that are $\approx 0.05-0.1$~dex higher than those of the thermal model at $M_* \gtrsim 10^{9.5}~\mathrm{M_\odot}$.

    \item At $z=0.1$, the \gsmfs{} of the thermal and hybrid models again become indistinguishable, while the normalization of the \sfms{} is about $0.1$~dex lower in the hybrid model. Interestingly, the trend in quenched fractions reverses: the hybrid model now exhibits a higher quenched fraction, reaching $\approx 35$ per cent at $10^{9} \lesssim M_*/\mathrm{M_\odot} \lesssim 10^{10}$, compared to $\approx 20$ per cent in the thermal model. This is a consequence of the stronger AGN feedback produced by the hybrid model, which extends into the dwarf galaxy regime \citep[see][]{2026MNRAS.547ag324H} and will be studied in detail in future work. At higher stellar masses ($M_* \gtrsim 10^{11}~\mathrm{M_\odot}$), the quenched fractions in both models converge, with the hybrid model predicting marginally higher values. At even greater masses ($M_* \gtrsim 10^{11.5}~\mathrm{M_\odot}$), the quenched fractions in both models decline to around 50~per cent, suggesting that neither thermal energy injection nor kinetic jets can fully offset the strong cooling flows in the most massive galaxies. We note, however, that the quenched fractions at the high-mass end are particularly sensitive to uncertainties in stellar mass due to Eddington bias, which may artificially elevate their values (see Appendix \ref{appendix: effect of eddington bias}).
\end{itemize}

The hybrid model exhibits significantly lower quenched fractions at high redshifts due to weaker AGN feedback, resulting mainly from slower BH growth at high $z$, as well as from the fact that jets in the hybrid model require longer time-scales to quench the host galaxy (see fig. G1 in \citealt{2025arXiv251216208C}). Although both the thermal and hybrid AGN feedback models use the modified Bondi-Hoyle-Lyttleton formula (with turbulence and vorticity corrections from \citealt{Krumholz_et_al_2006}) to compute BH mass accretion rates, the latter further modifies the accretion rate by multiplying it by the accretion efficiency, a dimensionless parameter $0 \leq \varepsilon_{\rm accr} \leq 1$. Physically, this parameter reduces the accretion rate into the BH event horizon relative to that at the scales of the BH accretion disc to account for mass loss due to accretion disc winds \citep[see][for details]{2026MNRAS.547ag324H}. At very high accretion rates (Eddington fractions greater than 1), which are readily reached in the \colibre{} simulations at high redshift, $\varepsilon_{\rm accr}$ is set to 1 per cent, which slows down the BH mass growth in the hybrid model relative to the thermal model (see Hu\v{s}ko et al., in preparation, for further details).

Over time, the lower quenched fractions in the hybrid model, combined with slightly weaker SN feedback, allow more stellar mass to form -- resulting in a marginally elevated \gsmf{} in the hybrid model at $M_* \gtrsim 10^{9.5}~\mathrm{M_\odot}$ by $z=2$. By $z=0.1$, the \gsmfs{} in both models converge, as they were calibrated to match the same observational constraints. Mathematically, the low$-z$ quenching in the hybrid model is expected to be stronger than in the thermal model, to ensure that the same amount of stellar mass is formed by $z=0$. Physically, this stronger quenching arises from the hybrid model’s more efficient AGN feedback, which at low redshifts is dominated by the kinetic jet mode, which experiences less radiative losses compared to thermal AGN feedback.

To better understand these differences between the two models, Fig.~\ref{fig:thermal_vs_hybrid_agn_evolution} shows the time evolution of the CSMD (left panel) and the number density of massive quiescent galaxies (right panel). The CSMD is plotted using two different apertures and galaxy stellar mass selection criteria, as in Fig.~\ref{fig:CSMD_evolution}, while the number density of massive quiescent galaxies is shown for three different stellar mass thresholds, as in Fig.~\ref{fig:n_passive_vs_z}. In Fig.~\ref{fig:thermal_vs_hybrid_agn_evolution}, we use a criterion for quiescent galaxies of $\mathrm{sSFR} < 0.2/t_{\rm H}(z)$, which is roughly equivalent to a fixed threshold of $\mathrm{sSFR} < 10^{-10}~\mathrm{yr^{-1}}$ at $3 < z < 7$, but approaches $\mathrm{sSFR} < 10^{-11}~\mathrm{yr^{-1}}$ at $z=0$. The top row shows the predictions from the simulations using the thermal and hybrid models, along with their comparison to observational data. The bottom row presents the relative differences between the models, using the thermal model as the baseline.

Both models form similar amounts of stellar mass at $z \gtrsim 7$. Between $z = 7$ and $z = 2$, the CSMDs in the two models diverge, with the differences peaking at $z \approx 2$, where the hybrid model has formed $\approx 25$~per cent more stellar mass. From $z = 2$ to $0$, the CSMDs converge, with the differences approaching zero by $z \approx 0$, as both models were calibrated to reproduce the same observed $z = 0$ \gsmf.

The behaviour of the number density of quiescent galaxies is in line with the stronger quenching predicted by the thermal model at high redshifts, as shown in Fig.~\ref{fig:thermal_vs_hybrid_agn}. The thermal model predicts a significantly higher number density of quiescent galaxies at $z \gtrsim 2$ across all three stellar mass bins, allowing it to better match the \textit{JWST} data compared to the hybrid model. By contrast, quenching in the hybrid model becomes as strong as that in the thermal model by $z \approx 1$ and exceeds it by about 30~per cent by $z = 0$ for galaxies with $M_* > 10^{10}~\mathrm{M_\odot}$. The number density of galaxies with $M_*> 10^{10.5}~\mathrm{M_\odot}$ remains similar between the two models at $0 < z < 1$, while the abundance of the most massive objects ($M_*> 10^{11}~\mathrm{M_\odot}$) is slightly higher in the hybrid model at $z < 2$, but converges to that of the thermal model by $z = 0$.

\section{Conclusions}
\label{section: conclusions}

In this work, we have studied the evolution of stellar masses and star formation rates in new \colibre{} cosmological hydrodynamical simulations of galaxy formation  \citep{2026COLIBREproject,2026COLIBREcalibration}. We analyzed a suite of fiducial \colibre{} simulations covering a range of cosmological volumes and resolutions: $(400~\mathrm{cMpc})^3$ and $(200~\mathrm{cMpc})^3$ at m7 resolution with a gas particle mass of $m_{\rm gas} = 1.47 \times 10^7~\mathrm{M_\odot}$; $(200~\mathrm{cMpc})^3$ at m6 resolution with $m_{\rm gas} = 1.8 \times 10^6~\mathrm{M_\odot}$; and $(100~\mathrm{cMpc})^3$, $(50~\mathrm{cMpc})^3$, and $(25~\mathrm{cMpc})^3$ at m5 resolution with $m_{\rm gas} = 2.3 \times 10^5~\mathrm{M_\odot}$. Our main focus was the evolution of the galaxy stellar mass function (\gsmf), which we predicted over $0 < z < 17$. Additionally, we investigated the evolution of the stellar-to-halo mass relation (\shmr), star-forming main sequence (\sfms), galaxy quenched fraction, cosmic star formation rate density (CSFRD), cosmic stellar mass density (CSMD), and the number density of massive quiescent galaxies at high redshift. To facilitate a fair comparison with observations, where inferred stellar masses have non-negligible errors, the \gsmf, \sfms, galaxy quenched fraction, and number density of massive quiescent galaxies were computed using stellar masses of simulated galaxies that include a lognormal scatter with a standard deviation given by equation (\ref{eq: random_scatter}). Finally, we studied the differences between two AGN feedback implementations that are available in \colibre{}: the fiducial model with thermal energy injections and a hybrid model combining thermal and kinetic jet feedback. Our key findings are as follows:

\begin{itemize}
    \item The \colibre{} \gsmf{} is consistent with the observationally inferred data over the entire redshift range where the observations are available (\mbox{$0 < z < 12$}), while demonstrating very good convergence with resolution over a factor of $64$ in gas particle mass (Figs. \ref{fig:gsmf_evolution_lowz} and \ref{fig:gsmf_evolution_highz}). The largest systematic deviations from the observations, $\approx 0.3$~dex, occur at $z \approx 2-4$. At much higher redshifts ($12 < z < 17$), which lack observational measurements, \colibre{} predicts a smoothly evolving \gsmf, with the normalization changing by $\approx 2$~dex between $z = 12$ and $z = 17$ (Fig.~\ref{fig:gsmf_evolution_highz}). At the highest redshift probed by the simulations, $z = 17$, the combined \gsmf{} predicted by the \colibre{} m5 and m6 resolutions still spans more than 2~dex in stellar mass.
   
    \item \colibre{} predicts that the \shmr{} for central subhaloes peaks at $M_{\rm halo} \sim 10^{12}~\mathrm{M_\odot}$ with a value of $M_* / M_{\rm halo} \approx 0.025$ and logarithmic slopes to the left and to the right of the peak of, respectively, $\approx 1.0$ and $-0.5$ (Fig.~\ref{fig:\shmr{}_evolution}). At all resolutions, the shape and normalization of the \shmr{} only mildly change with redshift and at $z=0.1$ are consistent with semi-empirical models from \citet{2018MNRAS.477.1822M} and \citet{2019MNRAS.488.3143B}.
    
    \item The CSMD is consistent with observationally inferred values at $0<z<12$ (Fig.~\ref{fig:CSMD_evolution}). At $z=0$, \colibre{} predicts a total CSMD of $\approx 3 \times 10^{8}~\mathrm{M_\odot \, cMpc^{-3}}$, with $\approx 90$ per cent residing in subhaloes of $M_* \geq 10^8~\mathrm{M_\odot}$ within $50$~pkpc of their centres.
    
    \item The CSFRD reproduces observationally inferred values reasonably well for $0 < z < 18$ (Fig.~\ref{fig:CSFRD_evolution}). In particular, at $8 < z < 18$, \colibre{} predictions for the CSFRD from subhaloes with $M_* \geq 10^7~\mathrm{M_\odot}$ are consistent with \textit{JWST} UV measurements reported by \citet{2023ApJS..265....5H}, \citet{2024MNRAS.533.3222D}, and \citet{2025arXiv250706292W}. At $z \approx 2$, \colibre{} captures some observational trends while underestimating others by $0.2-0.5$~dex, reflecting the well-known tension between intrinsic and some observed SFRs during this epoch. By $z = 0$, the predictions are again consistent with observations, yielding a total CSFRD of $\approx 0.012 -0.013~\mathrm{M_\odot \, yr^{-1} \, cMpc^{-3}}$.
    
    \item Subhaloes near the peak of the \shmr, characterized by $M_{\rm halo} \sim 10^{12}~\mathrm{M_\odot}$ and $M_* \sim 10^{10}~\mathrm{M_\odot}$, dominate the CSFRD from $z = 0$ to $z \approx 3$. At $z \gtrsim 6$, lower-mass subhaloes ($M_{\rm halo} \lesssim 10^{9.5}~\mathrm{M_\odot}$) become the primary contributors to the CSFRD. Meanwhile, the contribution from the most massive objects ($M_{\rm halo} \gtrsim 10^{13.5}~\mathrm{M_\odot}$) remains negligible at all redshifts (Fig.~\ref{fig:SFH_per_mass}).
    
    \item Similarly to the CSFRD, \colibre{} reproduces the observed \sfms{} reasonably well at $0 < z < 1$ and $z > 4$, but underpredicts the data at intermediate redshifts (Fig.~\ref{fig:sfr_evolution}). Specifically, between $1 < z < 4$, \colibre{} predicts an \sfms{} that is lower by $\approx 0.2-0.6$~dex compared to the values reported by \citet{2021MNRAS.505..540T} and \citet{2023MNRAS.519.1526P}, although it agrees much better with the measurements by \citet{2022ApJ...936..165L}. At $z < 5$, higher-resolution simulations yield an \sfms{} with a slightly yet systematically lower normalization, leading to somewhat worse agreement with observations, whereas at higher redshifts, the convergence with resolution is generally good.

    \item \colibre{} predicts a gradual increase in the fraction of quenched galaxies with decreasing redshift, consistent with observations (Fig.~\ref{fig:qf_evolution}). At $z \approx 0$, galaxies with $M_* \sim 10^9 - 10^{10}~\mathrm{M_\odot}$ exhibit the lowest quenched fraction ($\approx 10-20$ per cent), while those with $M_* \sim 10^{11.5}~\mathrm{M_\odot}$ have the highest ($\approx 75-90$ per cent). However, for even more massive galaxies ($M_* \gtrsim 10^{11.5}~\mathrm{M_\odot}$), the quenched fraction decreases to $\approx 40-60$ per cent, suggesting that AGN feedback does not completely suppress cooling flows in these massive haloes. At fixed stellar mass, \colibre{} predicts higher quenched fractions for satellites compared to centrals due to environmental effects, consistent with comparison data (Fig.~\ref{fig:qf_central_sat_difference}).

    \item \colibre{} is consistent with recent measurements by \textit{JWST} of the number density of massive quiescent galaxies at $2 < z < 7$ (Fig.~\ref{fig:n_passive_vs_z}). At $z \approx 7$, \colibre{} predicts a comoving number density of galaxies with $M_* > 10^{10}~\mathrm{M_\odot}$ and $\mathrm{sSFR} < 10^{-10}~\mathrm{yr^{-1}}$ of $\sim 10^{-6.5}~\mathrm{cMpc^{-3}}$, which rises to $\sim 10^{-3.5}~\mathrm{cMpc^{-3}}$ by $z\approx2$. Moreover, in the largest \colibre{} volume ($400^3~\mathrm{cMpc}^3$), quiescent galaxies with $M_* > 10^{11}~\mathrm{M_\odot}$ are found out to $z = 7.5$, with a number density of $\sim 10^{-8}~\mathrm{cMpc}^{-3}$ at $z \approx 7.5$ and $\sim 10^{-6}~\mathrm{cMpc}^{-3}$ at $z \approx 5$. Predictions from the largest available m5 simulation ($100^3~\mathrm{cMpc}^3$) are available only over a very narrow redshift range ($5 < z < 6$), owing to the relatively small cosmological volume and because the simulation has thus far progressed only down to $z = 4.6$. At a fixed redshift and stellar mass bin, the m6 model predicts a somewhat lower number density than the m7 model, with differences ranging between $\approx 0.2$ and $1$~dex.
    
    \item The \colibre{} model with hybrid AGN feedback predicts significantly weaker quenching by AGN at $z \gtrsim 2$ compared to the thermal model, resulting in a $\approx 1-2$~dex lower number density of massive quiescent galaxies at these redshifts (Fig.~\ref{fig:thermal_vs_hybrid_agn_evolution}). This weaker AGN quenching, combined with the slightly weaker SN feedback in the hybrid model, leads to an increase in the CSMD at $0.5 < z < 7$ and an elevated \gsmf{} around $z = 2$ relative to the thermal model (Figs. \ref{fig:thermal_vs_hybrid_agn} and \ref{fig:thermal_vs_hybrid_agn_evolution}). By contrast, at $z = 0$, the hybrid model predicts stronger quenching, with the CSMD and \gsmf{} converging to those of the thermal model.
\end{itemize}

In closing, we highlight that the \colibre{} simulations are consistent with both the low- and high-$z$ observationally inferred \gsmf, including the most recent measurements from \textit{JWST}. The simulations achieve this agreement using a standard universal stellar IMF \citep{2003PASP..115..763C} and a standard $\Lambda$CDM cosmology, predicting a nearly redshift-independent baryon conversion efficiency in DM haloes, with a peak value of $\approx 16$~per cent. This result challenges recent claims based on \textit{JWST} data suggesting that efficiencies much greater than $\sim 10$~per cent are necessary to explain the observations of high-mass galaxies at redshifts $z \gtrsim 7$ \citep[e.g.][]{2024MNRAS.533.1808W, 2025A&A...695A..20S}. However, we caution that the stellar masses for \textit{JWST} galaxies were inferred using a similar IMF to the one assumed in \colibre{} and that rest-frame UV luminosity function comparisons between \colibre{} and observations have not yet been made. 

We also note that, given the likely presence of systematic uncertainties in the observational analyses, we cannot exclude the possibility that some of the agreement between \colibre{} and the data found in this work is fortuitous. In future work, we will perform forward modelling to compare \colibre{} predictions with data in observer space, providing an important consistency check. Further follow-up work will investigate early quenching mechanisms of massive galaxies in \colibre{} and compare low- and high-redshift luminosity functions predicted by the simulations with observational data, to provide further theoretical insights into the evolution of galaxy stellar mass and star formation rates in the observable Universe.

\section*{Acknowledgements}

We thank Yannick M. Bah\'{e}, Simon Driver, Victor J. Forouhar Moreno, Anna de Graaff, Claudia del P. Lagos, Christopher Lovell, Kai Wang, John Weaver, and the anonymous referee for their useful comments. This work used the DiRAC@Durham facility managed by the Institute for Computational Cosmology on behalf of the STFC DiRAC HPC Facility (www.dirac.ac.uk). The equipment was funded by BEIS capital funding via STFC capital grants ST/K00042X/1, ST/P002293/1, ST/R002371/1 and ST/S002502/1, Durham University and STFC operations grant ST/R000832/1. DiRAC is part of the National e-Infrastructure. This project has received funding from the Netherlands Organization for Scientific Research (NWO) through research programme Athena 184.034.002. ABL acknowledges support by the Italian Ministry for Universities (MUR) program `Dipartimenti di Eccellenza 2023-2027' within the Centro Bicocca di Cosmologia Quantitativa (BiCoQ), and support by UNIMIB’s Fondo Di Ateneo Quota Competitiva (project 2024-ATEQC-0050). CSF acknowledge support from European Research Council (ERC) Advanced Grant DMIDAS (GA 786910). SP acknowledges support from the Austrian Science Fund (FWF) through project V 982-N. Data analysis was carried out using \textsc{swiftsimio} \citep{Borrow2020simio,2021arXiv210605281B}.
%%%%%%%%%%%%%%%%%%%%%%%%%%%%%%%%%%%%%%%%%%%%%%%%%%
\section*{Data Availability}

All \gsmf{} data presented in this work will be made available on \href{https://colibre-simulations.org/}{https://colibre-simulations.org/} upon publication. The remaining data underlying this article will be shared on reasonable request to the corresponding author. The public version of the \textsc{Swift} simulation code can be found on \href{http://www.swiftsim.com}{www.swiftsim.com}. The \textsc{Swift} modules related to the \colibre{} galaxy formation model will be integrated into the public version after the public release of \colibre.

\bibliographystyle{mnras}
\bibliography{main} 

\appendix

\section{Convergence with resolution and volume}
\label{appendix:convergence}

\begin{figure*}
    \centering
    \includegraphics[width=0.49\textwidth]{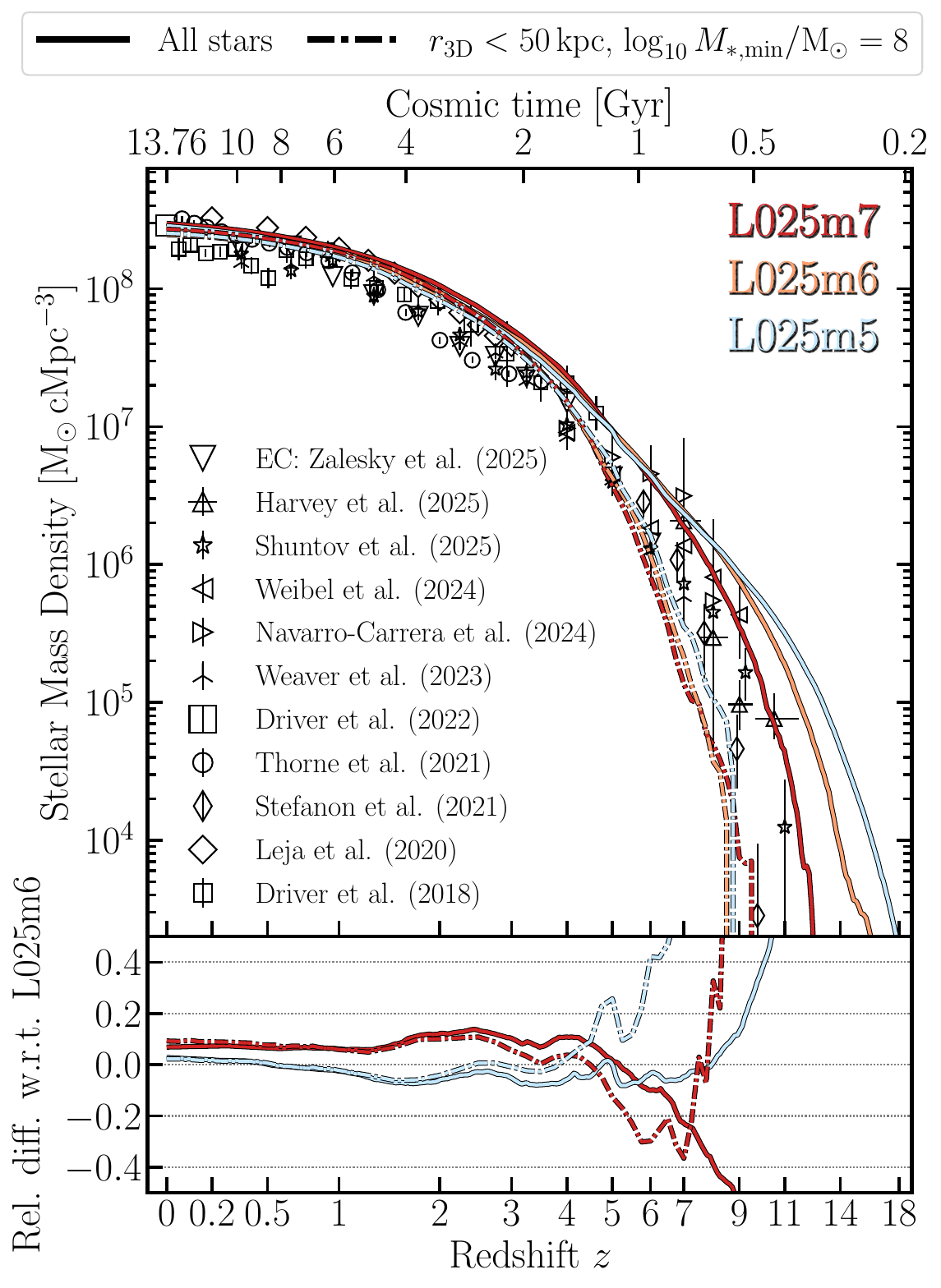}
    \includegraphics[width=0.49\textwidth]{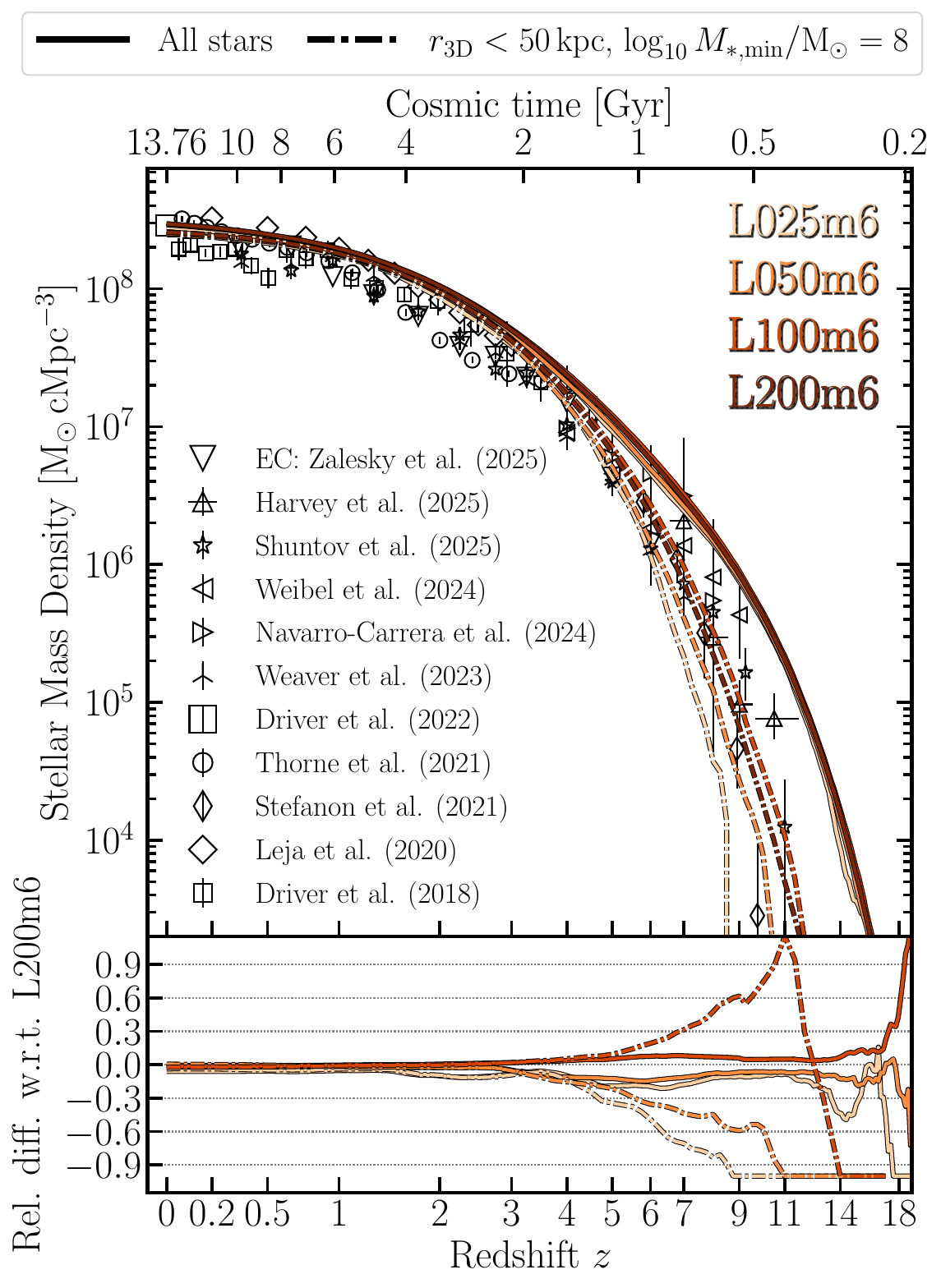}
    \caption{The impact of numerical resolution (\textit{left}) and cosmological volume (\textit{right}) on the evolution of the cosmic stellar mass density (CSMD). The left-hand panels compare the m5, m6, and m7 resolution simulations in the ($25~\mathrm{cMpc}$)$^3$ volume, while the right-hand panels compare ($25~\mathrm{cMpc}$)$^3$, ($50~\mathrm{cMpc}$)$^3$, ($100~\mathrm{cMpc}$)$^3$, and ($200~\mathrm{cMpc}$)$^3$ volumes at m6 resolution. The layout of the panels is the same as in Fig.~\ref{fig:CSMD_evolution}. The CSMD exhibits very good convergence with resolution (cosmological volume) at $z<5$ ($z<4$), with relative differences remaining below $\approx 20$ per cent.}
    \label{fig:numerical_convergence} 
\end{figure*}

In this section, we test the convergence of the \colibre{} simulations with numerical resolution and with cosmological volume. Because some subgrid parameter values in the \colibre{} models differ slightly between resolutions (see table 1 in \citealt{2026COLIBREproject}, for a summary of these differences), convergence with resolution should be understood as `weak convergence', following the terminology of \citet{2015MNRAS.446..521S}. 

We emphasize that in \colibre{}, the strength of SN and AGN feedback was optimized to reproduce the observed $z=0$ \gsmf{} and the size -- stellar mass relation in the stellar mass range $10^9 < M_* / \mathrm{M}_\odot < 10^{11.3}$ \citep[see][for details]{2026COLIBREcalibration}. Furthermore, the coupling efficiency of AGN feedback was tuned to match the observed $z=0$ BH mass -- stellar mass relation of massive galaxies. The calibration was performed independently at each resolution, and no constraints at higher redshift were applied. Consequently, the \colibre{} predictions at higher redshifts or outside the mass range used for calibration are not guaranteed to converge between different resolution levels, particularly for statistics other than those used in the calibration.

\subsection{Convergence of the cosmic stellar mass density}

Fig.~\ref{fig:numerical_convergence} shows the evolution of the CSMD where the left-hand panels compare the \colibre{} m5, m6, and m7 simulations, all in a ($25~\mathrm{cMpc}$)$^3$ volume, while the right-hand panels compare ($25~\mathrm{cMpc}$)$^3$, ($50~\mathrm{cMpc}$)$^3$, ($100~\mathrm{cMpc}$)$^3$, and ($200~\mathrm{cMpc}$)$^3$ volume simulations, all at m6 resolution. The layout of the panels is the same as in Fig.~\ref{fig:CSMD_evolution}.

\begin{figure}
    \centering
    \includegraphics[width=0.49\textwidth]{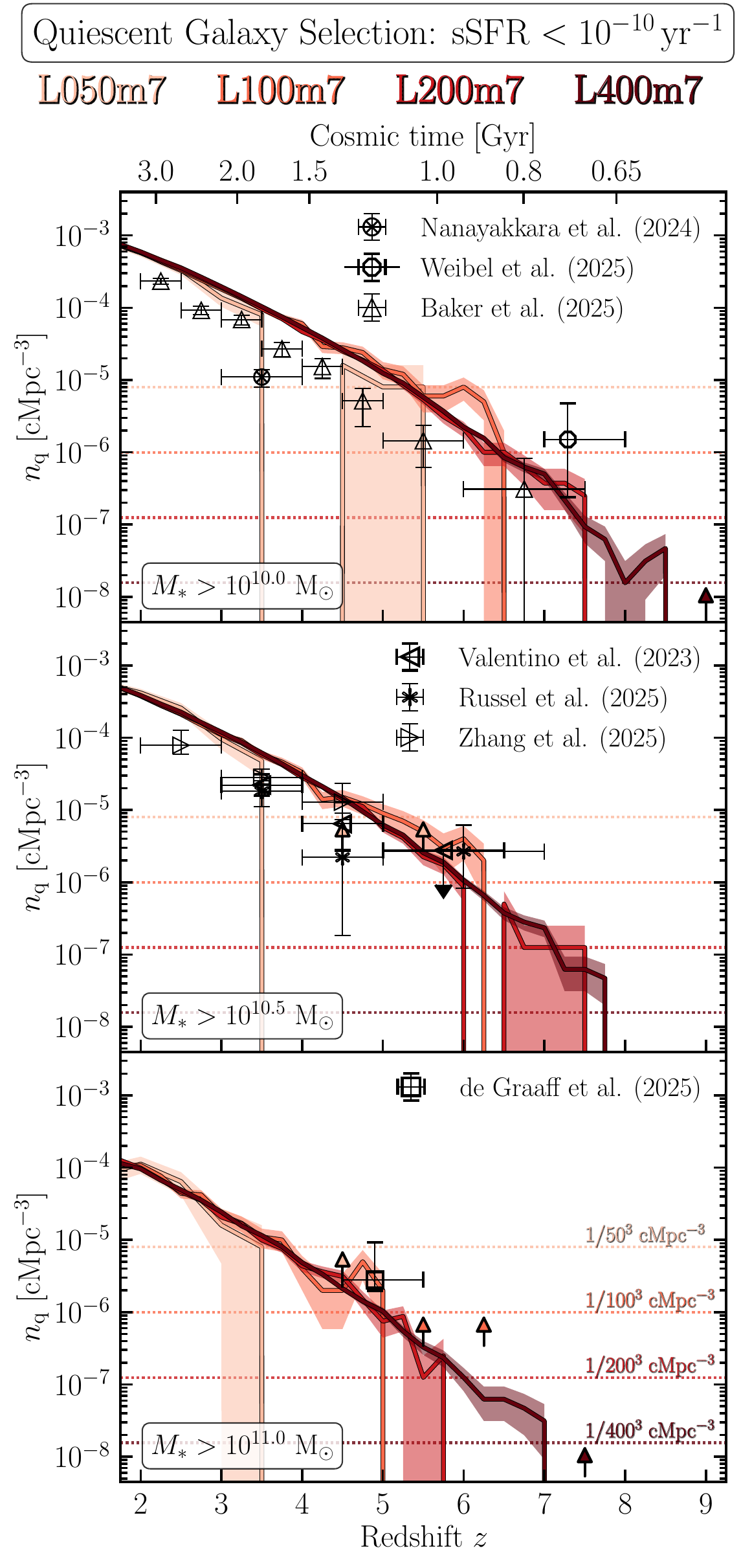}
    \caption{As Fig.~\ref{fig:n_passive_vs_z}, but shows the comoving number density of massive quiescent galaxies predicted by the \colibre{} m7 simulations in different cosmological volumes: ($\mathrm{50~cMpc}$)$^3$, ($\mathrm{100~cMpc}$)$^3$ ($\mathrm{200~cMpc}$)$^3$, and ($\mathrm{400~cMpc}$)$^3$, indicated by different shades of red. For reference, the thin horizontal dotted lines mark the minimum non-zero number density of quiescent galaxies that each volume can resolve. While predictions from volumes of $(100~\mathrm{cMpc})^3$ and larger converge at $z \lesssim 5$, the $(50~\mathrm{cMpc})^3$ volume is too small to provide meaningful predictions at $z \gtrsim 3.5$.}
    \label{fig:n_passive_vs_z_boxsize}
\end{figure}

The CSMD exhibits very good convergence with resolution at $z\lesssim 5$, with relative differences between adjacent resolutions within $\approx 20$ per cent. This convergence holds both for the total CSMD, computed using all stellar particles in the simulation (solid curves), and for the CSMD contributed only by subhaloes with stellar masses $M_* \geq 10^{8}~\mathrm{M_\odot}$ (dash-dotted curves). At higher redshifts ($z\gtrsim 6$), the total CSMDs begin to diverge, with higher-resolution simulations predicting systematically higher values of the CSMD. This trend is expected, as the total CSMD at these high redshifts is dominated by SFRs of the least massive subhaloes (see Fig.~\ref{fig:SFH_per_mass}), whose abundance is sensitive to numerical resolution. 

The CSMD also shows very good convergence with cosmological volume at fixed resolution for $z \lesssim 4$, which holds for both the total CSMD and that contributed only by $M_* \geq 10^{8}~\mathrm{M_\odot}$ subhaloes. At higher redshifts, convergence remains good for the total CSMD up to $z \approx 14$, with relative differences between simulations in different volumes remaining within $\approx 30$~per cent. This behaviour is expected, as at high redshifts the total CSMD is dominated by the least massive objects ($M_* \lesssim 10^{6.5}~\mathrm{M_\odot}$; see Fig.~\ref{fig:SMD_per_mass}), which are abundant in all four volumes. In contrast, the CSMD from $M_* \geq 10^{8}~\mathrm{M_\odot}$ subhaloes depends sensitively on cosmological volume above $z\approx 4$. On average, larger volumes yield higher CSMDs at high redshifts, as they are more likely to host $M_* \geq 10^{8}~\mathrm{M_\odot}$ galaxies. One exception to this trend is the ($100~\mathrm{cMpc}$)$^3$ volume, which predicts a slightly higher CSMD than that in the larger ($200~\mathrm{cMpc}$)$^3$ volume due to a rare overdensity in its ICs that forms a massive protocluster by $z\sim 5$, atypical for a ($100~\mathrm{cMpc}$)$^3$ volume.

\begin{figure*}
    \centering
    \includegraphics[width=0.99\textwidth]{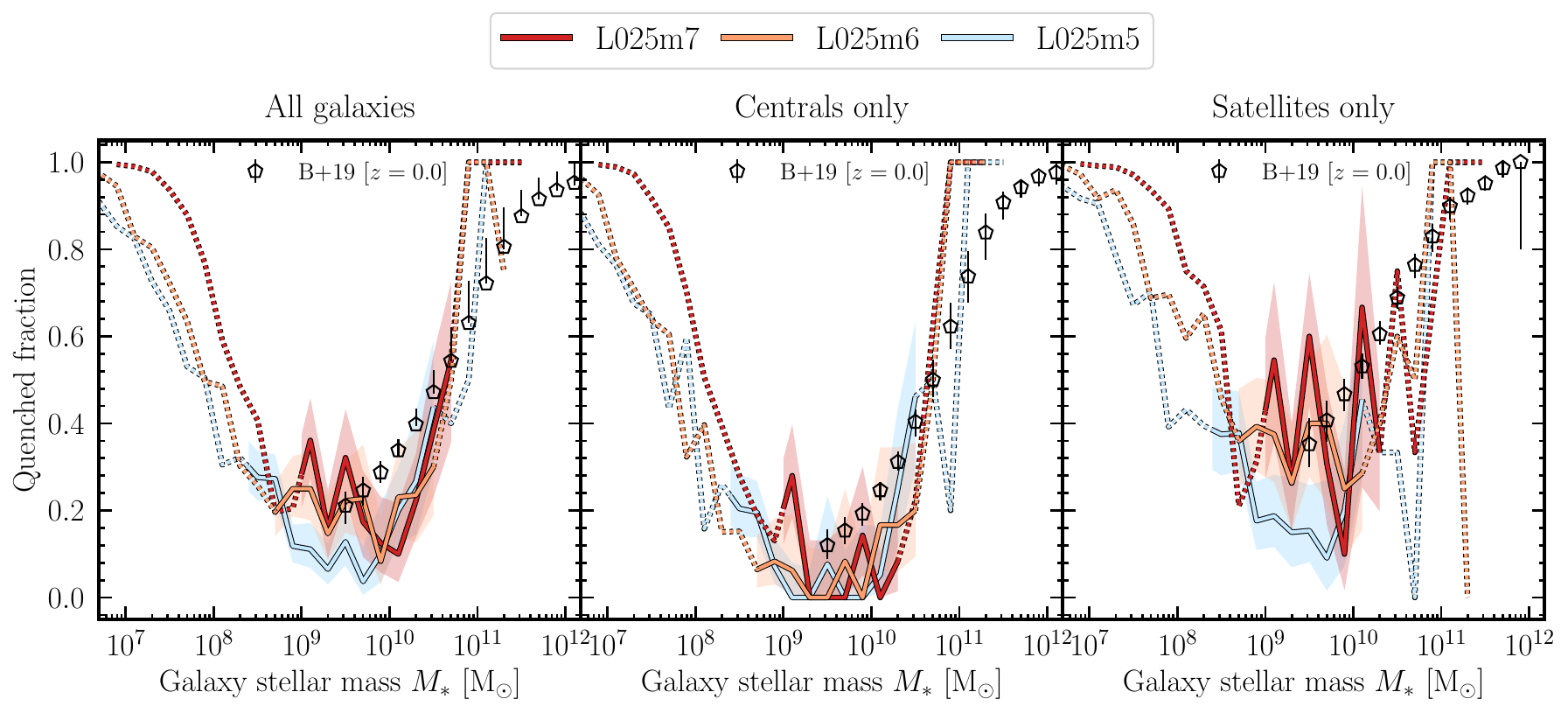}
    \caption{As in Fig.~\ref{fig:qf_central_sat_difference}, but with all simulations using the same cosmological volume of ($25~\mathrm{cMpc}$)$^3$. Adopting the same cosmological volume at all \colibre{} resolutions improves the convergence of the quenched fraction for satellite galaxies, compared to Fig.~\ref{fig:qf_central_sat_difference}, where the m5 simulation uses a much smaller volume than the m6 and m7 counterparts.}
   \label{fig:numerical_convergence_qf} 
\end{figure*}

\subsection{Convergence of the number density of massive quiescent galaxies}

Fig.~\ref{fig:n_passive_vs_z_boxsize} shows the evolution of the comoving number density of massive quiescent galaxies predicted by the \colibre{} m7 simulations in different cosmological volumes: $(50~\mathrm{cMpc})^3$, $(100~\mathrm{cMpc})^3$, $(200~\mathrm{cMpc})^3$, and $(400~\mathrm{cMpc})^3$, coloured in progressively darker shades of red for larger volumes. The comparison data (black symbols) are the same as in Fig.~\ref{fig:n_passive_vs_z}, and the number densities are computed for the same three stellar mass bins: $M_* > 10^{10}~\mathrm{M_\odot}$ (top panel), $M_* > 10^{10.5}~\mathrm{M_\odot}$ (middle panel), and $M_* > 10^{11}~\mathrm{M_\odot}$ (bottom panel). For reference, thin horizontal dotted lines indicate the minimum non-zero number density of quiescent galaxies resolvable in each cosmological volume.

While predictions from volumes of $(100~\mathrm{cMpc})^3$ or larger converge at $z \lesssim 5$, the $(50~\mathrm{cMpc})^3$ volume is too small to yield meaningful predictions above $z \approx 3.5$. In particular, its minimum resolvable number density of $\sim 10^{-5}~\mathrm{cMpc}^{-3}$ makes it unsuitable for comparison with high-redshift ($z \gtrsim 4$) observations, which typically report number densities comparable to or below this level. The $(100~\mathrm{cMpc})^3$ volume extends the predictions of non-zero number densities out to $z \approx 6.5$ for the adopted sSFR and $M_*$ cuts, with a minimum resolved number density of $10^{-6}~\mathrm{cMpc}^{-3}$; however, its predictions at $z \gtrsim 4$ are subject to small-number statistics.

\subsection{Convergence of the galaxy quenched fraction}

Fig.~\ref{fig:numerical_convergence_qf} shows the $z=0$ quenched fraction for all, central, and satellite galaxies predicted by the simulations at m7, m6, and m5 resolutions, all using the $(25~\mathrm{cMpc})^3$ cosmological volume. We find that the convergence of the satellite quenched fraction between m5 and the two lower resolutions improves significantly when the same cosmological volume is used, compared to Fig.~\ref{fig:qf_central_sat_difference}, which shows the quenched fraction in the simulations at different resolutions in different volumes.

\section{The effect of the choice of aperture}
\label{appendix: apertures}

\begin{figure}
    \centering
    \includegraphics[width=0.49\textwidth]{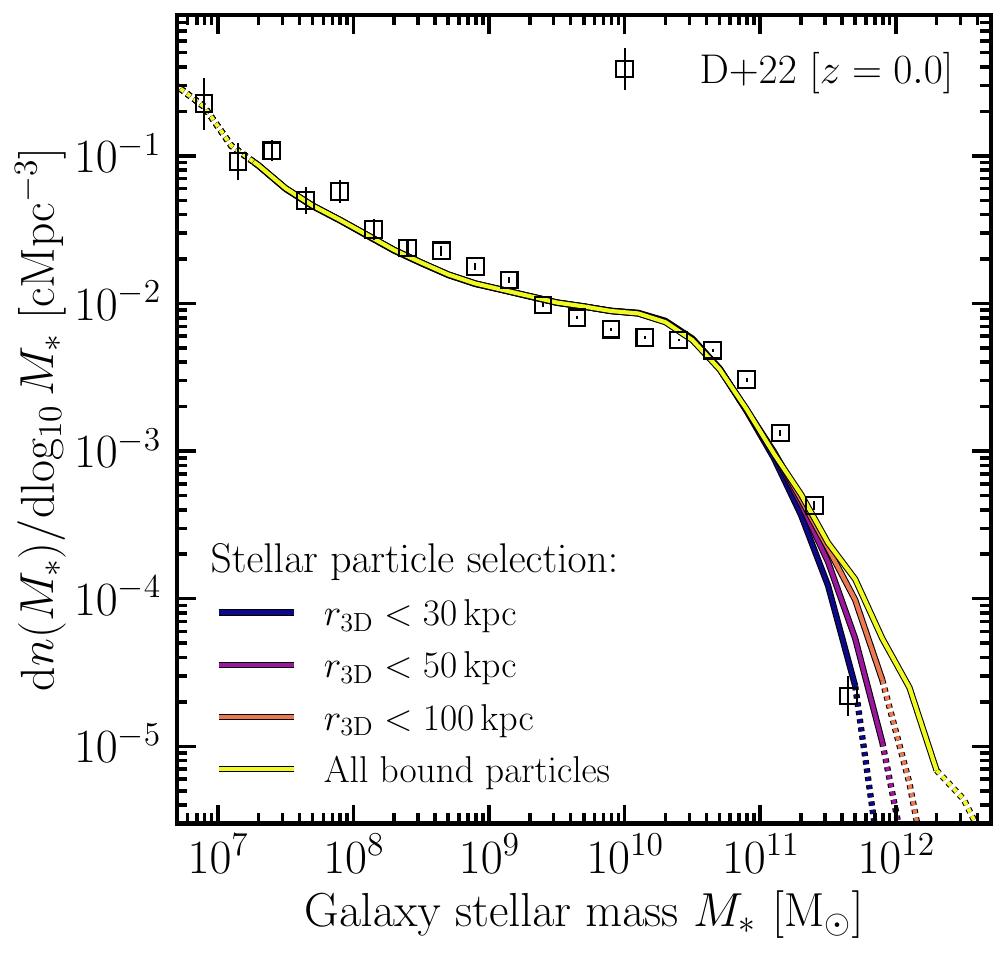}
    \caption{The $z=0$ galaxy stellar mass function (\gsmf) in the \colibre{} L200m6 simulation for four different aperture choices used to calculate subhalo stellar masses. The dark-blue, violet, and orange curves correspond to 3D spherical apertures with radii of $30$, $50$, and $100$~pkpc, respectively, while the yellow curve represents the case where no aperture cut is applied. In all cases, stellar masses are computed using only stellar particles gravitationally bound to subhaloes. Unlike in previous figures, here we do not add lognormal scatter to galaxy stellar masses to account for Eddington bias. For reference, the observed $z=0$ \gsmf{} from \citet{2022MNRAS.513..439D} is shown as black squares. The choice of aperture has a negligible impact on the \gsmf{} for stellar masses $M_* \lesssim 10^{11}~\mathrm{M_\odot}$, but becomes increasingly important at higher $M_*$.}
    \label{fig:effect_of_apertures} 
\end{figure}

In this section, we explore the impact of the aperture choice on the \colibre{} \gsmf{}. As described in $\S$\ref{subsection: halo_finding_and_galaxy_properties}, subhalo stellar masses in this work are computed by summing the masses of individual stellar particles that (i) are gravitationally bound to their respective subhaloes and (ii) are located within 50 pkpc 3D apertures. The aperture is centred on the particle (stars, gas, DM, or BHs) with the largest (i.e. most negative) gravitational binding energy within each subhalo.

Fig.~\ref{fig:effect_of_apertures} shows the $z=0$ \gsmf{} in the \colibre{} L200m6 simulation for different aperture choices. The dark-blue, violet, and orange curves correspond to stellar masses computed using bound stellar particles within $30$, $50$, or $100$ pkpc 3D apertures, respectively, with $50$ pkpc being our fiducial choice. The yellow curve represents the case where stellar masses are calculated by summing all bound stellar particles within subhaloes, without imposing an aperture constraint. Unlike in Figs.~\ref{fig:gsmf_evolution_lowz} and \ref{fig:gsmf_evolution_highz}, here we do not add lognormal scatter to galaxy stellar masses to account for Eddington bias, as this would introduce additional scatter in the \gsmf{} between different aperture cases, making their comparison less straightforward.

We find that the \colibre{} \gsmf{} is independent of the aperture choice for $M_* \lesssim 10^{11}~\mathrm{M_\odot}$. However, stellar masses of the most massive objects in the volume ($M_* \gtrsim 10^{11.5}~\mathrm{M_\odot}$) can vary by up to $\approx 0.5$~dex depending on the aperture, with the larger apertures resulting in higher $M_*$. These results are consistent with a similar analysis done for the \textsc{eagle} simulations \citep[see fig. 6 in][]{2015MNRAS.446..521S}.

\section{Average star formation rates}
\label{appendix_sfr}

\begin{figure}
    \centering
    \includegraphics[width=0.49\textwidth]{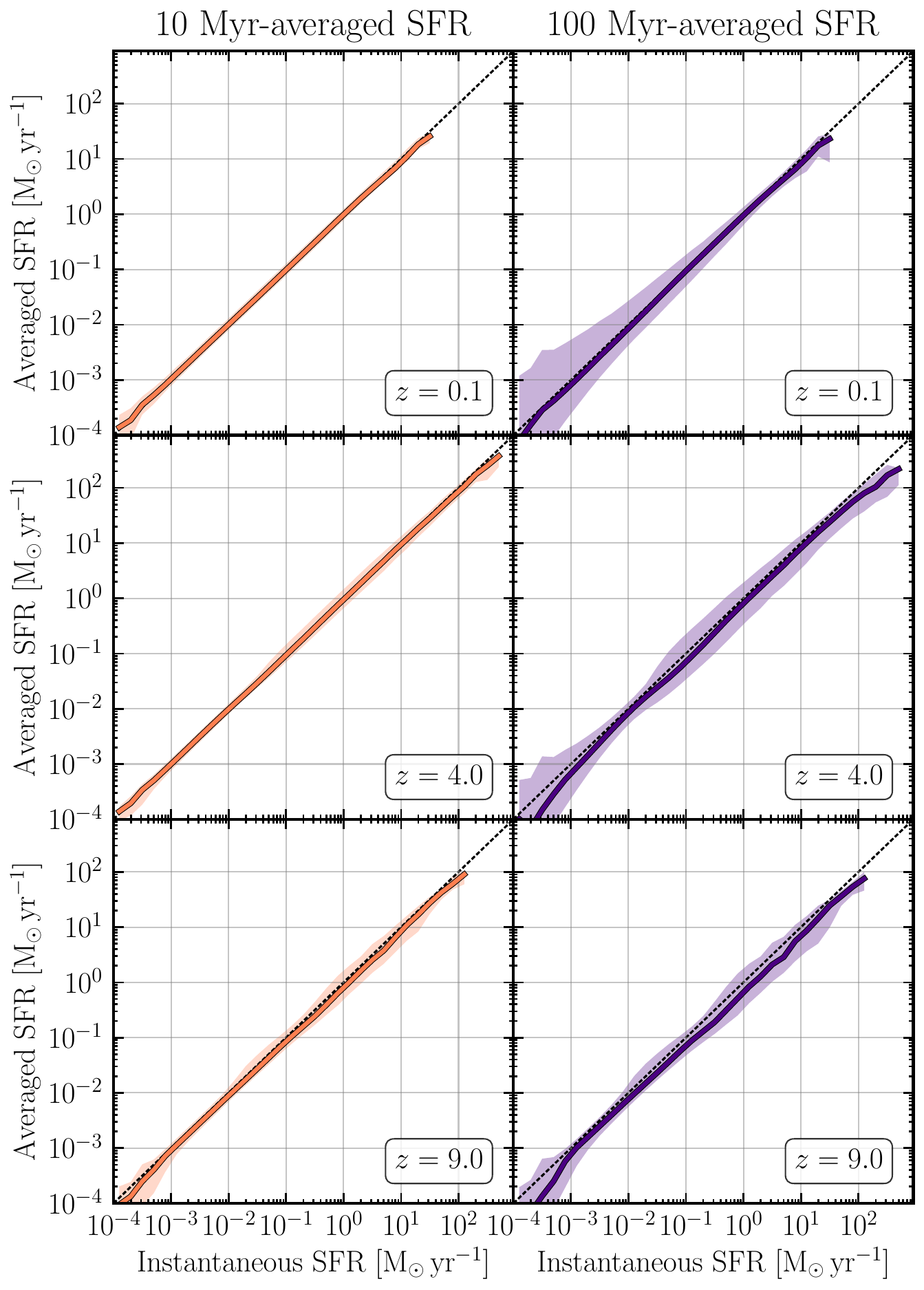}
    \caption{The galaxy SFR averaged over 10 Myr (\textit{left column}) and 100 Myr (\textit{right column}) of lookback time, plotted against the instantaneous galaxy SFR in the \colibre{} L200m6 simulation at redshifts $z=0.1$ (\textit{top row}), $z=4$ (\textit{middle row}), and $z=9$ (\textit{bottom row}). Solid lines represent the median relations, while shaded regions denote the 5$^{\rm th}$ to 95$^{\rm th}$  percentile scatter. For reference, the black diagonal line shows the $y=x$ relation. On average, we find little to no systematic deviation between the instantaneous and averaged SFRs, while the scatter increases with the length of the averaging interval.}
    \label{fig:average_vs_instantaneous_sfr}
\end{figure}

In this work, gas star formation rates (SFRs) are calculated as instantaneous SFRs. By contrast, observational measurements typically trace SFRs averaged over longer time-scales: $\sim 10~\mathrm{Myr}$ using recombination lines such as H $\upalpha$, and $\sim 100~\mathrm{Myr}$ from UV continuum emission \citep[e.g.][]{2014ARA&A..52..415M}. In this section, we explore the impact of time-averaging on the predicted SFRs and assess the validity of using instantaneous SFRs in comparisons with observational data.

The left (right) column of Fig.~\ref{fig:average_vs_instantaneous_sfr} shows the correlation between the SFR averaged over the past 10 (100)~Myr of lookback time and the instantaneous SFR. Results are presented for the \colibre{} L200m6 simulation at redshifts $z=0.1$ (top row), $z=4$ (middle row), and $z=9$ (bottom row). All SFRs are computed using gas particles bound to subhaloes and located within 50~pkpc of the subhaloes' centres. Solid lines represent the median of the averaged SFRs computed in $0.2$-dex bins of instantaneous SFR, while shaded regions denote the $5^{\rm th}$ to $95^{\rm th}$ percentile scatter. For reference, the black diagonal line indicates the $y=x$ relation.

We find no systematic deviations between the instantaneous and averaged SFRs, except in the $z=9$ bin, where the 100-Myr averaged SFR is marginally lower than the instantaneous value, by up to $\approx 0.1$~dex. As expected, the scatter is larger for the 100-Myr time-scale than for the 10-Myr one, since the longer average captures star formation events further in the past. The close agreement among all three SFR definitions justifies our use of instantaneous SFRs throughout the main part of this work.

\section{The effect of Eddington bias}
\label{appendix: effect of eddington bias}

As outlined in $\S$\ref{subsection:assumptions}, random errors in observationally inferred stellar masses, combined with the exponential decline in the number of galaxies at the high-mass end of the \gsmf, lead to Eddington bias, i.e. the dominance of upscatter over downscatter between adjacent galaxy mass bins. Because observed stellar masses inherently carry non-negligible uncertainties, this bias affects all galaxy statistics based on stellar mass -- including the \gsmf, the \sfms, and the galaxy quenched fraction -- unless explicitly corrected for \citep[e.g.][]{2018MNRAS.474.5500O,2020ApJ...893..111L}. In contrast, this bias is absent in statistics derived from raw simulation data, where galaxy stellar masses are precisely known. 

To ensure more consistent comparisons between \colibre{} and observations, throughout this work we assumed a lognormal scatter in galaxy stellar masses of \mbox{$\sigma_{\rm random}(z) = \min(0.1 + 0.1 \, z, 0.3)$ dex} to represent random errors on stellar masses (see equation \ref{eq: random_scatter}). In this section, we examine in detail how incorporating this scatter affects the \gsmf, the \sfms, the galaxy quenched fraction, and the number density of massive quiescent galaxies predicted by \colibre.

\subsection{The effect of Eddington bias on the GSMF, SFMS, and quenched fraction at redshift 0}

\begin{figure}
    \centering
    \includegraphics[width=0.49\textwidth]{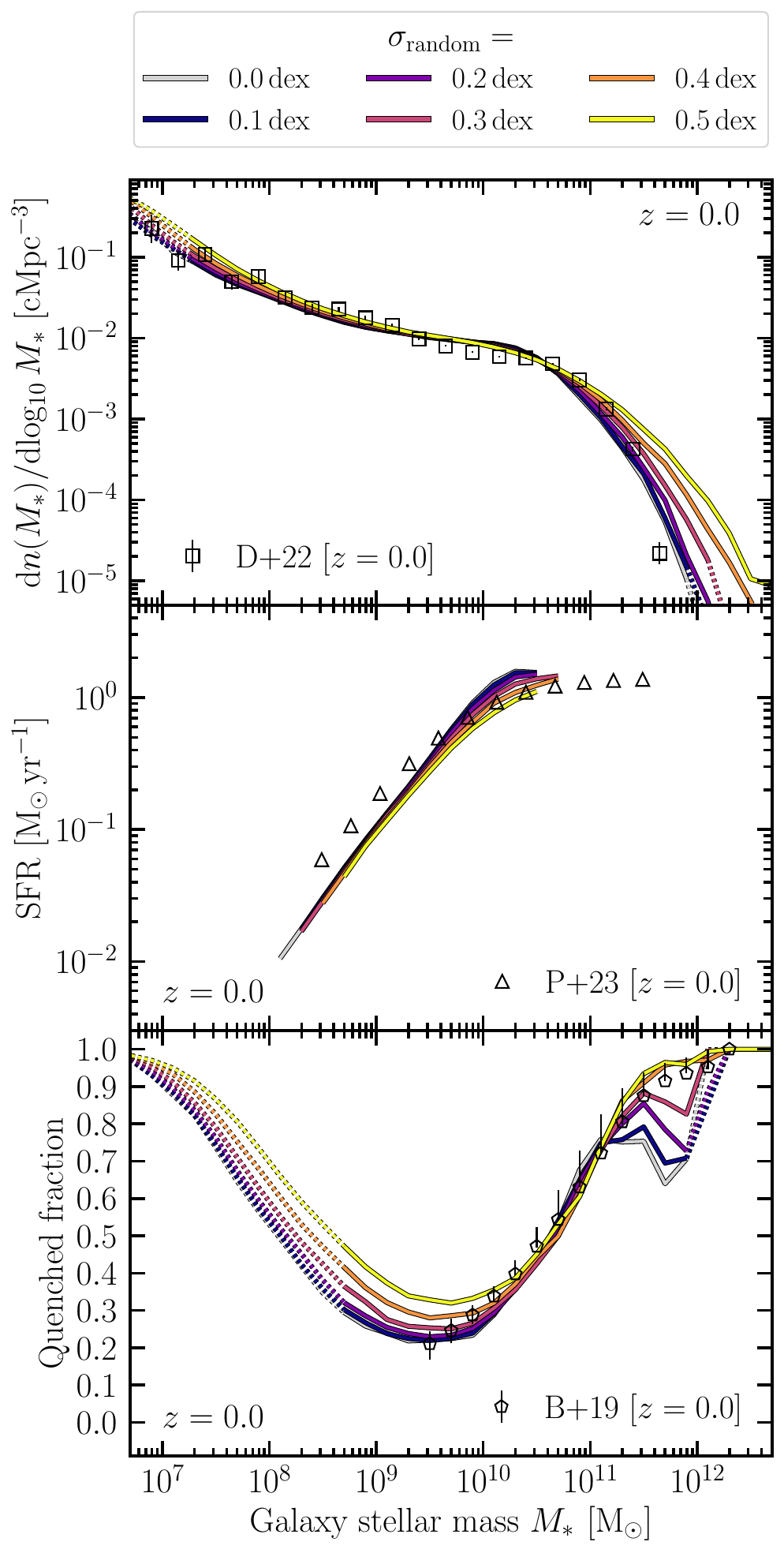}
    \caption{The \gsmf{} (\textit{top}), the \sfms{} (\textit{middle}), and the galaxy quenched fraction (\textit{bottom}) in the \colibre{} L200m6 simulation at $z=0$ after adding different amounts of lognormal scatter to galaxy stellar masses to account for Eddington bias (colours). The black symbols in the top, middle, and bottom panels indicate the data from, respectively, \citet{2022MNRAS.513..439D}, \citet{2023MNRAS.519.1526P}, and \citet{2019MNRAS.488.3143B}. The grey-coloured lines correspond to the case with no scatter (i.e. $\sigma_{\rm random} = 0.0$~dex). Due to the Eddington bias, the scatter affects all three relations, particularly at $M_* \gtrsim 10^{11}~\mathrm{M_\odot}$, where the effect becomes significant for $\sigma_{\rm random} \geq 0.3$~dex.}
    \label{fig:eddington_bias_plot}
\end{figure}

Fig.~\ref{fig:eddington_bias_plot} presents the $z=0$ \gsmf{} (top panel), \sfms{} (middle panel), and quenched fraction (bottom panel) from the \colibre{} m6 simulation in the ($200~\mathrm{cMpc}$)$^3$ volume. The coloured lines depict scenarios where a lognormal scatter with zero mean and a standard deviation ranging from $\sigma_{\rm random} = 0.1$~dex (dark blue) to $0.5$~dex (yellow) is added to galaxy stellar masses. The grey lines represent the case without added scatter. The black symbols indicate either observational measurements or predictions from semi-empirical models constrained by observations. As in previous figures showing the \sfms{} and quenched fractions, the \sfms{} is defined as the median SFR of galaxies with $\mathrm{SFR} > 0.1 \, \times$ the SFR at the \sfms, while the quenched fraction is defined as the fraction of galaxies with $\mathrm{sSFR} < 0.2 / t_{\rm H}(z)$. 

We find that adding lognormal scatter to galaxy stellar masses significantly affects the \gsmf{} and quenched fraction, while the impact on the \sfms{} is relatively minor. The most pronounced differences occur at $M_* \gtrsim 10^{11}~\mathrm{M_\odot}$ and for $\sigma_{\rm random} \geq 0.3$~dex. For example, a lognormal scatter with a standard deviation of $0.4$~dex increases the quenched fraction at $M_* \sim 3 \times 10^{11}~\mathrm{M_\odot}$ from $\approx 70$ to $\approx 90$~per cent and raises the \gsmf{} by $\approx 0.4$~dex. In contrast, adopting a scatter with $\sigma_{\rm random} = 0.1$~dex -- the value used in \colibre{} at $z=0$ -- yields negligible differences in both the \gsmf{} and quenched fraction.

\subsection{The effect of Eddington bias on the GSMF at high redshifts}

\begin{figure}
    \centering
    \includegraphics[width=0.49\textwidth]{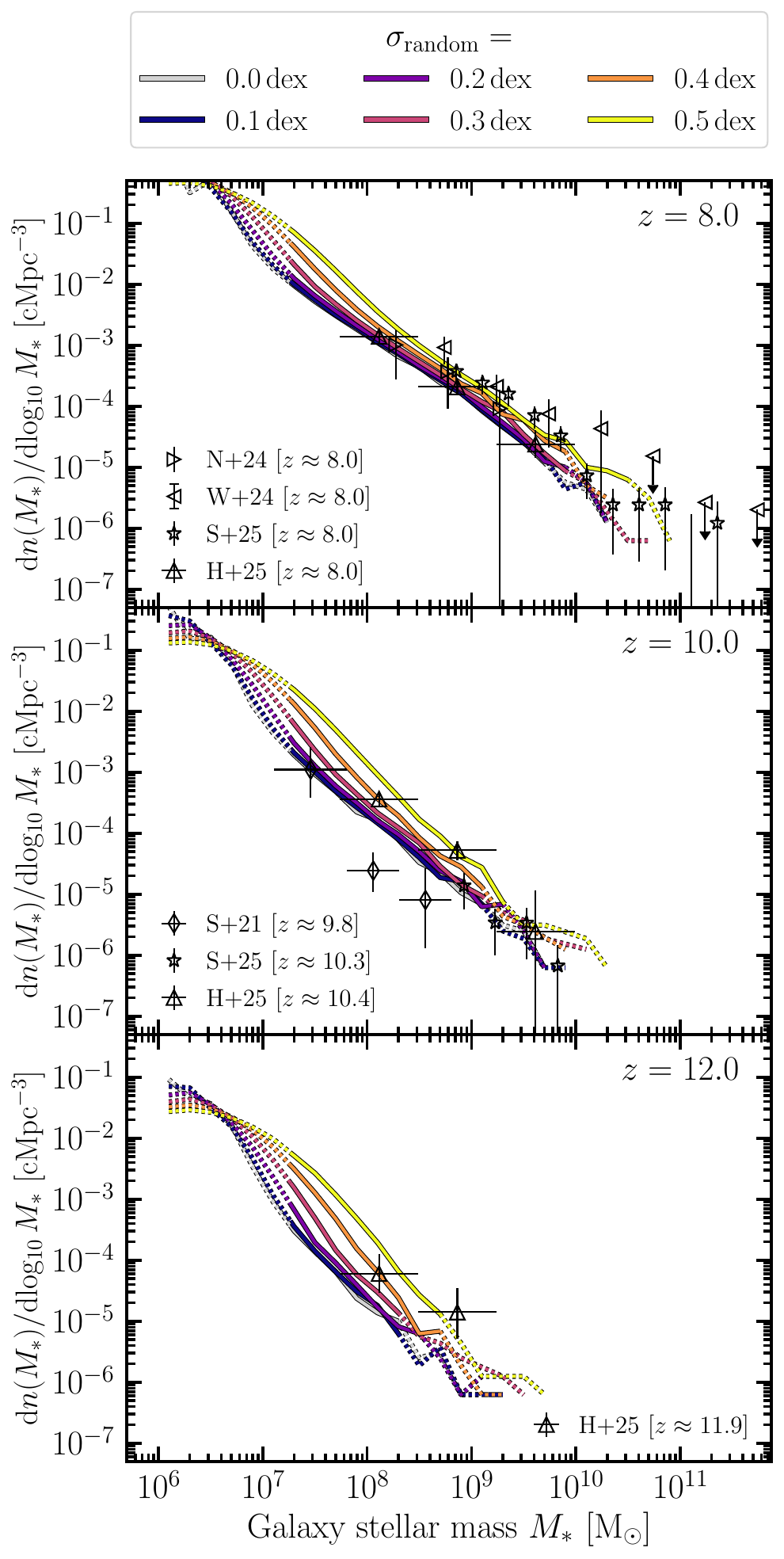}
    \caption{As Fig.~\ref{fig:eddington_bias_plot}, but shows only the \gsmf, with each panel corresponding to a different redshift: $z=8$ (\textit{top}), $z=10$ (\textit{middle}), and $z=12$ (\textit{bottom}). The observational data (black symbols) are the same as in Fig. \ref{fig:gsmf_evolution_highz} at the corresponding redshifts. The Eddington bias is important in bringing the simulation predictions (solid curves) into better agreement with the observational data at the high-mass end, particularly at $z=12$.}
\label{fig:eddington_bias_plot_multile_z}
\end{figure}

Fig.~\ref{fig:eddington_bias_plot_multile_z} extends the upper panel of Fig.~\ref{fig:eddington_bias_plot} by showing the effect of Eddington bias on the \gsmf{} at three additional redshifts: $z=8$ (top), $z=10$ (middle), and $z=12$ (bottom). Unlike at $z=0$, the Eddington bias significantly affects the \gsmf{} already above $M_* \sim 10^7~\mathrm{M_\odot}$, as the \gsmf{} is steeper in this stellar mass range at high redshifts compared to $z=0$. Adding a lognormal scatter improves the agreement between the simulation and observations at the high-mass end. In particular, adopting $\sigma_{\rm random} = 0.5$~dex at $z=12$ brings the \colibre{} \gsmf{} to within $1\sigma$ of the value in the higher-mass bin reported by \citet{2025ApJ...978...89H}.

\subsection{The effect of Eddington bias on the number density of massive quiescent galaxies}

\begin{figure}
    \centering
    \includegraphics[width=0.49\textwidth]{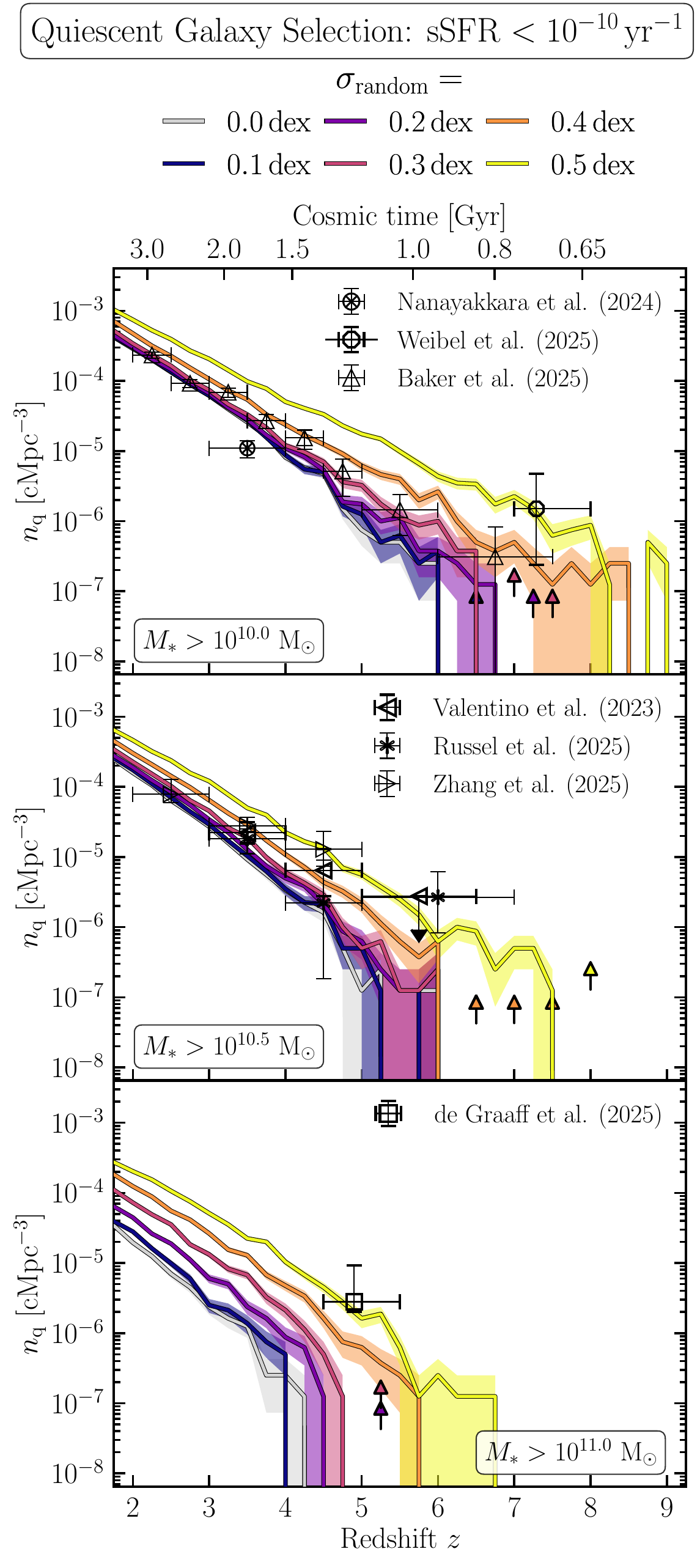}
    \caption{As Fig.~\ref{fig:eddington_bias_plot}, but shows the effect of Eddington bias on the comoving number density of massive quiescent galaxies. The layout of the figure is the same as Fig. \ref{fig:n_passive_vs_z}, with different panels corresponding to different stellar mass bins. The Eddington bias plays an important role in bringing the simulation predictions (solid curves) into better agreement with the observational data (black symbols), especially at higher redshifts and in higher stellar mass bins.}
\label{fig:eddington_bias_n_passive}
\end{figure}

Fig.~\ref{fig:eddington_bias_n_passive} shows the effect of Eddington bias on the evolution of the comoving number density of massive quiescent galaxies ($n_{\rm q}$) in the L200m6 simulation. The figure layout matches Fig.~\ref{fig:n_passive_vs_z}, including the observational data (black symbols). As in Fig.~\ref{fig:n_passive_vs_z}, different panels correspond to different stellar mass bins for quiescent galaxies contributing to $n_{\rm q}$: $M_* > 10^{10}~\mathrm{M_\odot}$ (top), $M_* > 10^{10.5}~\mathrm{M_\odot}$ (middle), and $M_* > 10^{11}~\mathrm{M_\odot}$ (bottom). In the simulation, quiescent galaxies satisfy $\mathrm{sSFR} < 10^{-10}~\mathrm{yr}^{-1}$. Note that while curves in Fig.~\ref{fig:eddington_bias_n_passive} correspond to fixed $\sigma_{\rm random}$ values (as indicated in the legend), Fig.~\ref{fig:n_passive_vs_z} uses $\sigma_{\rm random}$ from equation~(\ref{eq: random_scatter}), which depends on redshift but saturates at $0.3$~dex for $z\geq 2$. Adding lognormal scatter affects both galaxy stellar masses and their sSFRs, as stellar mass appears in the definition of sSFR.

Fig.~\ref{fig:eddington_bias_n_passive} demonstrates that the number density of massive quiescent galaxies is highly sensitive to the added lognormal scatter, with larger standard deviations of the scatter generally yielding higher $n_{\rm q}$ at fixed redshift and stellar mass bin. The effect strengthens at higher redshifts and/or for higher stellar mass bins. For instance, adopting $\sigma_{\rm random}=0.5$~dex increases $n_{\rm q}$ in the $M_* > 10^{10}~\mathrm{M_\odot}$ and $M_* > 10^{11}~\mathrm{M_\odot}$ bins by more than an order of magnitude at $z > 4.5$ and $z > 2$, respectively. All observational data shown in Fig.~\ref{fig:eddington_bias_n_passive} can be matched within $\approx 1\sigma$ for at least one value of $\sigma_{\rm random}$ between $0$ and $0.5$~dex.

\section{Binned galaxy stellar mass function from COLIBRE}
\label{appendix: binned_gsmf}

To facilitate comparisons between \colibre{} and other studies, Tables \ref{table: gsmf_lowz} and \ref{table: gsmf_highz} present the binned \gsmf{} values from the \colibre{} L200m6 simulation with thermal AGN feedback (the \msixcolor{} lines in Figs. \ref{fig:gsmf_evolution_lowz} and \ref{fig:gsmf_evolution_highz}). The left-most column lists the central values of $0.2$~dex stellar mass bins, while the remaining columns provide \gsmf{} values at different redshifts: from $z=0$ to $z=7$ in Table \ref{table: gsmf_lowz} and from $z=8$ to $z=17$ in Table \ref{table: gsmf_highz}. These values account for Eddington bias as the \gsmf{} is calculated
based on galaxy masses including random errors, as described by equation~(\ref{eq: random_scatter}). 

\begin{table*}
\caption{The binned \gsmf{} from the \colibre{} simulation in the ($200$ cMpc)$^3$ volume at m6 resolution with thermal AGN feedback. The left-most column shows the central values of $0.2$ dex stellar mass bins, while the remaining columns present \gsmf{} values at redshifts from $z=0$ to $z=7$. These values account for Eddington bias by the inclusion of lognormal random errors in galaxy stellar masses, as described by equation~(\ref{eq: random_scatter}).}
	\centering
	\begin{tabular}{lrrrrrrrrrrrrrr} % four columns, alignment for each
   \hline
   $\log_{10}(M_* / \mathrm{M_\odot})$ & \multicolumn{10}{c}{ $\log_{10} (\Phi / [\mathrm{cMpc^{-3} dex^{-1}}])$}   \\
	    \hline
 & $z=0.0$  & $z=0.1$  & $z=0.5$  & $z=1$ & $z=2$ & $z=3$ & $z=4$ & $z=5$ & $z=6$ & $z=7$ \\ \hline 
 6.5 & $-0.28$ & $-0.26$ & $-0.18$ & $-0.05$ & $0.19$ & $0.29$ & $0.28$ & $0.19$ & $0.06$ & $-0.09$ \\ 
 6.7 & $-0.53$ & $-0.51$ & $-0.43$ & $-0.33$ & $-0.08$ & $0.01$ & $-0.01$ & $-0.09$ & $-0.21$ & $-0.36$ \\ 
 6.9 & $-0.73$ & $-0.71$ & $-0.64$ & $-0.55$ & $-0.35$ & $-0.30$ & $-0.34$ & $-0.44$ & $-0.56$ & $-0.71$ \\ 
 7.1 & $-0.90$ & $-0.88$ & $-0.82$ & $-0.74$ & $-0.59$ & $-0.58$ & $-0.65$ & $-0.77$ & $-0.92$ & $-1.09$ \\ 
 7.3 & $-1.06$ & $-1.04$ & $-0.98$ & $-0.91$ & $-0.79$ & $-0.81$ & $-0.90$ & $-1.06$ & $-1.24$ & $-1.45$ \\ 
 7.5 & $-1.21$ & $-1.19$ & $-1.13$ & $-1.06$ & $-0.97$ & $-1.00$ & $-1.12$ & $-1.30$ & $-1.52$ & $-1.77$ \\ 
 7.7 & $-1.33$ & $-1.32$ & $-1.26$ & $-1.20$ & $-1.12$ & $-1.18$ & $-1.31$ & $-1.51$ & $-1.75$ & $-2.03$ \\ 
 7.9 & $-1.43$ & $-1.42$ & $-1.37$ & $-1.32$ & $-1.27$ & $-1.34$ & $-1.48$ & $-1.70$ & $-1.96$ & $-2.26$ \\ 
 8.1 & $-1.53$ & $-1.52$ & $-1.48$ & $-1.44$ & $-1.40$ & $-1.48$ & $-1.65$ & $-1.87$ & $-2.14$ & $-2.46$ \\ 
 8.3 & $-1.63$ & $-1.62$ & $-1.58$ & $-1.55$ & $-1.53$ & $-1.63$ & $-1.79$ & $-2.03$ & $-2.33$ & $-2.67$ \\ 
 8.5 & $-1.72$ & $-1.71$ & $-1.68$ & $-1.66$ & $-1.65$ & $-1.74$ & $-1.92$ & $-2.17$ & $-2.48$ & $-2.86$ \\ 
 8.7 & $-1.80$ & $-1.79$ & $-1.77$ & $-1.75$ & $-1.74$ & $-1.85$ & $-2.05$ & $-2.32$ & $-2.67$ & $-3.03$ \\ 
 8.9 & $-1.86$ & $-1.85$ & $-1.83$ & $-1.82$ & $-1.82$ & $-1.94$ & $-2.17$ & $-2.47$ & $-2.84$ & $-3.25$ \\ 
 9.1 & $-1.91$ & $-1.90$ & $-1.88$ & $-1.86$ & $-1.89$ & $-2.04$ & $-2.31$ & $-2.64$ & $-3.04$ & $-3.44$ \\ 
 9.3 & $-1.94$ & $-1.95$ & $-1.91$ & $-1.91$ & $-1.96$ & $-2.16$ & $-2.46$ & $-2.81$ & $-3.22$ & $-3.67$ \\ 
 9.5 & $-1.99$ & $-1.98$ & $-1.96$ & $-1.96$ & $-2.05$ & $-2.28$ & $-2.62$ & $-3.00$ & $-3.41$ & $-3.92$ \\ 
 9.7 & $-2.02$ & $-2.02$ & $-2.00$ & $-2.01$ & $-2.15$ & $-2.43$ & $-2.76$ & $-3.18$ & $-3.61$ & $-4.17$ \\ 
 9.9 & $-2.05$ & $-2.05$ & $-2.05$ & $-2.09$ & $-2.26$ & $-2.56$ & $-2.94$ & $-3.37$ & $-3.85$ & $-4.26$ \\ 
 10.1 & $-2.07$ & $-2.08$ & $-2.12$ & $-2.17$ & $-2.38$ & $-2.70$ & $-3.10$ & $-3.58$ & $-4.04$ & $-4.65$ \\ 
 10.3 & $-2.13$ & $-2.13$ & $-2.19$ & $-2.26$ & $-2.51$ & $-2.87$ & $-3.30$ & $-3.76$ & $-4.28$ & $-4.97$ \\ 
 10.5 & $-2.24$ & $-2.25$ & $-2.29$ & $-2.38$ & $-2.65$ & $-3.05$ & $-3.53$ & $-4.01$ & $-4.62$ & $-5.20$ \\ 
 10.7 & $-2.43$ & $-2.44$ & $-2.46$ & $-2.54$ & $-2.85$ & $-3.29$ & $-3.77$ & $-4.31$ & $-4.90$ & $-5.73$ \\ 
 10.9 & $-2.70$ & $-2.70$ & $-2.72$ & $-2.81$ & $-3.09$ & $-3.55$ & $-4.10$ & $-4.67$ & $-5.36$ & $-$  \\ 
 11.1 & $-3.00$ & $-2.99$ & $-3.05$ & $-3.16$ & $-3.43$ & $-3.93$ & $-4.51$ & $-4.88$ & $-$  & $-5.90$ \\ 
 11.3 & $-3.32$ & $-3.35$ & $-3.44$ & $-3.58$ & $-3.81$ & $-4.38$ & $-4.88$ & $-5.43$ & $-6.20$ & $-$  \\ 
 11.5 & $-3.72$ & $-3.73$ & $-3.87$ & $-4.07$ & $-4.22$ & $-4.88$ & $-5.73$ & $-6.20$ & $-$  & $-$  \\ 
 11.7 & $-4.23$ & $-4.26$ & $-4.45$ & $-4.66$ & $-4.81$ & $-5.36$ & $-6.20$ & $-$  & $-$  & $-$  \\ 
 11.9 & $-4.82$ & $-4.88$ & $-5.36$ & $-5.30$ & $-5.36$ & $-6.20$ & $-6.20$ & $-$  & $-$  & $-$  \\ 
 12.1 & $-6.20$ & $-5.73$ & $-6.20$ & $-5.90$ & $-6.20$ & $-6.20$ & $-$  & $-$  & $-$  & $-$  \\ 
    \hline
\end{tabular}
\label{table: gsmf_lowz}
\end{table*}

\begin{table*}
\caption{As Table \ref{table: gsmf_lowz}, but at redshifts ranging from $z=8$ to $z=17$.}
	\centering
	\begin{tabular}{lrrrrrrrrrrrrrr} % four columns, alignment for each
   \hline
   $\log_{10}(M_* / \mathrm{M_\odot})$ & \multicolumn{10}{c}{ $\log_{10} (\Phi / [\mathrm{cMpc^{-3} dex^{-1}}])$}   \\
	    \hline
  & $z=8$ & $z=9$ & $z=10$ & $z=11$ & $z=12$ & $z=13$ & $z=14$ & $z=15$ & $z=16$ & $z=17$ \\ \hline 
 6.5 & $-0.30$ & $-0.53$ & $-0.80$ & $-1.11$ & $-1.45$ & $-1.83$ & $-2.24$ & $-2.69$ & $-3.18$ & $-3.68$ \\ 
 6.7 & $-0.55$ & $-0.77$ & $-1.04$ & $-1.33$ & $-1.67$ & $-2.03$ & $-2.44$ & $-2.90$ & $-3.36$ & $-3.89$ \\ 
 6.9 & $-0.89$ & $-1.11$ & $-1.37$ & $-1.65$ & $-1.98$ & $-2.32$ & $-2.72$ & $-3.15$ & $-3.66$ & $-4.10$ \\ 
 7.1 & $-1.28$ & $-1.51$ & $-1.76$ & $-2.04$ & $-2.36$ & $-2.68$ & $-3.08$ & $-3.52$ & $-3.98$ & $-4.55$ \\ 
 7.3 & $-1.68$ & $-1.93$ & $-2.21$ & $-2.49$ & $-2.80$ & $-3.13$ & $-3.53$ & $-3.95$ & $-4.42$ & $-4.88$ \\ 
 7.5 & $-2.04$ & $-2.33$ & $-2.64$ & $-2.98$ & $-3.31$ & $-3.64$ & $-3.99$ & $-4.52$ & $-5.09$ & $-5.12$ \\ 
 7.7 & $-2.32$ & $-2.66$ & $-3.03$ & $-3.39$ & $-3.81$ & $-4.10$ & $-4.57$ & $-5.16$ & $-5.30$ & $-$  \\ 
 7.9 & $-2.59$ & $-2.95$ & $-3.39$ & $-3.77$ & $-4.21$ & $-4.62$ & $-5.16$ & $-5.36$ & $-5.90$ & $-$  \\ 
 8.1 & $-2.81$ & $-3.25$ & $-3.64$ & $-4.05$ & $-4.59$ & $-4.97$ & $-5.43$ & $-6.20$ & $-5.90$ & $-6.20$ \\ 
 8.3 & $-3.06$ & $-3.45$ & $-3.91$ & $-4.43$ & $-5.06$ & $-5.73$ & $-6.20$ & $-6.20$ & $-$  & $-$  \\ 
 8.5 & $-3.25$ & $-3.70$ & $-4.22$ & $-4.65$ & $-5.20$ & $-5.90$ & $-6.20$ & $-6.20$ & $-$  & $-6.20$ \\ 
 8.7 & $-3.48$ & $-3.93$ & $-4.42$ & $-4.90$ & $-5.51$ & $-5.73$ & $-$  & $-$  & $-$  & $-$  \\ 
 8.9 & $-3.66$ & $-4.21$ & $-4.74$ & $-5.30$ & $-5.73$ & $-$  & $-$  & $-$  & $-$  & $-$  \\ 
 9.1 & $-3.92$ & $-4.52$ & $-5.03$ & $-5.90$ & $-$  & $-$  & $-6.20$ & $-$  & $-$  & $-$  \\ 
 9.3 & $-4.23$ & $-4.76$ & $-5.25$ & $-5.90$ & $-6.20$ & $-$  & $-$  & $-$  & $-$  & $-$  \\ 
 9.5 & $-4.35$ & $-4.93$ & $-5.51$ & $-6.20$ & $-6.20$ & $-$  & $-$  & $-$  & $-$  & $-$  \\ 
 9.7 & $-4.61$ & $-5.36$ & $-6.20$ & $-$  & $-$  & $-$  & $-$  & $-$  & $-$  & $-$  \\ 
 9.9 & $-5.12$ & $-5.51$ & $-6.20$ & $-6.20$ & $-$  & $-$  & $-$  & $-$  & $-$  & $-$  \\ 
 10.1 & $-5.60$ & $-$  & $-$  & $-$  & $-$  & $-$  & $-$  & $-$  & $-$  & $-$  \\ 
 10.3 & $-5.51$ & $-6.20$ & $-$  & $-$  & $-$  & $-$  & $-$  & $-$  & $-$  & $-$  \\ 
 10.5 & $-5.90$ & $-6.20$ & $-$  & $-$  & $-$  & $-$  & $-$  & $-$  & $-$  & $-$  \\ 
 10.7 & $-6.20$ & $-$  & $-$  & $-$  & $-$  & $-$  & $-$  & $-$  & $-$  & $-$  \\ 
 \hline
\end{tabular}
\label{table: gsmf_highz}
\end{table*}

\section{The impact of particle selection on the CSMD and CSFRD}
\label{appendix: particle_selection}

\begin{figure*}
    \centering
    \includegraphics[width=0.49\textwidth]{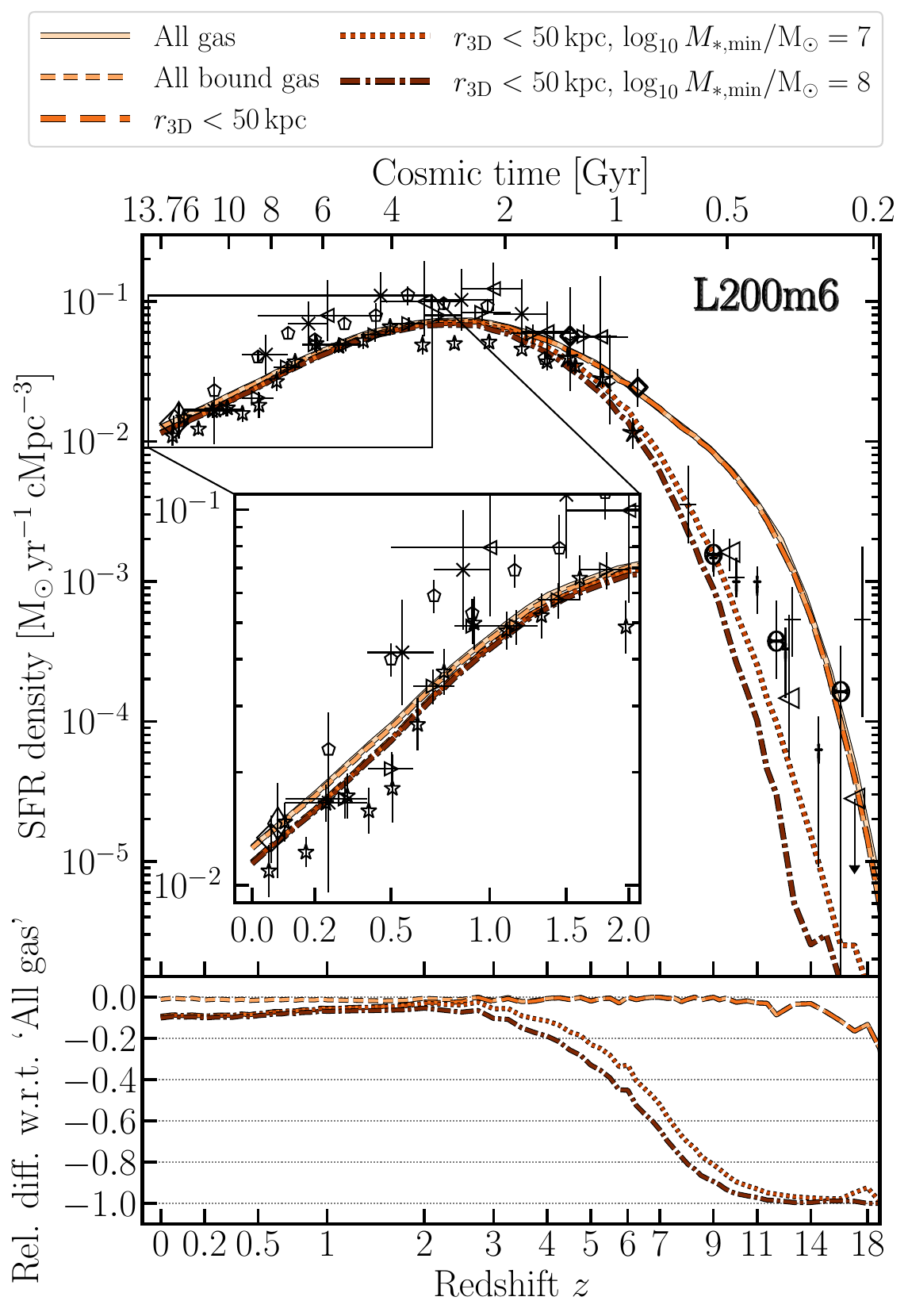}
    \includegraphics[width=0.49\textwidth]{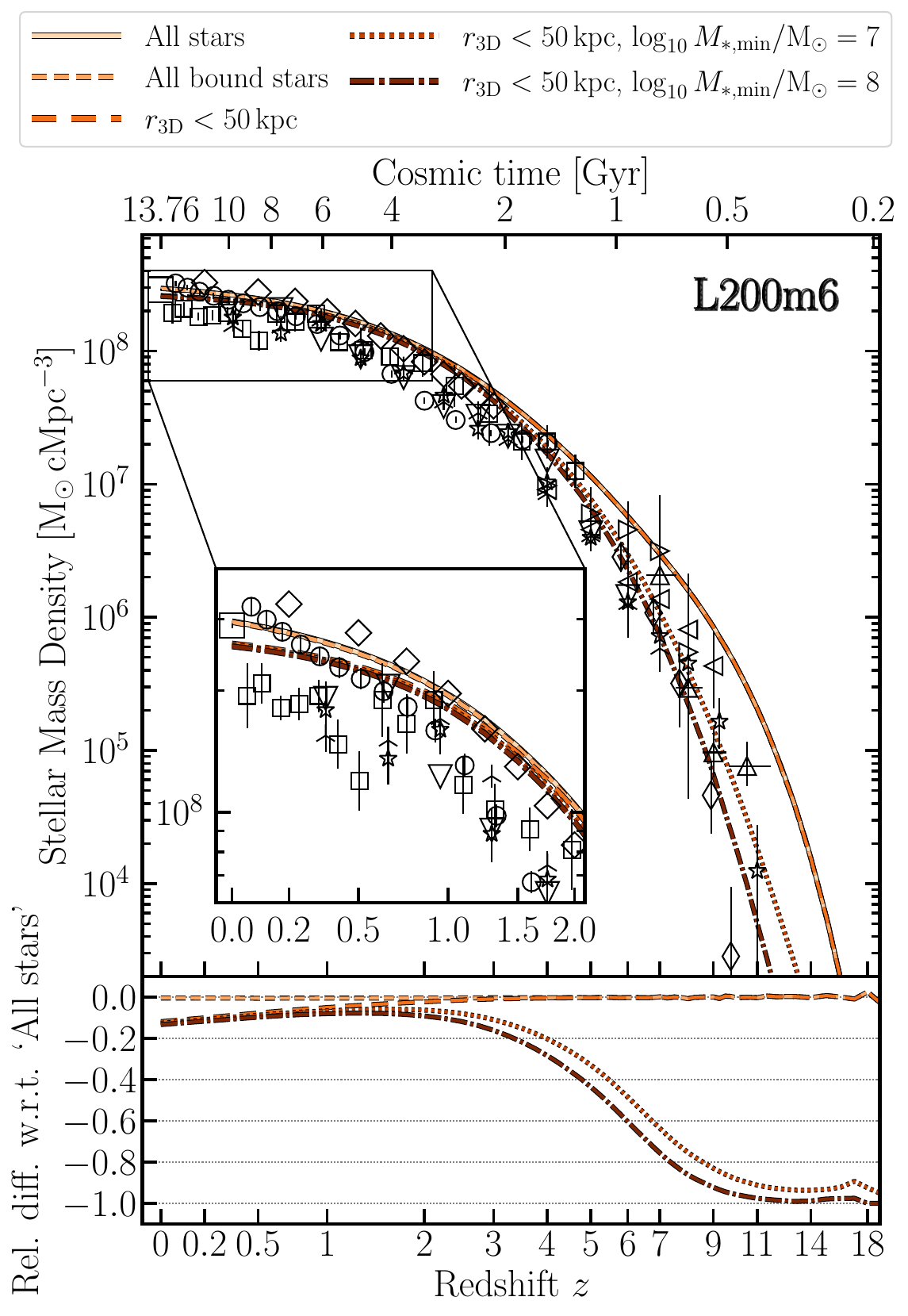}
    \caption{Evolution of the cosmic star formation rate density (CSFRD; \textit{left}) and cosmic stellar mass density (CSMD; \textit{right}) in the \colibre{} L200m6 simulation, calculated using different particle selection criteria (lines with varying styles and colours). The CSFRD (CSMD) based on all gas (stellar) particles in the simulated volume is shown by the solid lines, while the short-dashed lines indicate the selection of all bound gas (stellar) particles. The long-dashed, dotted, and dash-dotted lines correspond to selections that restrict the calculation of the CSFRD and CSMD to particles that are both bound and within 50 pkpc 3D apertures, with the dotted and dash-dotted lines further limiting the calculations to subhaloes with $M_* \geq 10^7~\mathrm{M_\odot}$ and $M_* \geq 10^8~\mathrm{M_\odot}$, respectively. The layouts of the left- and right-hand panels match those in Figs. \ref{fig:CSFRD_evolution} and \ref{fig:CSMD_evolution}, respectively, including the same set of comparison data (black symbols). Additionally, an inset in the bottom left corner shows a zoom-in on the evolution of CSFRD and CSMD at $0 < z <2$. The restrictions to subhaloes with $M_* \geq 10^7~\mathrm{M_\odot}$ and $M_* \geq 10^8~\mathrm{M_\odot}$ result in significant, though similar, differences at $z > 4$, while the 50 pkpc aperture constraint has a small but non-negligible impact at $z < 2$.}
    \label{fig:csfrd_csmd_with_different_selection} 
\end{figure*}

Here we examine how different particle selection criteria affect the CSFRD and CSMD predicted by the simulations. Fig.~\ref{fig:csfrd_csmd_with_different_selection} presents the evolution of CSFRD and CSMD in the \colibre{} L200m6 simulation. The left- and right-hand panels display, respectively, the CSFRD and CSMD computed using the following criteria:

\begin{enumerate}
    \item CSFRD and CSMD based on all star-forming gas particles and all stellar particles in the simulated volume, respectively (solid curves);
    \item CSFRD and CSMD based on only those particles that are bound to subhaloes (short-dashed curves);
    \item as (ii) but the particles (gas for CSFRD and stars for CSMD) are additionally restricted to lie within 50 pkpc 3D apertures (long-dashed curves);
    \item as (iii) but only subhaloes with stellar mass $M_* \geq 10^7~\mathrm{M_\odot}$ contribute to the CSFRD and CSMD (dotted curves);
    \item as (iii) but only subhaloes with stellar mass of $M_* \geq 10^8~\mathrm{M_\odot}$ contribute to the CSFRD and CSMD (dash-dotted curves).
\end{enumerate}
We note that of these, criteria (i) and (iv) were used in Fig.~\ref{fig:CSFRD_evolution}, and criteria (i) and (v) in Fig.~\ref{fig:CSMD_evolution}.

Fig.~\ref{fig:csfrd_csmd_with_different_selection} shows that different particle selection criteria influence the CSFRD and CSMD in various ways. First, the difference between the CSFRD (CSMD) computed using all gas (stellar) particles and that based only on bound gas (stellar) particles remains negligible across all redshifts. Second, restricting gas (stellar) particles to lie within $50$ pkpc apertures reduces the CSFRD (CSMD) by $\approx 10$ per cent at $z=0$, compared to using all bound particles. This difference diminishes with increasing redshift and becomes negligible at $z>3$ for both the CSFRD and CSMD. Finally, restricting the CSFRD and CSMD to subhaloes with $M_* \geq 10^7~\mathrm{M_\odot}$ and $M_* \geq 10^8~\mathrm{M_\odot}$, which is relevant for comparisons with high-redshift observations, has little impact at $0 < z < 3$, but becomes increasingly significant at higher redshifts. The choice of minimum stellar mass threshold -- $M_* \geq 10^7~\mathrm{M_\odot}$ versus $M_* \geq 10^8~\mathrm{M_\odot}$ -- has a negligible effect below $z=4$, but becomes progressively more important at higher $z$: with the differences in the CSFRD (CSMD) increasing from $\approx 0.06$ ($\approx 0.07$)~dex at $z=5$ to $\approx 0.5$ ($\approx 0.6$)~dex by $z=12$.

\section{Stellar mass density from subhaloes in different mass bins}
\label{appendix: SMD_per_mass}

\begin{figure*}
    \centering
    \includegraphics[width=0.99\textwidth]{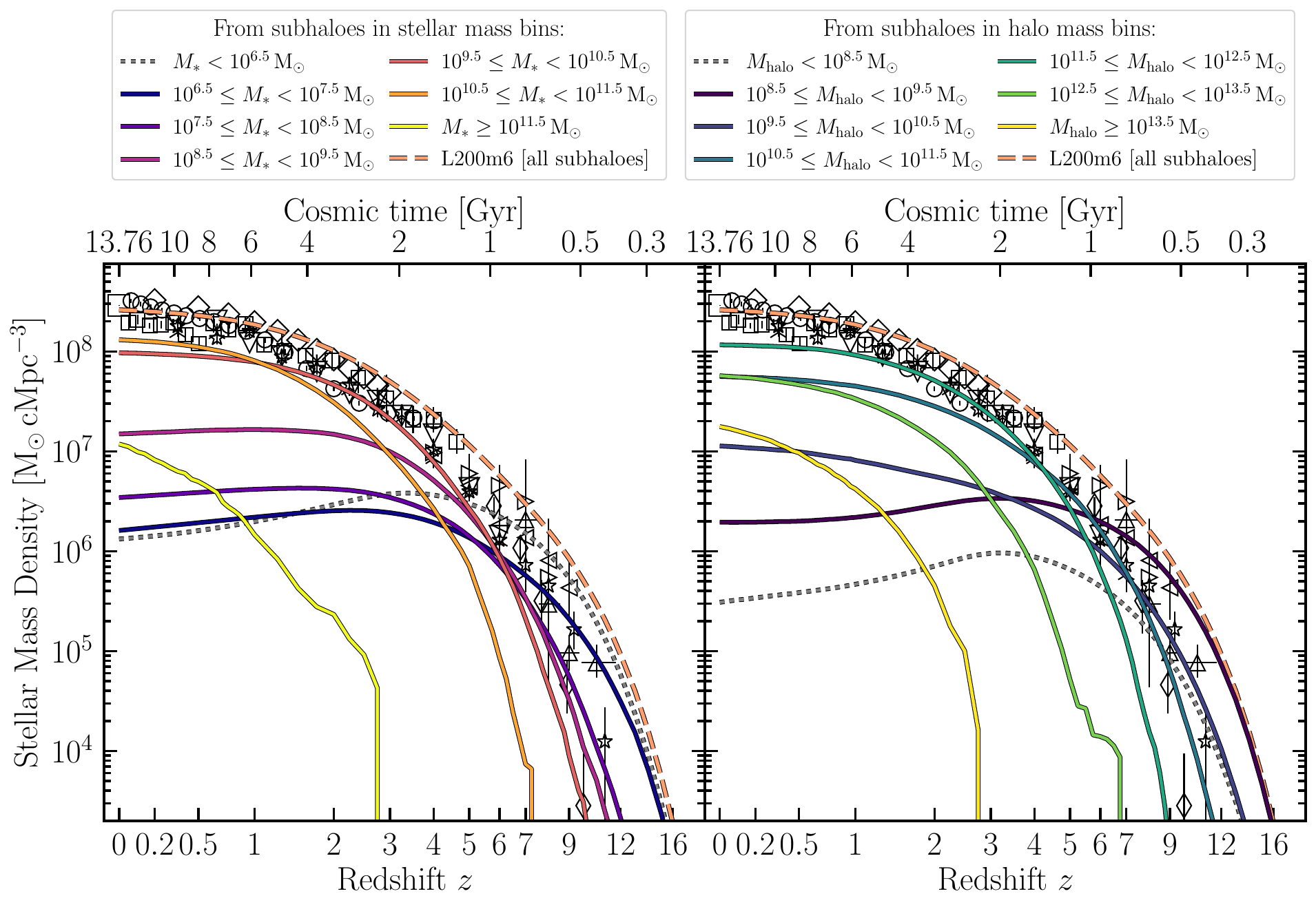}
    \caption{As Fig.~\ref{fig:SFH_per_mass}, but showing the cosmic stellar mass density (CSMD) instead of the cosmic star formation rate density. The observational data (black symbols) are as in Fig. \ref{fig:CSMD_evolution}. Between $0 < z < 3$, the CSMD is dominated by galaxies with stellar masses of $M_* \sim 10^{10}-10^{11}~\mathrm{M_\odot}$. At higher redshifts ($3 < z < 5$), galaxies with $M_* \sim 10^9 \, \rm M_\odot$ begin to contribute significantly, and by $z > 5$, the largest contribution comes from the lowest-mass systems ($M_* \sim 10^6-10^7~\mathrm{M_\odot}$).}
    \label{fig:SMD_per_mass}
\end{figure*}

Fig.~\ref{fig:SMD_per_mass} shows the CSMD in the L200m6 simulation, decomposed into contributions from subhaloes in different mass bins. It follows the same structure as Fig.~\ref{fig:SFH_per_mass}, but displays the CSMD instead of the CSFRD. The left (right) panel shows the decomposition by stellar (halo) mass. The CSMD in each mass bin is computed by summing the masses of individual stellar particles that are gravitationally bound to subhaloes and located within 50~pkpc.

The $0<z<3$ CSMD is dominated by $M_* \sim 10^{10}-10^{11}~\mathrm{M_\odot}$ galaxies. From $z=3$ to $5$, $M_* \sim 10^{9}~\mathrm{M_\odot}$ galaxies also make a significant contribution to the total CSMD, and at $z>5$, the CSMD is dominated by the least massive objects ($M_* \lesssim 10^{6.5}~\mathrm{M_\odot}$, comparable to the mass of a few stellar particles at m6 resolution). By contrast, the galaxies with $M_* \gtrsim 10^{11.5}~\mathrm{M_\odot}$ never dominate the CSMD. 

In terms of halo mass, subhaloes near the peak of the \shmr{}, with $M_{\rm halo} \sim 10^{12}~\mathrm{M_\odot}$, dominate the CSMD at $0 < z < 3$. Between $z = 3$ and $5$, subhaloes with $M_{\rm halo} \sim 10^{11}~\mathrm{M_\odot}$ become similarly dominant. At $z > 5$, the CSMD is primarily contributed by $M_{\rm halo} \sim 10^{9}~\mathrm{M_\odot}$ subhaloes. Objects with $M_{\rm halo} \lesssim 10^{8.5}~\mathrm{M_\odot}$ and those with $M_{\rm halo} \gtrsim 10^{13.5}~\mathrm{M_\odot}$ never dominate the CSMD.

At $z=0$, about $50$ per cent of the CSMD is found in $10^{10.5} < M_* / \mathrm{M_\odot} < 10^{11.5}$ galaxies, $\approx 37$ per cent in $10^{9.5} < M_* / \mathrm{M_\odot} < 10^{10.5}$ galaxies, $10^{8.5} < M_* / \mathrm{M_\odot} < 10^{9.5}$ and $M_* > 10^{11.5}~ \mathrm{M_\odot}$ galaxies contribute $\approx 6$~per cent and $\approx 5$~per cent, respectively, and the remaining $\approx 2$~per cent of the CSMD is in $M_* < 10^{8.5}~\mathrm{M_\odot}$ objects. In terms of halo mass, $\approx 45$~per cent of the $z=0$ CSMD resides in subhaloes with $10^{11.5} < M_{\rm halo} / \mathrm{M_\odot} < 10^{12.5}$, while those with $10^{12.5} < M_{\rm halo} / \mathrm{M_\odot} < 10^{13.5}$ and $10^{10.5} < M_{\rm halo} / \mathrm{M_\odot} < 10^{11.5}$ each contribute about 22~per cent. The most massive subhaloes, in the mass bin $M_{\rm halo} > 10^{13.5}~\mathrm{M_\odot} $, contribute only $\approx 7$~per cent to the $z=0$ CSMD. Finally, subhaloes with $10^{9.5} < M_{\rm halo} / \mathrm{M_\odot} < 10^{10.5}$ account for $\approx 4$~per cent, while those with $M_{\rm halo} < 10^{9.5}~\mathrm{M_\odot}$ contribute the remaining fraction (less than $\approx 1$~per cent).

%%%%%%%%%%%%%%%%%%%%%%%%%%%%%%%%%%%%%%%%%%%%%%%%%%

% Don't change these lines
\bsp	% typesetting comment
\label{lastpage}
\end{document}